\documentclass[pdflatex,sn-mathphys-num]{sn-jnl}
\usepackage{graphicx}%
\usepackage{multirow}%
\usepackage{amsmath,amssymb,amsfonts}%
\usepackage{amsthm}%
\usepackage{mathrsfs}%
\usepackage{textcomp}%
\usepackage[title]{appendix}%
\usepackage{xcolor}%
\usepackage{manyfoot}%
\usepackage{booktabs}%
\usepackage[sort&compress]{natbib}
\usepackage{placeins}
\usepackage{multirow}

\makeatletter
\newcommand*{\rom}[1]{\expandafter\@slowromancap\romannumeral #1@}
\makeatother
\usepackage{enumitem}
\usepackage{listings}%
\usepackage{xcolor} 
\usepackage{comment}
\usepackage{algorithm}%
\usepackage{algorithmicx}%
\usepackage{algpseudocode}%
\usepackage{mathtools}
\usepackage{hyperref}    % load once, near the end
\hypersetup{
  colorlinks=true,
  linkcolor=black,   % ToC, figure refs, etc.
  citecolor=blue,   % bibliography citations
  urlcolor=blue     % URLs
}

\usepackage{listings}%
\definecolor{codegreen}{rgb}{0,0.6,0}
\definecolor{codegray}{rgb}{0.5,0.5,0.5}
\definecolor{codepurple}{rgb}{0.58,0,0.82}
\definecolor{backcolour}{rgb}{0.95,0.95,0.92}
\lstdefinestyle{mystyle}{
    backgroundcolor=\color{backcolour},   
    commentstyle=\color{codegreen},
    keywordstyle=\color{magenta},
    numberstyle=\tiny\color{codegray},
    stringstyle=\color{codepurple},
    basicstyle=\ttfamily\footnotesize,
    breakatwhitespace=false,         
    breaklines=true,                 
    captionpos=b,                    
    keepspaces=true,                 
    numbers=left,                    
    numbersep=5pt,                  
    showspaces=false,                
    showstringspaces=false,
    showtabs=false,                  
    tabsize=2
}
\title[Modularity maximization and community detection in complex networks through recursive and hierarchical annealing in the DWAVE Advantage quantum processing units]{Modularity maximization and community detection in complex networks through recursive and hierarchical annealing in the DWAVE Advantage quantum processing units}

\author*[1]{\fnm{Joan} \sur{Falc\'o-Roget}}\email{joan.falcoroget@gmail.com, kzajac@agh.edu.pl}

\author[2,3]{\fnm{Kacper} \sur{Jurek}}
\equalcont{These authors contributed equally to this work.}

\author[1,2]{\fnm{Barbara} \sur{Wojtarowicz}}
\equalcont{These authors contributed equally to this work.}

\author[1,2]{\fnm{Karol} \sur{Capa{\l}a}}

\author*[1,2,3]{\fnm{Katarzyna} \sur{Rycerz}}

\affil[1]{\orgname{Computational Neuroscience Group, Sano Centre for Computational Medicine}, \orgaddress{\street{Czarnowiejska 36}, \city{Krakow}, \postcode{30-054}, \country{Poland}}}

\affil[2]{ \orgname{AGH University of Krakow, Faculty of Computer Science}, \orgaddress{\street{Mickiewicza 30}, \city{Krak\'ow}, \postcode{30-059}, \country{Poland}}}

\affil[3]{\orgdiv{Quantum Computing Lab}, \orgname{Academic Computer Center Cyfronet AGH}, \orgaddress{\street{Nawojki 11}, \city{Krakow}, \postcode{30-950}, \country{Poland}}}

\abstract{Quantum adiabatic optimization has long been expected to outperform classical methods in solving NP-type problems. While this has been proven in certain experiments, its main applications still reside in academic problems where the size of the system to be solved would not represent an obstacle to any modern desktop computer. Here, we develop a systematic procedure to find the global optima of the modularity function to discover community structure in complex networks solely relying on pure annealers rather than hybrid solutions. We bypass the one-hot encoding constraints by hierarchically and recursively encoding binary instances of the problem that can be solved without the need to guess the exact penalties for the Lagrange multipliers. We study the variability and robustness of the annealing process as a function of network size, directness of connections, topology, and the resolution of the communities. We show how our approach produces meaningful and at least equally optimal solutions to state-of-the-art community detection algorithms while maintaining tractable and potentially scalable computing times. Lastly, due to its recursive nature, the annealing process returns intermediate subdivisions, thus offering interpretable rather than black-box solutions. These \textit{dendrograms} can be used to unveil normal and pathological hidden hierarchies in brain networks, hence opening the door to clinical workflows. Overall, this represents a first step towards an applicable practice-oriented usage of pure quantum annealing, potentially bridging two segregated communities in modern science and engineering: that of network science and quantum computing.}


\begin{document}
\maketitle

\clearpage
{\small
\tableofcontents
}
{\small
\listoffigures
}
{\small
\listoftables
}

\clearpage
\section{Introduction}
Real-life systems are composed of multiple individual components whose interactions, albeit generally describable in simple terms, lead to the emergence of non-trivial phenomena. These complex systems can be effectively modeled using complex networks composed of nodes and edges or links between them. A large body of work has been addressed to formally characterize these networks, focusing on both structural properties and the dynamic processes they support. Unraveling community structure in complex networks has proven to be a crucial problem for the accurate understanding of multiple real-life systems (e.g., neurological disorders). Community structure, as understood by the existence of highly connected groups of nodes, is usually attained by maximizing a modularity function \cite{Newman-Girvan_2004} that quantifies the connectivity between intra- and inter-modular nodes in the network \cite{Girvan2002}. 

\subsection{Related works (I): Algorithms to maximize the modularity}
The modularity function elegantly maps to an infinite-range spin-glass system \cite{Newman2006,Reichardt2006}. This caused a rapid emergence of methods based on statistical physics \cite{Massen2005,Reichardt2006}, and simulated annealing was used to illustrate the presence of close maxima in the modularity function~\cite{Guimera2004}. However, the most common solutions rely on mathematical, statistical, and computer science principles \cite{Girvan2002,Newman-Girvan_2004,Newman2004,Duch2005,Ziv2005,Newman2006,Blondel2008,Traag2019}, where the use of carefully thought-out heuristics allowed the algorithms to achieve competitive scores and outstanding computing times \cite{Danon2005}. Yet many of these approaches do not always return community structures with the highest modularity \cite{aref2022,aref2023,Good2010}. In fact, identifying community structure via modularity maximization is not always optimal~\cite{Rosvall2008,Peixoto2014,Damle2018,Ghasemian2020}, although in certain settings these approaches are formally equivalent to modularity maximization~\cite{Bickel2009,Zhao2012,Newman2016}. While we consider these alternative descriptions, we focus on maximizing the modularity.

\subsection{Related works (II): Maximizing the modularity with quantum annealing}
Crucially, maximizing the modularity is an NP-hard problem \cite{Brandes2008}, a scenario where quantum annealing (QA) is expected to excel at \cite{Farhi2001}. QA is grounded in the theoretical principle that, under ideal adiabatic conditions, the quantum processor will converge to the global minimum of the problem's energy landscape \cite{Johnson2011,Rajak2022}. This represents a solid argument to deviate from purely algorithmic attempts to find a community structure in complex networks. 

Unfortunately, one of the main obstacles to QA stems from the fact that qubits are necessarily represented with binary variables of a quadratic cost function. Thus, solving problems with more than two states (or assignments) requires specific encodings and constraints assuring the correctness of the solution~\cite{glover_quantum_2022,chancellor_domain_2019}. For example, one-hot encoding, while popular, needs constraints to ensure each variable takes exactly one value from its possible states ~\cite{Ushijima-Mwesigwa2017,Negre2020}. Then, the penalty weight is added through Lagrange multipliers, whose values must be empirically tuned for each specific problem ~\cite{mayowa_2022}. 

An alternative is to depart from the binary quadratic formulation and use a discrete quadratic model to encode several classes in a $q$-state Potts system~\cite{Wierzbinski2023}. While successful, this approach comes at a huge transparency cost, since the exact calculations are run entirely inside the D-Wave ecosystem and divided into classical and quantum processes in a manner that is not accessible or controllable to the user. 

\subsection{Summary of contributions}
In this work, we introduce a novel algorithm that leverages pure QA to maximize network modularity and uncover intrinsic community structures. The algorithm employs recursive subdivisions to detect an arbitrary number of communities, eliminating the need for one-hot encoding and its associated constraints. We extend the formalism in \cite{Newman2006} to a general case applicable to various types of networks, arbitrary values of the resolution parameter \cite{Reichardt2006}, and we show how the function to optimize via QA remains formally independent of all of these. Lastly, we rigorously evaluate the algorithm’s performance and scalability across different network topologies and modularity functions, highlighting its strengths compared to current state-of-the-art methods. These results mark a significant step toward the practical application of quantum computing in real-world scenarios, fostering greater integration across scientific disciplines.

\section{Materials and methods} \label{sec:methods}
\subsection{Community structure and modularity index}
A graph or network $G=(V,E)$ is a set of vertices or nodes $V$ and edges $E$. Moreover, a community in the graph is a subset of nodes $c\subseteq V$. A community structure is a set of non-overlapping communities $C=\{c_r | c_r \subseteq V, c_s \cap c_r=\varnothing, \text{ for } s \neq r \}$. The set of all possible community structures is denoted as  $\mathcal{C}$. Importantly, a community is not a subgraph as it doesn't require the deletion of inter-community edges. Lastly, the quality of a given community structure $C$ can be quantified with the modularity index \cite{Newman2004},
\begin{equation} \label{eq:modularity}
    Q(C) = \sum_{c \in C}\left[ \frac{L_c}{m} - \gamma \left(\frac{d_c}{2m}\right)^2\right],
\end{equation} where $L_c$ is the number of links between nodes in the $c$-th community, $d_c$ is the sum of the degrees of those nodes, and $m$ is the total number of links (or edges) in the network. Moreover, $\gamma$ is a resolution parameter that effectively accounts for different topological scales \cite{Reichardt2006} at which interactions between nodes happen \cite{Delvenne2010}. It also represents a resolution limit, where communities smaller than a given size cannot be detected by maximizing the modularity ~\cite{Fortunato2007,Kumpula2007}. A certain community structure is said to be modular if it exhibits a high modularity index when compared to multiple random instances of a given class of random networks (e.g., Erd\H{o}s-Renyi). 

Community detection algorithms aim to find a given community structure $C \in \mathcal{C}$ whose associated modularity function $Q$ lies at a global maximum -- or its opposite at a minimum, 
\begin{equation}\label{eq:min_mod}
    \hat{Q} \doteq \max_{\forall C \in \mathcal{C}} \left[Q(C)\right] =- \min_{\forall C \in \mathcal{C}} \left[-Q(C)\right].
\end{equation}
More precisely, $\hat{Q}$ is the modularity of the network, since it is obtained from the community structure that maximizes $Q$ in Eq. \ref{eq:modularity}. To maximize the modularity index, the first term in Eq. \ref{eq:modularity} suggests defining clusters containing a large number of edges. On the contrary, the second term aims to minimize the size of such clusters. This is important because any algorithm could be prone to isolating given nodes to boost the modularity index \cite{Wierzbinski2023}. Even though $\mathcal{C}$ is bounded, the number of possible elements it contains grows faster than exponentially with the size of the network \cite{Fortunato2007}.

Hereon, we denote a community structure of size $|C|=k$ as $C^k$. For a network of two communities (i.e., $k=2$), the modularity function in Eq. \ref{eq:modularity} can be mapped into an Ising system with variables $s_i = \{-1,1\}$ indicating that a certain node belongs to either one of the two communities present in the network~\cite{Newman2006,Reichardt2006}. For undirected networks, applying the transformation to Eq. \ref{eq:modularity} yields
\begin{equation} \label{eq:mod_ising_symm}
    Q(C^2) = \frac{1}{4m} \sum_{ij\in G} B_{ij} s_i s_j + \frac{(1-\gamma)}{2},
\end{equation} where
\begin{equation} \label{eq:und_mod_matrix}
    B_{ij} = A_{ij} - \gamma \frac{g_i g_j}{2m},
\end{equation} are the entries of the \textit{modularity} matrix, $A_{ij}$ are the entries of the symmetric adjacency matrix, $m = \frac{1}{2}\sum_{ij \in G}A_{ij}$ is the total edge-weight, and $g_i=\sum_j A_{ij}$ is the node weighted degree. For directed networks~\cite{Leicht2008}, applying the transformation to Eq. \ref{eq:modularity} yields
\begin{equation} \label{eq:mod_ising_dir}
    Q(C^2) = \frac{1}{2m} \sum_{ij\in G} B_{ij}^{d} s_i s_j + \frac{(1-\gamma)}{2},
\end{equation} where 
\begin{equation} \label{eq:dir_mod_matrix}
    B_{ij}^d = A_{ij} - \gamma \frac{g_i^{in} g_j^{out}}{m}
\end{equation} are the entries of the \textit{directed} modularity matrix, $A_{ij}$ are the entries of the asymmetric adjacency matrix,  $m=\sum_{ij}A_{ij}$ is the total edge-weight, $g_i^{in}=\sum_j A_{ij}$ is the node weighted in-degree, and $g_j^{out}=\sum_i A_{ij}$ is the node weighted out-degree. The superscript $d$ denotes a directed or asymmetric matrix. We detailed all the algebraic intermediate steps in the Supplementary Material. 

\subsection{Unconstrained hierarchical binary optimization}
To exploit QA without considering discrete variables and encodings, we wish to maximize the modularity by recursively subdividing each community into two. For that, let us define a process $\mathcal{P} \doteq (p_k)_{k=0}^{K}$ as a sequence of elements $p_k$ such that its series converges to a finite value,
\begin{equation} \label{eq:process_P}
    \hat{Q}^{\mathcal{P}} \doteq \sum_{k=1}^{K} p_k,
\end{equation} where $K$ is any positive integer indicating the number of detected communities. Minimizing $\hat{Q}^{\mathcal{P}}$ need not be straightforward, but optimizing each element is equivalent,
\begin{equation} \label{eq:min_process}
    \min \left[ \hat{Q}^{\mathcal{P}} \right] = \min\left[\sum_{k=1}^{K} p_k \right] = \sum_{k=1}^{K}  \min \left[ p_k \right]. 
\end{equation} Our heuristic is based on the additive nature of $Q$, where each community represents an isolated entity that independently contributes to its value \cite{Newman2004}. A natural way to exploit this property is to consider the increments in the modularity after recursive divisions~\cite{Newman2006}. Therefore, 
\begin{equation} \label{eq:p_k}
    p_k = -\left[ Q(C^{k+1}) - Q(C^{k}) \right]
\end{equation} where$Q(C^{k})$ is the modularity of a network with a community structure $C^k$ consisting of $k$ non-overlapping communities. It is straightforward to see that $\hat{Q}^\mathcal{P} \to Q(C^{K})$ with $C^{K}$ being a community structure of arbitrary size $K$. Recall that, from Eq. \ref{eq:modularity}, $C^1\equiv V$ and $Q(V)=0$.

Formally, there is no guarantee that a given \textit{optimized} process will univocally yield the partition with the highest modularity index in Eq. \ref{eq:min_mod}, i.e., $\min \left[ \hat{Q}^{\mathcal{P}} \right] \to \hat{Q}$. In what follows, we derive the Quadratic Unconstrained Binary Optimization (QUBO) cost function for each element of $\mathcal{P}$.

\subsection{Undirected networks}
From Eq. \ref{eq:modularity}, the modularity of an undirected network with $k$ communities can be decomposed as 
$$ Q(C^{k}) = \sum_{c \in C^{k-1}} \left[ \frac{L_c}{m} - \gamma \left(\frac{d_c}{2m}\right)^2\right] + \frac{L_k}{m} - \gamma \left(\frac{d_k}{2m}\right)^2. $$ Then, the modularity index of the same network after splitting the $k$-th community in two can also be decomposed as
\begin{align*}
Q(C^{k+1}) 
  &= \sum_{c \in C^{k-1}} 
     \left[ \frac{L_c}{m} - 
     \gamma \left(\frac{d_c}{2m}\right)^2 \right]\\
  &\quad \quad \quad \quad + 
     \sum_{r=1}^{2}\left[ \frac{L_{k_r}}{m} - \gamma \left(\frac{d_{k_r}}{2m}\right)^2 \right],
\end{align*}
where $k_1 \cup k_2 = k$. Using Eq. \ref{eq:mod_ising_symm} we obtain a compact expression for the elements of $\mathcal{P}$,
\begin{equation} \label{eq:pk_symm}
    \begin{split}
        p_k &= \frac{-1}{4m} \sum_{ij \in k} \left( B_{ij}s_i s_j - B_{ij}\right) \\
        &= \frac{-1}{4m} \sum_{ij \in k} \left( B_{ij} - \delta_{ij} \sum_{r\in k} B_{ir}\right)s_i s_j \\
        &= \frac{-1}{4m} \sum_{ij \in k} B_{ij}^{\text{k}} s_i s_j \\
        %&= \frac{-1}{m} \sum_{ij \in k} B_{ij}^{\text{k}} x_i x_j + \frac{1}{2m} \sum_{ij \in k} B_{ij}^{\text{k}} x_i + \frac{1}{2m} \sum_{ij \in k} B_{ij}^{\text{k}} x_j + \frac{1}{4m} \sum_{ij \in k} B_{ij}^{\text{k}} \\
        &= \frac{-1}{m} \sum_{ij \in k} B_{ij}^{\text{k}} x_i x_j,
    \end{split}
\end{equation} where we used $s_i^2=1$, the Kronecker delta function $\delta_{ij}$, and \textbf{B}\textsuperscript{$k$} is the generalized modularity matrix for the $k$-th community~\cite{Newman2006},
\begin{equation} \label{eq:gen_mod_matrix}
    B_{ij}^{\text{k}} = B_{ij} - \delta_{ij} \sum_{r\in k} B_{ir} \ \ \forall i,j \in k.
\end{equation} In the fourth row, we have applied a linear mapping from the Ising variables to a set of binary ones $s_i=2x_i-1$ to obtain the QUBO formulation. The last row follows from properties of the generalized modularity matrix(Supplementary Material S1).

Importantly, the first element of the process need not be considered separately. When $k=1$, taking into account that $x_i^2 = x_i$, $p_{k}$ reduces to the function to split the whole network in two, $$\frac{-1}{m} \left[ \sum_{ij\in G} B_{ij} x_i x_j + (1-\gamma)\sum_{i\in G} g_{i} x_i \right] = -Q(C^{2}) + 1 - \gamma.$$ The minimum of $p_{1}$ is independent of the terms  $1-\gamma$ and can be arbitrarily dropped; we only wrote them for completeness. Even more, $$\text{if } x_i=x_j \ \ \forall i,j \in k \implies p_k = 0,$$ which provides a natural stopping criterion for the hierarchical divisions without the need to predefine a maximum number of communities~\cite{Negre2020,Wierzbinski2023}. In practical terms, given that the network is symmetric, the summations don't need to cover all the $N^2$ entries but rather the diagonal and upper diagonal entries.

\subsection{Directed networks}
If the network has some degree of asymmetry, we could proceed identically as for undirected networks. The only difference would be the usage of the expressions for directed networks to obtain the elements $p_k$ of the process $\mathcal{P}$ (Supplementary Material S2) 

\begin{equation} \label{eq:pk_dir}
    p_k = \frac{-8}{m} \left[ \sum_{ij \in k} B_{ij}^{d,\text{k}} x_i x_j  - \frac{1}{2} \sum_{j\in k} x_j \sum_{i\in k} \left(B_{ij}^d - B_{ji}^d \right)\right].
\end{equation}
However, contrary to the undirected case, the first element of the process does not coincide with the modularity function to optimize for the full graph (i.e., $p_1 \neq Q(C^2)$). This is undesirable because it would mean that the first step of the hierarchy needs to be separated from the rest. Furthermore, the $ij$-th and $ji$-th entries would be associated with the product of binary variables $x_i x_j$. Then, the QPU would interpret the coupling of the two qubits as a sum of $B_{ij}^{d,\text{k}} + B_{ji}^{d,\text{k}}$, hence losing all the non-symmetric information. 

To address this, the two off-diagonal entries should be combined into a single descriptive interaction instead of two separate contributions. To leverage the desirable properties of the process \(\mathcal{P}\), we adopt an elegant solution from Leicht and Newman \cite{Leicht2008}. Notably, \(p_k\) in Eq. \ref{eq:p_k} is a scalar, so its transpose \(p_k^\top\) has the same global minima. Therefore, for directed networks, the linear combination \(\frac{1}{2} \left( p_k + p_k^\top \right)\) can be optimized recursively instead of optimizing \(p_k\) alone. In practice, this modification updates the generalized modularity matrix as follows:

\begin{equation} \label{eq:symmetrized_B_elements}
    B_{ij}^{\text{k}} = \frac{1}{2} \left[ B_{ij}^{d} + B_{ji}^{d} - \delta_{ij} \sum_{r \in k} \left( B_{ir}^{d} + B_{jr}^{d} \right) \right] \ \ \forall i,j \in k. 
\end{equation}

Importantly, all the desirable properties showed for the undirected scenario hold: 1) if $k=1$, the element \(\frac{1}{2} \left( p_1 + p_1^\top \right)\) reduces to the QUBO in Eq. \ref{eq:mod_ising_dir}; 2) $\text{if } x_i=x_j \ \ \forall i,j \in k \implies p_k = 0$; and 3) for undirected networks, this symmetrized version of the generalized modularity matrix reduces to Eq. \ref{eq:gen_mod_matrix}. Points 1 and 2 ensure that we remain within a QUBO formalism without linear terms and that the first binary division is obtained by optimizing the correct function. The latter point presents a convenient form from the software perspective.

\subsection{Algorithmic details}
We ran the hierarchical process $N_{runs}=20$ times independently (see Algorithm \ref{algo:HAnnealing}). Since, in practice, reaching $\max Q$ once is enough, this number ensures a competitive performance across all networks tested in this work (Fig. S2; see also Supplementary Material S3). More specifically, modeling each of the 20 outcomes as a single Bernoulli event, we observed probabilities of success higher than 40\%, and, in most cases, $\max Q$ was found with frequencies narrowing 100\%. In general, however, users should tailor this limit based on available computational resources and the behavior of the specific problem instance. As a rule of thumb, more runs translate into a more extensive exploration of the annealer's variability. In each of these runs, the returned modularity was kept only if its value was higher than the previously assigned one (line 11). The D-Wave Advantage sampler was initialized on every one of these runs. We provide a general description of the software package together with snippets of the high-level Python code developed in the Supplementary Material, enhancing the usability of quantum annealing resources for optimization purposes~\cite{lamza_qhyper_2024}.

As shown in the pseudocode for the Hierarchical annealing process in Algorithm \ref{algo:HAnnealing}, we have used a common recursive approach to divide and conquer types of algorithms. First, for each network, we computed a generalized modularity matrix (lines 20-21)  and built the appropriate QUBO (line 22) and loaded the corresponding clique embedding from the cache (line 23). Next, we sampled results from the quantum annealing process for that QUBO (line 24-26). In the {\tt Advantage\_system5.4} version of a D-Wave sampler, we chose the best result from $100$ different samples (i.e., \texttt{num\_reads}=100). The \texttt{annealing\_time} was set to $20.00 \pm 0.01 \ \mu s$, and the rest of the parameters were also left to their respective default values. Next, we recursively called the same procedure for both returned subcommunities unless one of them was empty (lines 27-34). 

\begin{algorithm} 
    \caption{Hierarchical quantum annealing (QA) for modularity maximization}
    \label{algo:HAnnealing}
    \begin{algorithmic}[1]
        \State $G(V,E) \leftarrow$ \texttt{Graph from a set of nodes and edges}   
        \State \texttt{Compute and cache clique embedding of size $\forall n\leq |V|$}
        \State $\gamma \leftarrow$ \texttt{Set the resolution parameter} 
        \State $N_{runs} \leftarrow$ \texttt{Number of modularity maximizations}
        \State \texttt{m $= 0$, $C=\varnothing$, dendrogram $=\varnothing$} 
        \\
        \State \textcolor{teal}{\# Repeated modularity maximization}
        \State \textbf{B}\texttt{ or} \textbf{B}\textsuperscript{\textit{d}}  \texttt{$\leftarrow$ Modularity matrix (Eqs. \ref{eq:und_mod_matrix}, \ref{eq:dir_mod_matrix})}   
        \For{\texttt{r$ \leq N_{runs}$}}
            \State \Call{$C^k$, $D$ $\leftarrow$ Binary\_\_QA}{\textbf{B} \texttt{or} \textbf{B}\textsuperscript{\textit{d}}, $V$, $\gamma$}
            \State \texttt{Compute $Q$ in Eq. \ref{eq:modularity} from $C^k$}
            \If{\texttt{m$\leq Q$}}
                \State \texttt{m, $C$, dendrogam $\leftarrow$ $Q$, $C^k$, $D$}
            \EndIf 
        \EndFor   
        \\
        \State \textcolor{teal}{\# Recursive quantum annealing routine  }
        \Procedure{Binary\_\_QA}{\textbf{B}, $V$, $\gamma$}            
            \State \textbf{B}\textsuperscript{$V$} \texttt{$\leftarrow$ Generalized Mod. matrix (Eq. \ref{eq:symmetrized_B_elements})}
            \State \texttt{QUBO $\leftarrow$ $k$-th element of $\mathcal{P}$ (Eq. \ref{eq:pk_symm})}
            \State \texttt{QUBO\textsubscript{E} $\leftarrow$ Load clique embedding of size $|V|$}
            \State $k_1$, $k_2 \leftarrow$ \Call{QA}{\texttt{QUBO\textsubscript{E}, num\_reads, annealing\_time}} 
            \If{$k_1 \neq \varnothing$ and $k_2 \neq \varnothing$} 
                \State \Call{Binary\_\_QA}{\textbf{B}, $k_1$, $\gamma$}
                \State \Call{Binary\_\_QA}{\textbf{B}, $k_2$, $\gamma$}
            \ElsIf{$^k_1 \neq \varnothing$}
                \State \Return{$k_1$}
            \Else
                \State \Return{$k_2$}
            \EndIf
        \EndProcedure
    \end{algorithmic} 
\end{algorithm} 

\subsection{Technical aspects of the D-Wave minor embedding}
The \texttt{Advantage\_system5.4} D-Wave QPU topology uses a Pegasus graph of degree 16. The QUBO graph must be mapped onto that architecture, which is achieved by minor-embedding, i.e., mapping each logical binary variable to a group of physical qubits on the actual machine so all required connections are realized. As the QUBOs are dense, we utilized a predefined routine for complete graph embeddings (clique embedding). To reduce the time, we cached the embeddings for all clique sizes only once and reused them accordingly in Algorithm \ref{algo:HAnnealing}. Crucially, although Hierarchical annealing relies on pure quantum processes, the embedding remains a classical step. The chain strength was set using D-Wave’s default strategy, which determines the value based on the number of logical qubits and the variance in connectivity of the physical qubits. Broken chains were resolved using majority voting, assigning each logical qubit the most common spin value along its chain. % \cite{dattani_pegasus_2019} \cite{choi_minor-embedding_2011} ~\cite{raymond2020improving}

\subsection{Computational complexity and running times}
For a problem of size $n$, the Hierarchical annealing procedure comprises two main stages: (i) computation of the modularity matrix and (ii) recursive binary QA optimization. The first stage is executed only once before entering the recursion. Using big-$\mathcal{O}$ notation, scales as $\mathcal{O}(n^2)$ due to an outer product between degree vectors. The second stage requires a more nuanced analysis, as each recursion step invokes a QA optimization.

At each step of the recursion, the generalized modularity matrix needs to be computed, followed by a QA instance. The former is done in linear time $\mathcal{O}(n)$, while the latter depends on the specific problem to be solved. For convenience, we denote the computational time required to perform a single binary QA step as $f(n)$, and the total accumulated annealing time as $\mathcal{K}_n$. Recall that the final complexity should consider \textbf{B}, thus yielding $\mathcal{O}(n^2 + \mathcal{T}(n))$ where $\mathcal{T}(n)$ is the complexity of the recursion tree.

Certain instances of the recursion tree grow linearly with the size of the network $n$, but these are uncommon as they imply that each subdivision yields isolated nodes. Instead, a more reasonable outcome is a dynamic and unbalanced recursive tree for which the Akra-Bazzi theorem provides an asymptotic solution to the expected computational complexity (Supplementary Material S4). Briefly, the theorem states that recurrences of the form
\begin{equation} \label{eq:akra-bazzi}
\mathcal{T}(n) = \sum_{i=1}^s a_i \mathcal{T}(b_i n + h_i(n)) + g(n),
\end{equation} have an asymptotic solution, 
\begin{equation}\label{eq:akra-bazzi_solution}
    \mathcal{T}(n) \sim n^p \left(1 + \int_{n_0}^n \frac{g(u)}{u^{p+1}} du \right).
\end{equation} The parameter $p$ satisfies the characteristic equation $\sum_{i=1}^s a_i b_i^p - 1 = 0,$ with $a_i > 0$, $0 < b_i < 1$ for all $i = 1, \ldots, s$, and $g(n)$ is a continuous, positive, and regular function for all sufficiently large $n$. The functions $h_i(n)$ represent small perturbations in the arguments.

In general, a community $k$ is split into two communities $k_1, k_2$ of arbitrary sizes $|k_1|=n_k/a_k$ and $|k_2|=n_k(1-1/a_k)$, where $n_k$ is the number of nodes in community $k$ and $1/a_k$ is the fraction of nodes in $k$ assigned to $k_1$. The ratio $a_k$ will differ for every community $k$, but it can be modeled as a random number: $a_k=a+\delta_k+O(\delta_k^2)$ with $\delta_k \sim U(-a,L)$, and $1\leq a \leq L$ for some finite $L$ depending on the average ratio along the tree. Then, the community sizes can be expressed as,
\begin{equation}
    \begin{split}
        &\frac{n_k}{a_k} = \frac{n_k}{a}\left(\frac{1}{1+\delta_k/a}\right) = \frac{n_k}{a}+h_1(n_k)\\
        &n_k\left(1-\frac{1}{a_k}\right) = n_k\left(1-\frac{1}{a}\frac{1}{(1+\delta_k/a)}\right)=n_k\left(1-\frac{1}{a}\right)+h_2(n_k)
    \end{split}
\end{equation} where $h_1(n_k)$ and $h_2(n_k)$ represent bounded perturbations stemming from variable community sizes and rounding errors (e.g., floor and ceiling functions $[\cdot]$). Additionally, the resolution limit guarantees that the recursion will halt no later than at some minimum effective size $n_0 \in [1, n - a]$, where $\mathcal{T}(n_0) = \mathcal{O}(1)$.

Formally, the time to compute the whole recursion for a community $k$ is a function of the two attached sub-problems and the corresponding operations taking place at the same level,
\begin{equation} \label{eq:recursion_time} \small
    \begin{split}
        \mathcal{T}(n_k) = \quad &\mathcal{T}\left(\left[\frac{n_k}{a}\right]+h_1(n_k)\right)+\\
        & \quad \mathcal{T}\left(\left[n_k\left(1-\frac{1}{a}\right)\right]+h_2(n_k)\right) + n + f(n),
    \end{split}
\end{equation} \normalsize which satisfies Eq.~\ref{eq:akra-bazzi} if $s=2$, $a_1=a_2=1$, $b_1=\frac{1}{a}$, $b_2=1-\frac{1}{a}$, $g(n)=n+f(n)$, and $p=1$ (see Supplementary Material). Crucially, the QA complexity $f(n)$ seems exponential at worst \cite{Albash2018}, which still satisfies the conditions of regularity. Solving for Eq.~\ref{eq:akra-bazzi_solution} gives a complexity of $\mathcal{T}(n)\sim \mathcal{O}(n\log n + \mathcal{K}_n)$ with 
\begin{equation} \label{eq:complexity_quantum}
    \mathcal{K}_n \doteq n\int_{n_0}^n\frac{f(u)}{u^2}du.
\end{equation} While $f$ is generally unknown, it can be empirically measured and compared with theoretical assumptions (Table \ref{tab:complexity}). Lastly, while the resolution parameter $\gamma$ influences tree depth through the resolution limit, it does not change the computational complexity of the recursion. However, it might impact the annealing times because it modifies the underlying Hamiltonian~\cite{Reichardt2006,Traag2011}. 

\begin{table}[h] 
    \centering
    \begin{tabular}{l||c|c}
    $f(n)$ & $\mathcal{K}_n$ & Total complexity $\mathcal{T}(n)$ \\
    \hline
    $c$ & $O(1)$ & $\mathcal{O}(n \log n)$ \\
    $\log n$ & $O(\log n)$ & $\mathcal{O}(n \log n + \log n) \sim \mathcal{O}(n \log n)$ \\
    $n$ & $O(n \log n)$ & $\mathcal{O}(n \log n)$ \\
    $n^s, \ s > 1$ & $O(n^s)$ & $\mathcal{O}(n \log n + n^s) \sim \mathcal{O}(n^s)$ \\
    $e^{n^s}, \ s \in \mathbb{R}^+$ & \textit{n.c.f.} & $\mathcal{O}(n e^{n^s})$ \\
    \end{tabular}
    \caption{\textbf{Asymptotic behavior of $\mathcal{K}_n$ and $\mathcal{T}(n)$ for different $f(n)$ growth rates.} The parameter $c$ is any positive and finite constant. 'n.c.f.' stands for \textit{No Closed Form} (see section S4.3 in the Supplementary Material).}
    \label{tab:complexity}
\end{table}

The total execution time comprised of four components: (i) local operations, such as creation of Python objects; (ii) retrieval of cached embeddings; (iii) network communication with the Advantage solver; and (iv) service time on the QPU\footnote{\url{https://docs.dwavequantum.com/en/latest/quantum_research/operation_timing.html}}. The first is implementation-dependent and not intrinsic to the algorithm. The second can be measured using canonical Python tools and, albeit dependent on hardware specifics, could depend non-trivially on the problem's size and topology. Network communication includes problem transmission and result retrieval, whereas the service time comprises queueing, preprocessing, and QPU execution. For user convenience, network communication and service are provided jointly as a single \textit{Sample} function in the DWAVE API, a name we use throughout the rest of the paper for readability.  Although the QPU access time accounts for all \texttt{numreads}, this definition reflects practical usage, as obtaining a viable solution typically requires multiple samplings.

We evaluated the scalability of the Hierarchical annealing procedure by computing the maximum modularity for networks of increasing size and varying resolution parameters, recording the corresponding elapsed times. For each experiment, we collected three metrics: QPU access, cache-read embedding, and combined sampling–communication time (which included QPU access but excluded embedding retrieval). Measurements were obtained at each recursion step and aggregated to compute the total execution time per problem. Because Algorithm \ref{algo:HAnnealing} was run iteratively, we considered the running times of the solution that yielded the maximum modularity.

Lastly, because paid usage agreements typically govern access to quantum computing resources, we also quantified the official computing time consumed in the D-Wave Leap Dashboard. This enabled a fair comparison between standard D-Wave approaches, such as discrete quadratic models (DQM), and our custom implementation, providing practical guidance for estimating resource consumption when planning D-Wave computational grants. We manually recorded the available usage time immediately before and after each full execution of the modularity maximization routine, as well as for the DQM runs (Fig. S3).

\subsection{Network generation, power-law exponent, scale-free networks, and stochastic block models}
We generated random networks using three different models. First, the Erdős-Rényi model \cite{erdds1959random} connected any pair of initially disconnected nodes with a constant probability $p$, producing random networks without clear statistical organization. Despite this, non-trivial geometrical behaviors can still emerge, making it a useful null model for study. Second, the Barabasi-Albert model \cite{Barabasi1999} added new nodes dynamically, where the probability of a new node $i$ connecting to node $j$ was proportional to $j$'s degree, $p_j \propto k_j$. This preferential attachment mechanism produced networks with robust scaling properties and enhanced resilience. The Holme-Kim model \cite{Holme2002} incorporated preferential attachment and a triad formation step to boost clustering. This produced highly clustered, power-law networks. To test our approach in directed networks, we also used the preferential attachment mechanism applied to directed versions of the graphs \cite{bollobas2003}. This yielded power-law and scale-free networks with directed edges. The parameters of this model were left as the default values, both in Bollob\'as, \textit{et al.}, 2003 \cite{bollobas2003}, and in the NetworkX Python package \cite{hagberg2008}.

We fitted a discrete power-law distribution to their node degrees present in a given network \cite{alstott2014powerlaw}. A network is said to have a power-law distribution or structure if the probability $P(k)$ of a given node having degree $k$ follows an exponential distribution,
$$P(k)\sim k^{-\alpha},$$
where $\alpha$ is the power-law exponent. Furthermore, if this exponent lies between 2 and 3, the network exhibits a scale-free distribution, where the probability of highly connected nodes does not vanish even for high degrees $k$ \cite{Barabasi1999}.

Lastly, we used the stochastic block model (SBM) to evaluate the Hierarchical annealing procedure against known ground truths~\cite{abbe2018}. Networks consisted of $N=100$ nodes assigned to three planted communities of sizes 50, 40, and 10. Following~\cite{Karrer2011}, nodes belonging to communities $r,s$ were connected with a probability
\begin{equation}\label{eq:SBM_mixed}
\omega_{rs} = \lambda\omega^{\text{planted}}_{rs} + (1-\lambda) \xi,
\end{equation} where $\lambda\in [0,1]$ is a mixing parameter, $\xi\sim U(0,1)$, and 
\begin{equation} \label{eq:SBM}
\boldsymbol{\omega}^{\text{planted}}=
\begin{pmatrix}
p_{in} & p_{out} & p_{out} \\
p_{out} & p_{in} & p_{out} \\
p_{out} & p_{out} & p_{in}
\end{pmatrix}.
\end{equation} $p_{in}$ and $p_{out}=\tfrac{1}{2}(1-p_{in})$ controlled the intra- and inter-community connection probabilities, respectively. We generated 10 networks for each $\{p_{in},\lambda \}$ pair, with $p_{in}\in\{0.6,0.7,0.8,0.9\}$ and $\lambda\in [0,0.9]$ ins steps of 0.1. For $\lambda=0$, this is equivalent to the planted partition model, enabling a principled comparison between modularity maximization and probabilistic inference methods~\cite{Newman2016}, with task difficulty increasing monotonically with $\lambda$. %~\cite{condon2001}

\subsection{Other algorithms and multiple solutions} 
We evaluated results using several classical algorithms: Louvain~\cite{Blondel2008}, Leiden~\cite{Traag2019}, Bayan~\cite{aref2022} --- known to return exact solutions for networks up to a considerable size~\cite{aref2023}, Hierarchical Gurobi, spectral clustering~\cite{Damle2018}, simulated annealing [see \href{https://github.com/perrygeo/simanneal?tab=readme-ov-file}{code}], Infomap~\cite{Rosvall2008}, and degree-corrected stochastic block model inference~\cite{Karrer2011,Peixoto2014}; as well as DQM~\cite{Wierzbinski2023}, a hybrid quantum-classical method. Each algorithm was run a certain number of times per network, and the community structure with the highest modularity was retained as the "true" structure. 

Unless stated otherwise, Louvain, Leiden, Infomap, spectral clustering, and simulated annealing were each run 100 times per network, whereas SBM inference was run 1000 times to ensure. Hierarchical Gurobi followed the same recursion as Hierarchical annealing but used the classical Gurobi solver; both it and Bayan were run for 20 instances due to licensing constraints. DQM was run five times, as its solutions were highly stable. Both DQM and simulated annealing required a predefined maximum number of communities: for DQM, we used the resolution limit ($\max|C|\to\sqrt{n}+1$), while for simulated annealing, we used the number of communities found by Hierarchical annealing, as performance otherwise degraded to near random. For simulated annealing, we used \texttt{T\textunderscore{max}}=10, \texttt{T\textunderscore{min}}=0.001, and 50 updates.

Community detection algorithms often yield varying structures due to the use of heuristics or inherent solution degeneracy \cite{fornito2016fundamentals}. We evaluated the variability between communities identified by two algorithms through 1) consensus classification matrices and 2) the Dice-S{\o}rensen score. For each network, an $N \times N$ consensus coclassification matrix recorded how often nodes $i$ and $j$ were grouped into the same community across algorithm runs \cite{Lancichinetti2012}. Perfect agreement between algorithms would result in matrix entries $M_{ij}$ of 0 or 1 (or 0\% and 100\%), indicating stable or well-defined community structures. Conversely, a range of intermediate $M_{ij}$ values suggests inconsistent community assignments. While informative, this matrix does not assess community structure optimality, necessitating the additional use of a modularity index. To compare community structures directly, we computed the Dice score between pairs of communities as:
\begin{equation} \label{eq:dicescore}
    D(C^{A_1}_i,C^{A_2}_j) = 2\frac{|C^{A_1}_i \cap C^{A_2}_j|}{|C^{A_1}_i|+|C^{A_2}_j|},
\end{equation}
where $C^{A_k}_i$ is the $i$-th community from algorithm $k$, and $|X|$ is the size of set $X$. This score was computed for all community pairs across algorithms (e.g., Louvain and Hierarchical annealing) to quantify the similarity of their outputs.

We measured the improvement of the classical algorithm w.r.t. the Hierarchical annealing by computing the relative increase. For that, we extracted the maximum modularity from the pool of solutions and computed the relative error percentage-wise. Thus,

\begin{equation} \label{eq:increase}
    \%RI = 100 \cdot \frac{\text{Max } Q_{HA} - \text{Max } Q_{Z}}{Q_{Z}},  
\end{equation} where $Z$ is in place for any of the classical alternatives described.

When node assignments were pre-defined, we compared the recovered communities to the true structure using the variation of information, split–join distance, normalized mutual information, and the adjusted Rand index; all implemented in igraph~\cite{csardi2006igraph}.
%When node assignments were pre-defined, we compared the recovered communities to the true structure using the variation of information~\cite{meilua2003}, split–join distance~\cite{van2000performance}, normalized mutual information~\cite{Danon2005}, and the adjusted Rand index~\cite{hubert1985comparing}; all implemented in igraph~\cite{csardi2006igraph}.

\section{Results}
\subsection{Hierarchical annealing in simple networks}
\begin{figure}
    \centering
    \includegraphics[width=.9\linewidth]{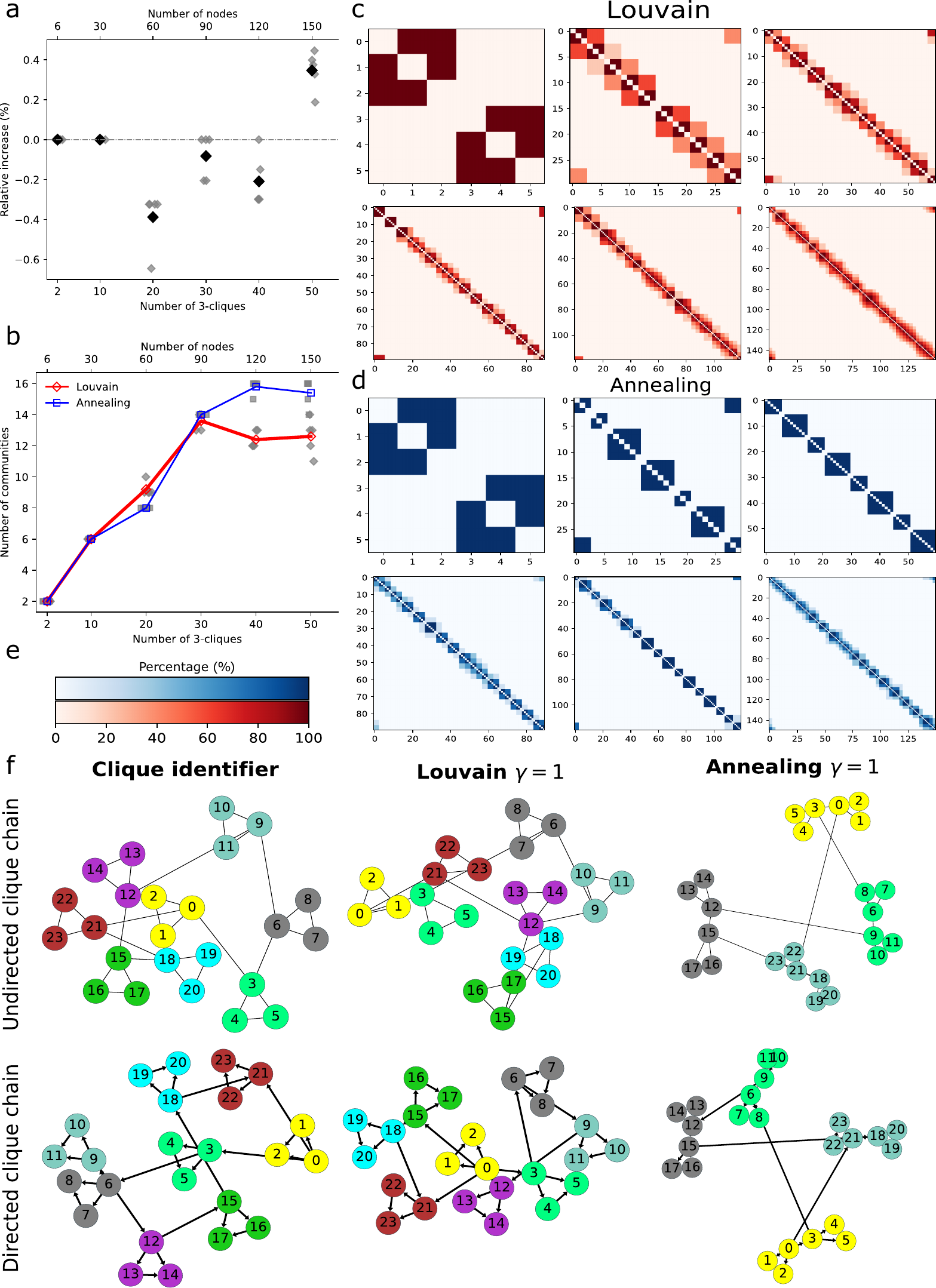}
    \caption{\textbf{Communities detected with Annealing and Louvain in chains of 3-cliques.} \textbf{a} Modularity of the community structure found by the two algorithms measured as the relative increase (\%) w.r.t. the Louvain solution. Individual points (in gray) and the mean increase (black diamonds). \textbf{b} Number of communities found with each algorithm. \textbf{c, d} Consensus matrices constructed with 5 different runs for each clique chain. \textbf{e} Percentage colorbar in \textbf{c} and \textbf{d}. \textbf{f} Output of the two algorithms and the generated graph (left) for a chain of 8 3-cliques. Both the Louvain and the hierarchical annealer solutions have the exact modularity.}
    \label{fig:clique-chains}
\end{figure}

As a first benchmark, we generated chains of 3-cliques for both undirected and directed examples, treating each one of these 3-cliques as a proxy for true communities. Hierarchical annealing and Louvain algorithms were executed 5 times per network, and their performance was compared based on returned modularities (Fig. \ref{fig:clique-chains}a), number of found communities (Fig. \ref{fig:clique-chains}b), and variability of the algorithms in terms of their consensus classification matrices (Fig. \ref{fig:clique-chains}c-d; see Methods).  For these experiments, the resolution parameter $\gamma$ was left to 1. For networks with up to 30 nodes, both algorithms achieved identical modularities in all 5 runs. The Hierarchical annealing algorithm maintained stable and optimal performances as the network size increased, achieving the same maximum modularity as Louvain in all but one network. Notably, it even surpassed Louvain's maximum modularity by 0.5\% for a network of 150 nodes (Fig. \ref{fig:clique-chains}a). Interestingly, for a network of 60 nodes, the annealing algorithm produced a slightly less optimal solution, yet demonstrated no variability. This suggests that the inability to explore the state space of possible solutions may lead to suboptimal behaviors. Overall, the annealing algorithm exhibited slightly less variability than Louvain without compromising solution optimality. 

We show a graphic example in Fig. \ref{fig:clique-chains}f for an undirected and directed network of 8 3-cliques (i.e., 24 nodes) where, despite finding slightly different partitions, the modularities are identical (i.e., 0.625 and 0.6295 for the undirected and directed chains, respectively). Similar results hold for a chain of 1-cliques (i.e., nodes), where the absence of any recursive structure does not impede the algorithm from obtaining high modularity indices (Fig. S4).

The next example was the Zachary Karate Club network, a well-known dataset encoding social interactions. The current agreement for this system describes 4 communities with a maximum modularity of 0.44904 for $\gamma=1$ (Table \ref{tab:karate}). As opposed to the clique chains studied before, the topological and geometrical structure of this network does not suggest a well-defined hierarchical structure, thus not explicitly favoring the recursive nature of our solution. We also tested a recent quantum-classical hybrid \cite{Wierzbinski2023}, known to correctly resolve this network. The solution obtained through pure Hierarchical quantum annealing was entirely equivalent.

\begin{table}[h]
    \centering
    \resizebox{\columnwidth}{!}{%
        \begin{tabular}{l||c|c|c|c}
            \textbf{Algorithm} & \textbf{Max} $\mathbf{Q}$ & \textbf{Frequency} (\%) & \textbf{Communities} & \textbf{Time} (s) [mean $\pm$ SEM]\\
            \hline
            \textit{H. annealing} & 0.444904 & 100 & 4 & \textit{n.a.} \\
            \textit{H. Gurobi} & 0.444904 & 100 & 4 & 0.110 $\pm$ 0.002 \\
            DQM & 0.444904 & 100 & 4 & 12.80 $\pm$ 0.04 \\
            Louvain & 0.444904 & 40 & 4 & 0.0037 $\pm$ 0.0004 \\
            Leiden & 0.444904 & 100 & 4 & 0.00248 $\pm$ 0.00008 \\
            Bayan & 0.444904 & 100 & 4 & 2.04 $\pm$ 0.06 \\
            Spectral clustering & 0.444904 & 100 & 4 & 0.00436 $\pm 2\cdot 10^{-5}$ \\
            Simulated annealing & 0.444904 & 2 & 4 & 3.1772 $\pm$ 0.01731 \\
            Infomap & 0.434521 & 100 & 3 & 0.000664 $\pm 7\cdot 10^{-6}$ \\
            SBM Inference & 0.434521 & 0.01 & 3 & 0.0028 $\pm$ 0.0004 \\ 
        \end{tabular}
    }
    \caption{\textbf{Modularity of the Karate club network.} Maximum modularity resulting from the different algorithms tested. The maximum and frequencies are taken from a pool of 50 independent runs for all algorithms except for SBM inference, which required 1000 runs. In italics, methods that stem from our algorithm. 'n.a.' stands for not applicable since the computing times are analyzed separately.}
    \label{tab:karate}
\end{table}

\subsection{Hierarchical annealing as a function of network topology}
We studied the performance of the proposed annealing process as we varied the geometry and the topology of the networks. We evaluated the solutions of the Louvain, Leiden, simulated annealing, spectral clustering, and Hierarchical annealing algorithms for 3 different complex types of networks of increasing sizes. The DQM algorithm was ran in a subset of networks due to its high consumption of computing time. Specifically, for these experiments, DQM required approximately 300× more billed computing time than the Hierarchical annealing procedure, substantially impacting access to computational resources and the planning of computational grants (Table~S1). A similar argument holds for Hierarchical Gurobi and Bayan, where external computational resources needed to be acquired. Infomap and SBM inference were not considered here because they do not maximize modularity, but instead discover communities from different principles~\cite{Rosvall2008,Peixoto2014}.

\begin{figure*}
    \centering
    \includegraphics[width=1\linewidth]{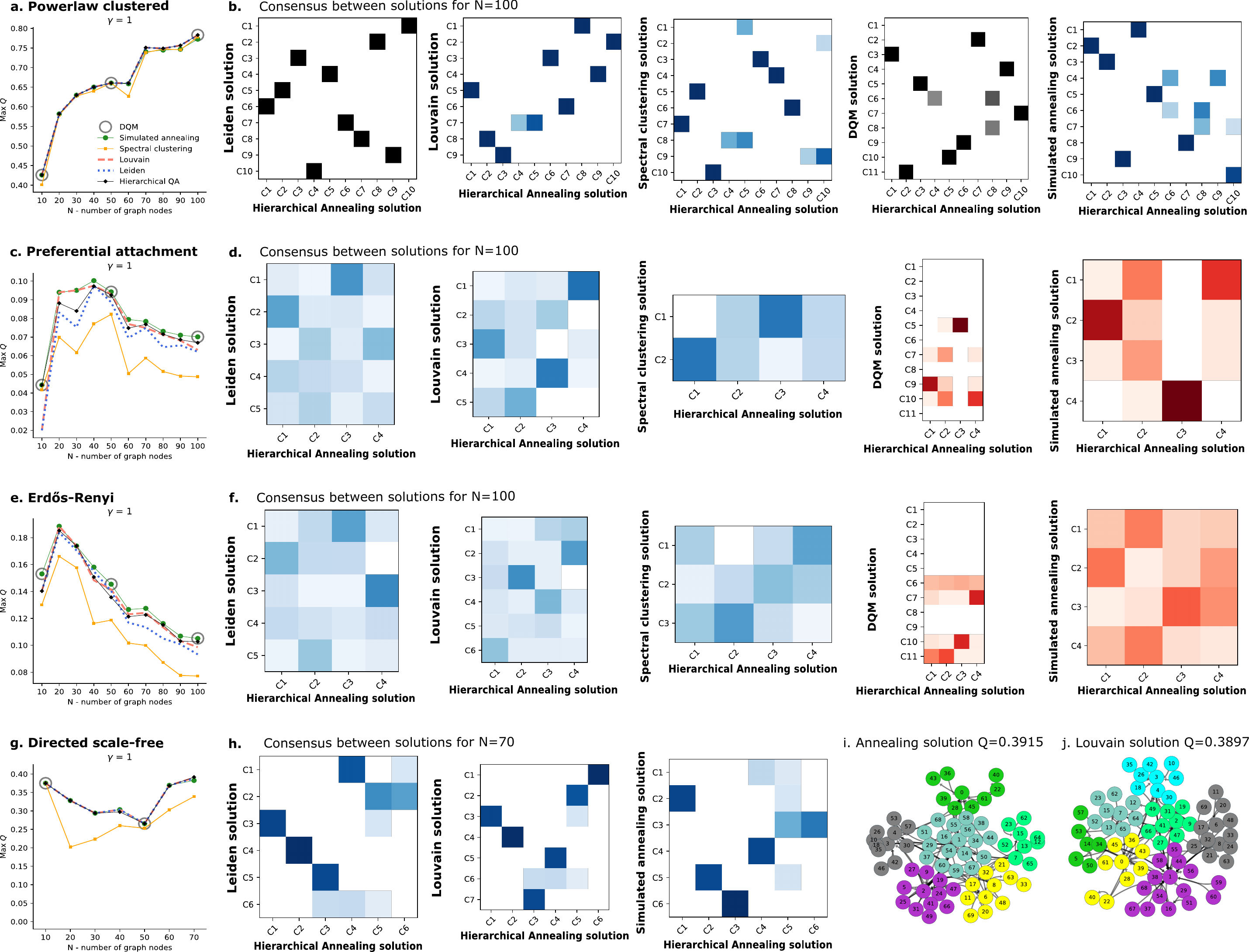}
    \caption{\textbf{Hierarchical annealing solutions in random networks.} \textbf{a} Maximum modularity ($\max Q$) of power-law–clustered networks as a function of network size. \textbf{b} Consensus matrices for the $N=100$ network in (\textbf{a}), comparing communities identified by the hierarchical annealing procedure with those obtained using alternative algorithms. Opacity is proportional to the Dice similarity score between pairs of communities. Black, blue, and red outlines indicate cases in which hierarchical annealing achieved equal, higher, or lower modularity, respectively, relative to the corresponding alternative solution. White entries denote zero Dice similarity. \textbf{c} Maximum modularity ($\max Q$) of Barabasi-Albert networks. \textbf{d} Consensus matrices of the Barabasi-Albert network ($N=100$) studied in (\textbf{c}). \textbf{e} Maximum modularity ($\max Q$) of Erd\H os-Renyi networks. \textbf{f} Consensus matrices of the Erd\H os-Renyi network ($N=100$) studied in (\textbf{e}). \textbf{g} Maximum modularity ($\max Q$) of directed scale-free. \textbf{h} Consensus matrices of the directed scale-free network ($N=70$) studied in (\textbf{g}). \textbf{i-j} Communities found the Hierarchical annealing (\textbf{i}) and Louvain (\textbf{j}) algorithms in the directed scale-free network in (\textbf{h}). See Fig. S5-S8 for the full set of consensus matrices.}
    \label{fig:random-nets}
\end{figure*}

The first networks were generated following preferential attachment and triad formation steps \cite{Holme2002}. This generated a network with scale-free and/or power-law properties, high clustering, and a good modular organization as the size of the network is increased. The number of edges to add and the probability of performing a triad formation step were kept at 1 and 0.1, respectively, for all the sizes. Hierarchical annealing displayed a robust, optimal, and highly congruent behavior when compared to the other algorithms (Fig. \ref{fig:random-nets}a-b).

Then, we generated networks using only preferential attachment (i.e., without triad formation) \cite{Barabasi1999}. However, depending on the number of edges added at each step, the final network may or may not display a modular organization. To test the robustness of Hierarchical annealing to discover communities in potentially modular networks, we kept the aforementioned parameter equal to 40\% of the network size. This generated poorly modular systems, but without random topologies. For these non-trivial networks, our method remained competitive across sizes, largely surpassing the Leiden algorithm (Fig. \ref{fig:random-nets}c-d). DQM and simulated annealing were robust but substantially more computationally intensive than the other methods. Moreover, simulated annealing required a pre-specified number of communities inferred by Hierarchical annealing, since when initialized at the resolution limit, its performance failed to pose any real challenge to the alternative solutions.

We also tested the Hierarchical annealing process in random networks \cite{erdds1959random}, where the probability of two nodes being connected is uniform. The topology of Erd\H{o}s-R\'enyi networks has been well characterized and is considered a valid null system for multiple graphs and topological metrics in biological systems. Again, our solution produced comparably optimal solutions for all algorithms (Fig. \ref{fig:random-nets}e-f). Similarly to the Barabasi-Albert systems, the DQM and simulated annealing solutions were competitive at the expense of computational tractability and the need for a priori sets of communities to detect.

Finally, we studied the performance of the hierarchical quantum annealing solution in directed scale-free networks generated using the preferential attachment mechanism \cite{bollobas2003}. Again, our solution resulted in high-quality community structures for multiple network sizes (Fig. \ref{fig:random-nets}g-j). Both DQM and simulated annealing could resolve these systems, but with a slight underperformance.

\begin{figure*}
    \centering
    \includegraphics[width=\linewidth]{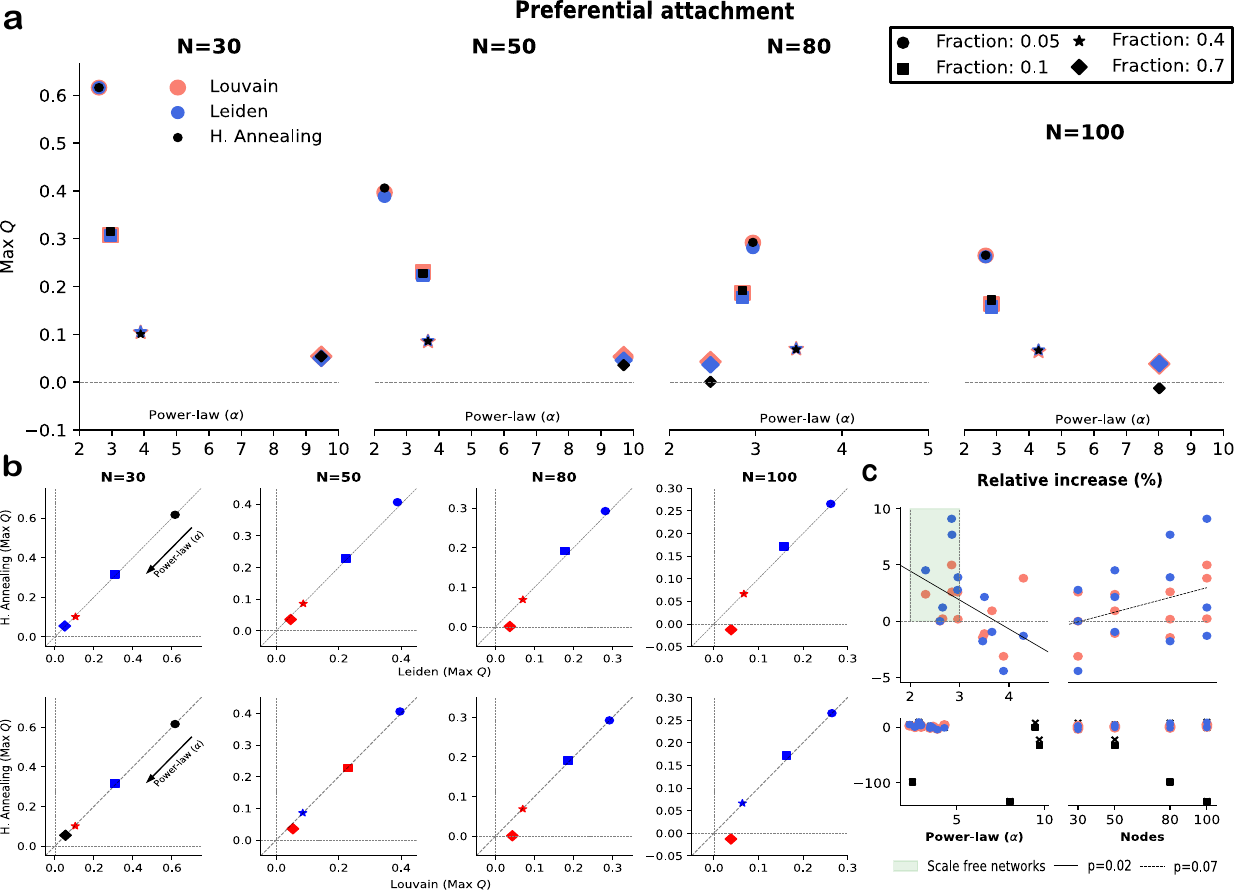}
    \caption{\textbf{Hierarchical annealing as a function of inherent hierarchies in random networks.} \textbf{a} Modularity obtained with the Leiden, Louvain, and Hierarchical Annealing algorithms across networks varying in size (panels: from left to right) and scale-freeness (marker shape: "Fraction" $f$), generated via preferential attachment \cite{Barabasi1999}. \textbf{b} For networks of increasing size (panels: left to right), comparison of Hierarchical Annealing with Leiden (panels: top) and Louvain (panels: bottom). Black markers located on the diagonal indicate identical modularities; blue/red markers located on the upper/lower triangular parts denote superior/inferior annealing solutions. Marker shapes correspond to the legend in (\textbf{a}). \textbf{c} Relative improvement of annealing solutions in (\textbf{a-b}) as a function of $\alpha$ and network size. The green shaded region (top left) corresponds to scale-free networks. Bottom panels are identical but explicitly show cases excluded from the regression analyses (black markers).}
    \label{fig:topology}
\end{figure*}

To further study the robustness of the quantum solutions, we studied the behavior of the Hierarchical annealing process as a function of the power-law exponent. For that, we generated multiple networks of different sizes following the preferential attachment mechanism described previously, each one with significantly different power-law structures and potential hierarchies. We compared the Hierarchical annealing process against the Louvain and Leiden algorithms, finding very similar solutions in terms of the modularity index. Even more, the solutions showed the same inverse relationship on the power-law structure as the classical counterparts (Fig. \ref{fig:topology}a). 

More specifically, the Hierarchical annealing process yielded 3 out of 36 equally optimal solutions, 16 out of 36 higher modularity scores, and 13 out of 36 worse solutions (Fig. \ref{fig:topology}b). Therefore, we fitted a bilinear model using an ordinary least squares procedure to explain the relative increase $(\%RI)$ of the quantum solution as a function of both parameters, $$ \%RI \sim 1 + \alpha + N,$$ where $\alpha$ denotes the power-law exponent, and $N$ is the number of nodes. This model carried explanatory power ($R^2=0.3680$, $p=0.0080$, $F$-statistic against a constant model, AIC=118.2572) and both coefficients significantly contributed to the prediction ($\alpha$: $ p=0.0.0116$, $N$: $ p=0.0462$). Additionally, we included an interaction term of the form $\alpha N$. This alternative model showed a slightly higher coefficient of determination ($R^2=0.3869$, $p=0.0144$, $F$-statistic against a constant model, AIC=119.5288) but the individual coefficients were non-significant ($\alpha$: $ p=0.0863$, $N$: $ p=0.7353$, $\alpha N$: $p=0.4416$), suggesting that independent rather than interaction effects better explain the performance of the algorithm. 

To visualize and confirm these trends, we fitted two independent linear models for each parameter. The coefficient of the power-law exponent maintained the significance, while the coefficient of the number of nodes didn't, despite the increasing trend in the relative increase (Fig. \ref{fig:topology}c). For scale-free networks (i.e., $\alpha \in [2,3]$), all the Hierarchical annealing solutions were either superior or equivalent to ones found by the classical counterparts (Fig. \ref{fig:topology}c, green shaded area). Importantly, we only considered solutions with a relative increase and power-law exponent that fell within the 95\% range (i.e., $|z_{score}|<1.96$). Using this criterion, 4 solutions out of 36 were discarded. Crucially, 3 out of these 4 outliers corresponded to extremely high power-law exponents.

\begin{figure*}
    \centering
    \includegraphics[width=\linewidth]{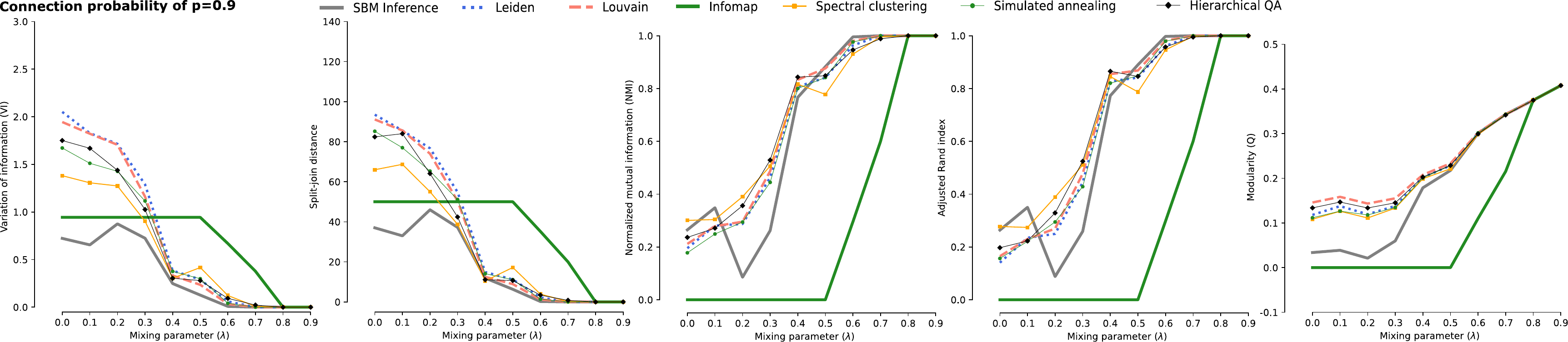}
    \caption{\textbf{Hierarchical annealing in stochastic block models (SBM).} Performance of benchmark methods relative to the hierarchical annealing procedure on an SBM with three communities ($p_{in}=0.9$, $p_{out}=0.05$). Recovery accuracy is assessed using four metrics against the ground truth, with the corresponding maximum modularity ($\max Q$) shown in the rightmost panel as a function of the mixing parameter $\lambda$.}
    \label{fig:sbm_mixing}
\end{figure*}

An alternative benchmark to evaluate community detection methods is the stochastic block model (SBM), in which communities are planted a priori and must then be recovered. Across a range of difficulty levels, the Hierarchical annealing procedure performed consistently well across several evaluation metrics (Fig.~\ref{fig:sbm_mixing}; see also Fig. S9). Importantly, these experiments were not intended to validate the consistency of modularity maximization as a community detection method itself~\cite{Bickel2009,Zhao2012,Ghasemian2020}, but to establish Hierarchical annealing as a valid approach to recover known structure by attaining $\max Q$~\cite{Newman2016}.

\subsection{Hierarchical annealing for different resolution parameters}

The binary hierarchical formalism we developed explicitly incorporated the resolution parameter $\gamma$ in the function to optimize, as opposed to earlier accounts, both classical \cite{Newman2006,Leicht2008} and quantum \cite{Ushijima-Mwesigwa2017,Negre2020}. Briefly, the process was formally independent of $\gamma$, which only appeared in the calculation of the generalized modularity matrix (Methods). However, since $\gamma$ determines both the QUBO coefficients, recursion depth, and physical constraints, we examined its influence. 

\begin{figure}
    \centering
    \includegraphics[width=1\linewidth]{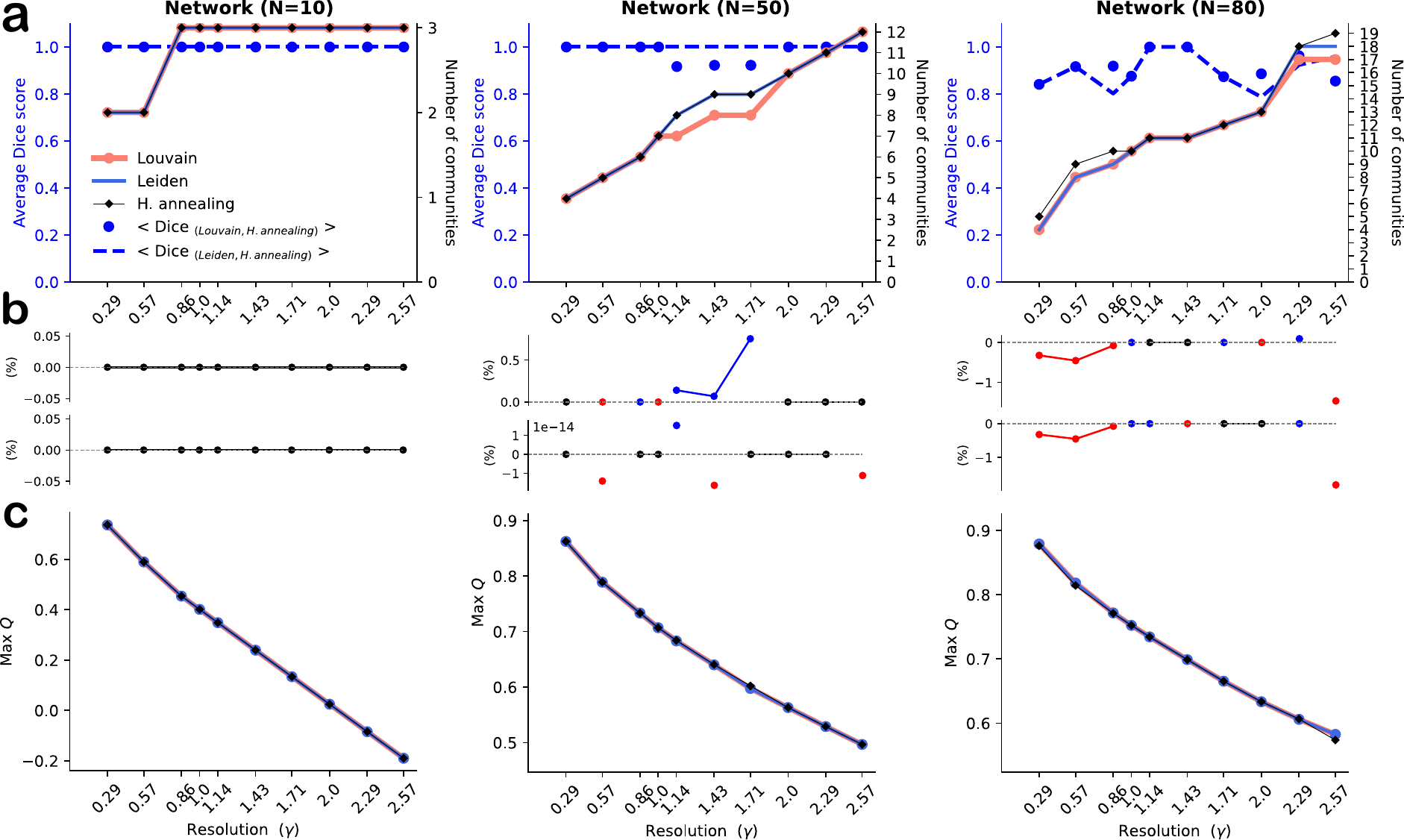}
    \caption{\textbf{Hierarchical annealing in different clustered power-law networks for different resolution parameters.} Each column depicts the results described below for a network of $N=10, 50, 80$ nodes, respectively. \textbf{a} Measure of the average overlap between the communities found by the Louvain and Hierarchical annealing algorithms (left axis, blue) and the number of communities (right axis, salmon and black) as a function of the resolution parameter $\gamma$. \textbf{b} Relative increase of the Hierarchical annealing measured w.r.t. the Louvain (top) and Leiden (bottom) solutions in (\textbf{a}). Black markers correspond to identical solutions, blue markers correspond to better solutions from the annealer, and red markers indicate worse performances than the Louvain alternative. \textbf{c} Maximum modularity per resolution value.}
    \label{fig:resolution}
\end{figure}

We generated 3 different networks with nodes $N=10, 50, 80$ using preferential attachment and triad formation steps to test whether the framework returned an optimal community structure. We then ran our Hierarchical annealing procedure for different values of the resolution parameter and compared it to the Leiden and Louvain alternatives since they were the fastest and most competitive while incorporating $\gamma$ in their formalism. The overlap between the different solutions was consistently close to 1, and never below 0.8, as measured with the average Dice score between all the communities (Fig. \ref{fig:resolution}a). This overlap, as expected, co-varied with the number of communities found. Furthermore, the performance was stable across the range of $\gamma$ studied, showing some slight increases and decreases for individual cases (Fig. \ref{fig:resolution}b-c). A similar situation held for both Erd\H{o}s-Renyi and Barabasi-Albert networks (Figs. S10-S12), with some slight but still comparable underperformance for $\gamma>2$. In contrast, for directed scale-free networks, the empirical performance of the hierarchical annealing algorithm degraded both for low and high resolution values as the size of the networks grew (Fig. S13).

\subsection{Hierarchical annealing: running times}
\begin{figure*}
    \centering
    \includegraphics[width=1\linewidth]{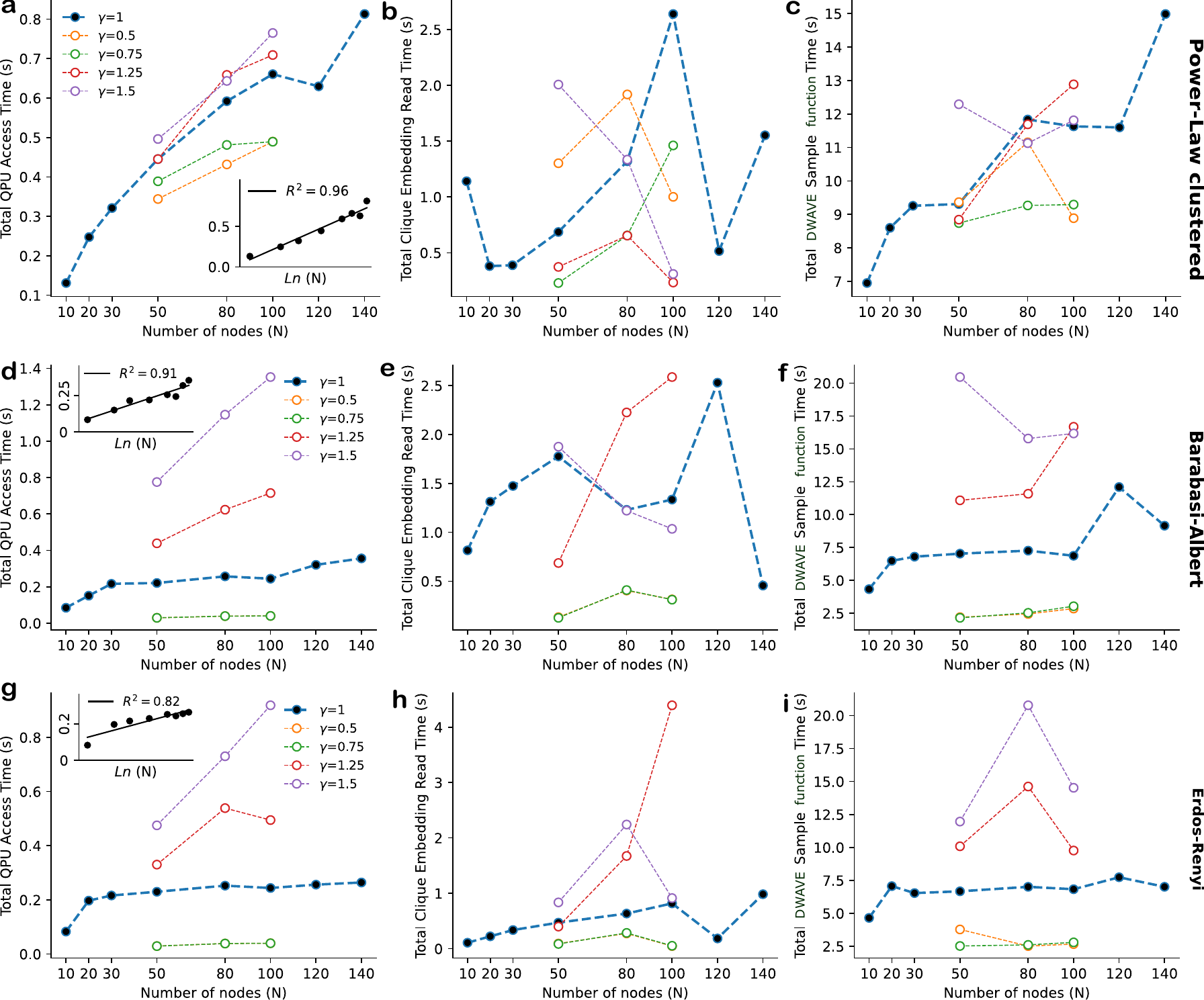}
    \caption{\textbf{Experimental measurements of QPU access, embeddings, and \emph{Sample} times.} \textbf{a} The total QPU access times as an empirical measure of $\mathcal{K}_n$ in Eq.~\ref{eq:complexity_quantum}, corresponding to the sum of the repeated steps of binary QA optimization in power-law clustered networks of increasing size and different resolution parameters, $\gamma$. The inset depicts linear fits between measured times and the natural logarithm of network size ($t \propto \ln N$). \textbf{b} The total embedding and reading times associated with mapping the power-law networks in (\textbf{a}) to the QPU topology in the Advantage system. \textbf{c} The total time required to execute the DWAVE \emph{Sample} function for the power-law networks in (\textbf{a}). \textbf{d-f} Identical to (\textbf{a-c}) but in Barabasi-Albert networks. \textbf{g-i} Identical to (\textbf{a-c}) but in Erd\H os -Renyi networks. See Supplementary Material for the exact numbers corresponding to the results shown here.}
    \label{fig:qpu_times}
\end{figure*}

We measured the computing times of the hierarchical annealing (Algorithm~\ref{algo:HAnnealing}) for the same types of networks (Methods). The number of nodes was gradually increased from 10 to 140 to assess how running times scale with problem size. For Power-law Clustered, Albert-Barabási, and Erd\H{o}s-Rényi networks, experimental times were recorded for a reduced set of resolution parameters and varying sizes. For the Karate Club network, the modularity index was maximized across multiple resolution parameters. As expected, QPU access times increased with the number of nodes for all networks (Fig.~\ref{fig:qpu_times}). For $\gamma=1$, we compared linear and logarithmic models to characterize the scaling, finding that the latter yielded slightly better agreements with empirical times (Table~\ref{tab:r-squared}). Higher resolution parameters increased QPU access times (see Tables S2-S7). 

In Power-law clustered networks, which exhibited inherent hierarchical structure, the effect of $\gamma$ was modest and scaled similarly to the $\gamma=1$ case (Fig.~\ref{fig:qpu_times}a). In contrast, for Barabási--Albert and Erd\H{o}s--Rényi networks, where no clear hierarchies existed, QPU access times behaved nontrivially, suggesting a stronger influence of $\gamma$ on the problem Hamiltonian. For the Karate Club network, total QPU access times ranged from 0.1062~(s) for $\gamma=0.5$ to 0.3771~(s) for $\gamma=2$, with an approximately linear dependence (Table~S8).

\begin{table}[h] 
    \centering
    \begin{tabular}{l||c|c}
    Network type & Logarithmic $t\sim \log n$ & Linear $t\sim n$ \\
    \hline
    Powerlaw clustered & $R^2=0.96$ & $R^2=0.94$ \\
    Barabasi-Albert & $R^2=0.91$ & $R^2=0.87$ \\
    Erd\H{o}s-Renyi & $R^2=0.82$ & $R^2=0.58$ \\
    \end{tabular}
    \caption{\textbf{Determination coefficient} $\mathbf{R^2}$ \textbf{for QPU access time fits as a function of network size.} Logarithmic and linear fits were used to model $\mathcal{K}_n$ (see also Fig. \ref{fig:qpu_times}).}
    \label{tab:r-squared}
\end{table}

Cache-read embedding times showed no clear functional dependence on the number of nodes, with variations likely due to internal processes. For Power-law networks, the resolution parameter had no effect (Fig.~\ref{fig:qpu_times}b), while in Barabási--Albert and Erd\H{o}s--Rényi networks, higher resolutions generally increased reading times (Fig.~\ref{fig:qpu_times}e,h). In contrast, for the Karate Club network, cache-read embedding times ranged from 0.2072~(s) for $\gamma=0.5$ to 3.7889~(s) for $\gamma=2$, indicating a dependence on the specific QUBO values and resulting embeddings.

Interestingly, communication and queueing times represented a substantial overhead relative to actual QPU processing, highlighting inefficiencies that could be mitigated by local annealers or alternative queueing schemes. For Power-law clustered networks, these times consistently increased (Fig.~\ref{fig:qpu_times}c), while they remained stable for Barabási--Albert and Erd\H{o}s--Rényi networks (Fig.~\ref{fig:qpu_times}f,i). The effect of the resolution parameter was inconsistent, likely due to the larger magnitude and unclear dependence of cached-embedding read times compared to QPU access times. For the Karate Club network, total sampling and communication times ranged from 7.1366~(s) for $\gamma=0.5$ to 30.8441~(s) for $\gamma=2$.

\subsection{Application of QA to structural brain connectivity}
\begin{figure}[h]
    \centering
    \includegraphics[width=1\linewidth]{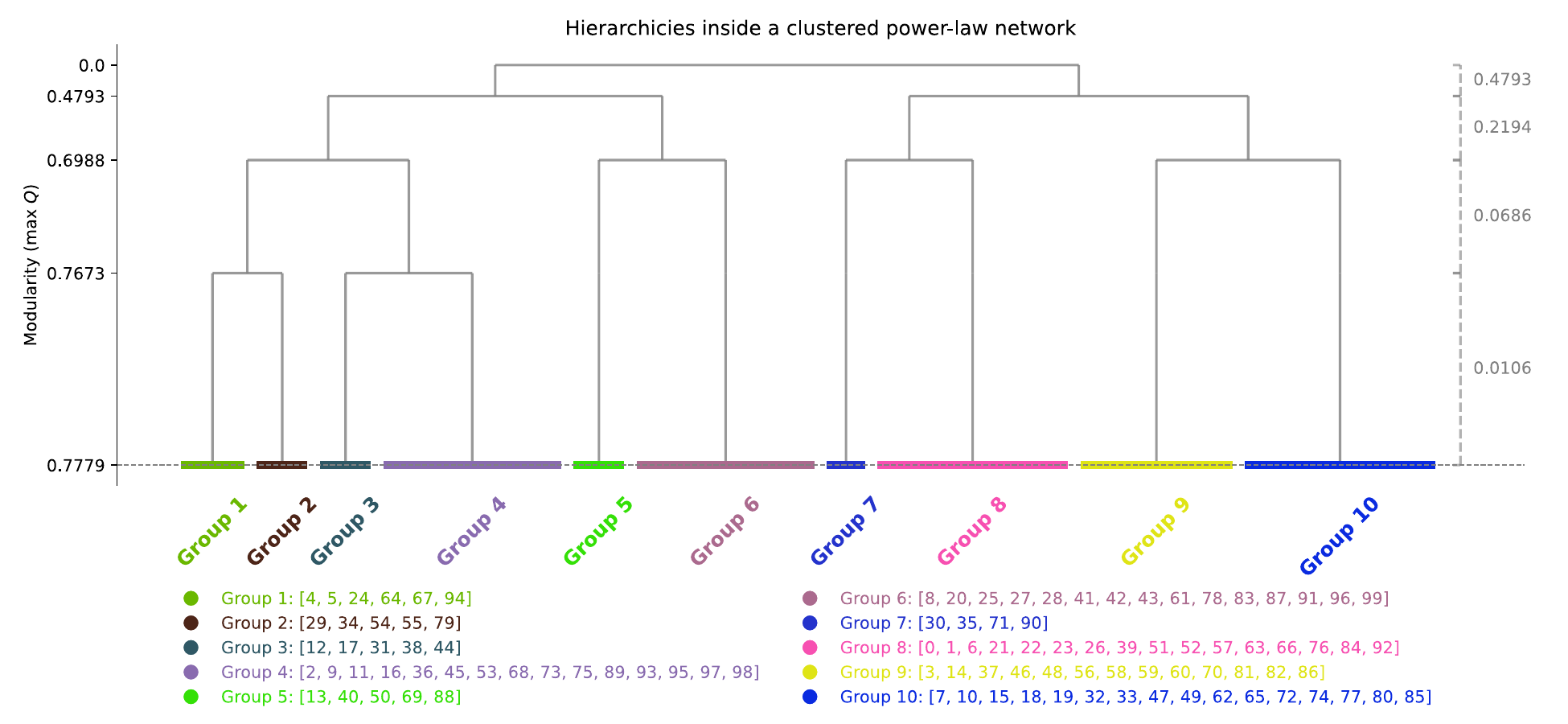}
    \caption{\textbf{Dendrogram of a clustered power-law network \cite{Holme2002} of N=100 nodes.} Hierarchical structure found by the H. annealing algorithm. The hierarchy corresponds to the one unraveled after maximizing the modularity of the network as described in the main text; that is, {\tt num\_runs = 15} and {\tt resolution = 1}. Both the Louvain and Leiden algorithms returned a maximum modularity of $Q=0.7779$. The left axis shows the modularity at each step of the process, while the right axis displays the corresponding increments. The bottom-most row represents the final output of the H. annealing algorithm.}
    \label{fig:dendro_pw}
\end{figure}

One of the main advantages of our proposal is that we could inspect the output process at every step of the process. We plotted the modularity and the community structure in the form of a \textit{dendrogram}. The final division was dependent on each earlier subdivision, thus being interpretable as a hidden community structure. Such hidden structures would be informative of certain network vulnerabilities or strengths. We applied this visualization technique to all the networks studied up until now, comparing the modularity of the final community structure with that of the same benchmark methods, achieving competitive and identical results (Fig. \ref{fig:dendro_pw}; see also Fig. S14-S18).

\begin{figure}[h!]
    \centering
    \includegraphics[width=0.75\linewidth]{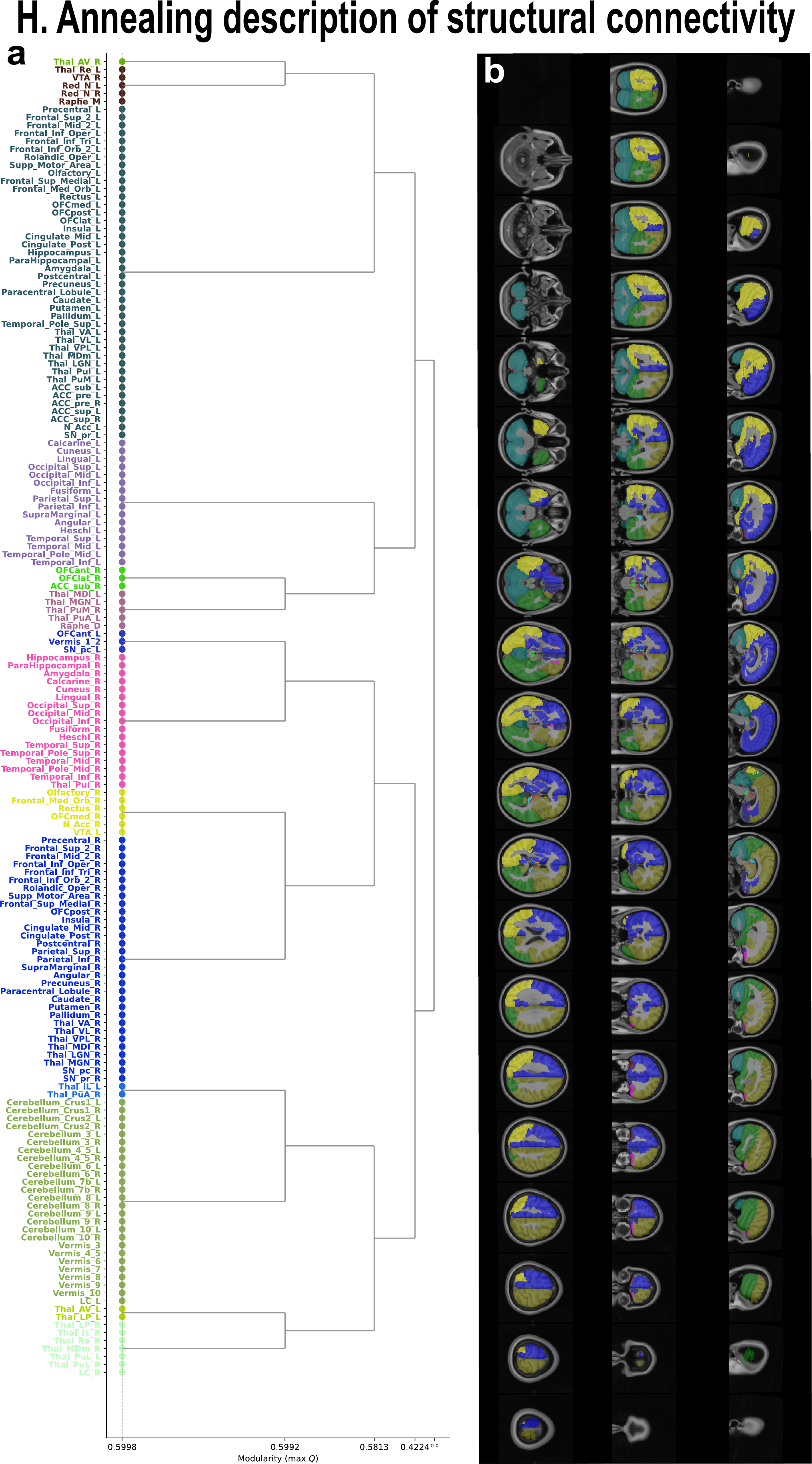}
    \caption{\textbf{Dendrogram of a real brain structural connectivity network \cite{falco2024functional}.} \textbf{a} Hierarchical structure found by the H. annealing algorithm. The hierarchy corresponds to the one unraveled after maximizing the modularity of the network as described in the main text; that is, $N_{runs}=10$ and $\gamma=1$. The Louvain and Leiden algorithms returned a maximum modularity of $Q=0.611$ and $Q=0.610$, respectively. The left axis shows the modularity at each step of the process, while the right axis displays the corresponding increments. \textbf{b} Overlay of the structural communities on the Montreal Neurological Institute (MNI) 152 template. Colors correspond
    to the communities in (\textbf{a}).}
    \label{fig:dendro_aal}
\end{figure}

As a last proof of concept, we analyzed real brain connectivity networks as derived from diffusion-weighted magnetic resonance imaging methods. These networks contained a total of 166 nodes and were densely connected while exhibiting somewhat scale-free structures \cite{falco2024functional}. We also inspected the anatomical positioning and organization of the found community structure as well as the hierarchical levels of the process (Fig. \ref{fig:dendro_aal}). Despite the large number of variables (i.e., qubits), the H. annealing solution was strikingly close to the solutions found by the Louvain and Leiden algorithms. Furthermore, the communities found by the quantum algorithm only differed slightly in the frontoparietal border, areas that are highly interconnected with U-fibers. The rest of the identified communities were largely in agreement between algorithms (Fig. S19-S20), hence setting a very promising set of baseline results for more advanced and sophisticated QA methods and algorithms.

\section{Discussion} \label{sec:discussion}

Here we present a quantum approach to discovering communities in complex networks. We unify the optimization process for binary, weighted, and directed networks and explicitly consider the resolution parameter to consider multiple topological scales present in the networks. Our experiments show that quantum annealing (QA) can be successfully used to maximize the modularity of a network and pinpoint meaningful groups of highly connected nodes. Its performance is robust across multiple experimental conditions and, at least comparable to alternative methods based on classical heuristics. However, the recursion in Algorithm~\ref{algo:HAnnealing} does not guarantee global optimality, nor do quantum annealing, simulated annealing, or other heuristics considered here for this NP-hard problem. From a theoretical standpoint, quantum and simulated annealing are closely related: both are probabilistic methods that sample Boltzmann-like distributions, with quantum fluctuations in QA playing a role analogous to thermal fluctuations in simulated annealing. While prior works study problem-dependent advantages of QA without universal superiority~\cite{Somma2007,Nishimori2015,Wauters2017}, results on current hardware remain mixed due to noise and finite annealing effects~\cite{Okada2019}. In this context, our results show that both simulated and hierarchical QA achieve competitive performance, with the latter offering practical advantages by avoiding an \emph{a priori} choice of the number of communities. Simulations of quantum hardware and processes could help assess whether any advantage persists at larger scales~\cite{King2021}, but they are computationally expensive and currently limited to systems with only a few physical qubits~\cite{Chen2022,morrell2024quantumannealing}.

Opposite to earlier attempts, we restrict ourselves to pure QPUs, avoiding hybrid solutions that obscure the algorithm results \cite{Negre2020,Wierzbinski2023} and provide an account of the possible origins of suboptimal performances. While many community detection algorithms rely on maximizing a quality function \cite{Fortunato2010}, not all are well-suited for every type of network~\cite{Ghasemian2020} and alternative definitions of community structure could be better suited~\cite{Rosvall2008,Peixoto2014}. Thus, the same could be said about our proposal, which explicitly relies on the existence of hierarchies. While we show how the Hierarchical annealing process performs well in networks without a clear topological organization, we note how and why exploiting inherent hierarchies could be beneficial rather than detrimental in the study of real-life complex systems. 

\subsection{Unveiling hierarchies in complex networks with quantum computing}
Biological and social networks are dynamic rather than static systems, even if we study their properties at a fixed point in time. Their evolution is thus constrained to the underlying mechanism that governs the interactions between their constituents. Barabasi and Albert \cite{Barabasi1999} showed that preferential attachment, where highly connected nodes attract more links, leads to scale-free networks with a slower decay of highly connected nodes than in random networks. Preferential attachment can lead to structural hierarchies, but the latter is not guaranteed \cite{Holme2002,barabasi2003scale}. Furthermore, life processes themselves are hierarchical \cite{oltvai2002life,alon2003biological}, inspiring specific methods to discover such organization \cite{SalesPardo-etal_2007}. In neuroscience, brain network structure is influenced by plasticity mechanisms occurring across temporal and spatial scales. Hebbian plasticity, in particular, has been shown to modulate preferential attachment, resulting in scale-free and clustered network structures \cite{Lynn2024}. It is therefore not unreasonable to assume some underlying hierarchical structure to design the algorithms. In the context of our work, we show how the Hierarchical annealing process performs exquisitely in these kinds of systems, even improving over classical alternatives (Fig. \ref{fig:topology}). %Additionally, network size did not significantly affect performance, suggesting the process is robust and scalable. 
The slightly suboptimal results we observed in some cases likely stem from the hierarchical process itself \cite{Arenas2008}, rather than issues with the annealing steps. Basically, in networks where the evidence of a scale-free, or at least, power-law degree distribution is weak \cite{Broido2019}, our Hierarchical annealing procedure is not guaranteed to produce the best community structure. Oppositely, hierarchical approaches should be the go-to since they are specially designed to exploit and interpret the results.

One of the main advantages of the hierarchical approach we studied is the direct interpretability of each one of the recursive divisions. We can construct \textit{dendrograms} of the sequence of divisions that yield the highest modularity indices for any given network. However, the correctness of such dendrograms depends entirely on the optimality of the represented solution. For that, it remains essential to compare the modularity of the community structure with solutions from other algorithms. As expected, the results obtained using H. annealing for all the networks studied in this work were extremely competitive, thus offering insights into the hidden structure of these networks (Fig. \ref{fig:dendro_pw}-\ref{fig:dendro_aal}). There exist some attempts to reveal hierarchies present in complex networks, most notably relying on bottom-up hierarchical clustering methods \cite{SalesPardo-etal_2007,Arenas2008} and the combination of resolution parameters \cite{lambiotte2013multi}, therefore offering a complementary rather than alternative description of fractal-like structure. A few other methods have been developed using the Louvain algorithm to unravel hierarchical relationships within complex systems. For example, each iteration of the Louvain algorithm can be interpreted as a step in a hierarchical process \cite{Blondel2008,meunier2009hierarchical}. Alternatively, separate instances of the Louvain algorithm may be used to discover community structures within sub-modules \cite{bassett2010efficient}. Instead, our approach considers the whole system at each step inside the hierarchy \cite{Newman2006,Leicht2008} (see also Eq. \ref{eq:p_k}). Because of that, and contrary to the Louvain solution, this guarantees that every step in the hierarchy is divided taking into account the whole system rather than local sub-modules, a desirable feature for strongly nested systems where our method is expected to outperform several alternatives (Fig. \ref{fig:topology}).

\subsection{Towards quantum-based multi-resolution community detection methods}
The specific case where the resolution parameter $\gamma$ is 1, for which we recover known results in the literature \cite{Newman2006}, simplifies several expressions throughout this work. However, discovering communities via maximizing the modularity function possesses an inherent limitation where communities smaller than a given size are likely to be omitted \cite{Fortunato2007}. Kumpula, \textit{et al.}, proved that this limit persisted for all $\gamma$ independently of the null model used in the definition of $Q$ \cite{Kumpula2007}. Although the resolution parameter was introduced to weigh the topological scale at which communities ought to be detected \cite{Reichardt2006}, it was elegantly shown that it could be employed to overcome the resolution limit \cite{Traag2011}, albeit disregarding the null model (e.g., $\frac{g_ig_j}{2m}$). Consequently, approaches that use different definitions for the modularity function and the parameter $\gamma$ also exist \cite{reichardt2004detecting,raghavan2007near,Arenas2008,ronhovde2010local}. However, they still fall within spin-glass systems; therefore, it is possible to tackle them using the same formalism we used. 

Nonetheless, Newman and Girvan's quality index \cite{Newman-Girvan_2004}, along with the explicit inclusion of a resolution parameter \cite{Reichardt2006}, remains, up to date, the most common choice for discovering community structure. Unfortunately, the exact value remains an arbitrary choice. In short, large values of the resolution will yield smaller communities, while smaller values of $\gamma$ will result in fewer communities of increased size. Hence, we rooted the parameter $\gamma$ inside our framework for two main reasons: 1) users need not tune the objective function to maximize, and 2) it easily adapts to multiresolution methods. We maintained $\gamma$ as a free parameter, and we derived all our expressions without specifying its value. This feature allows us to speculate that QA can effectively detect community structure using multiresolution methods, where $\gamma$ is gradually changed or sampled nonuniformly along the optimization process \cite{Jeub2018}. Further work should address whether this holds or not, the range for which the annealer cannot coordinate the large number of communities found, and the computational time required to obtain the large number of relatively small-sized communities expected for $\gamma > 3$.

Our experiments using multiple values of $\gamma$ within a meaningful range show that the Hierarchical annealing process is robust even in networks without inherent hierarchies (Fig. S10-S13). Interestingly, in directed scale-free networks, the hierarchical annealing algorithm yielded sub-optimal values for both low and high values of the resolution parameter but remained competitive in the neighborhood of $\gamma=1$ (Fig. S14). Given the stable results for directed networks at $\gamma=1$ (Fig.\ref{fig:random-nets}d) and for undirected networks across arbitrary $\gamma$, the symmetrization in Eq.\ref{eq:symmetrized_B_elements} might distort the topology and hierarchy of the symmetrized network relative to the original directed system. This effect is most pronounced at resolutions that favor either very large or very small communities, suggesting that a dedicated QA-based multi-resolution framework for directed networks may not be directly achievable through symmetric recursive approaches~\cite{Leicht2008}. Therefore, although $\gamma$ integrates naturally into the formalism, its value introduces subtle performance dependencies, possibly linked to tree depth, QUBO coefficients, or the physical requirements to embed a given problem, warranting further investigation.

\subsection{Alternative encoding strategies}
Quantum processing units (QPUs), regardless of type, are known to struggle with encoding large numbers of variables. In one-hot encoding, this number is of the order $N\times M$, where $N$ is the number of nodes in the graph and $M$ is the number of allowed communities. This is suboptimal because there is no guarantee that the network contains $M$ distinct communities, leading to wasted computational resources \cite{Negre2020}. Our approach, by contrast, limits the number of variables to the size of the network. Additionally, our method optimizes systems using discrete rather than binary variables, offering another advantage. Current approaches using one-hot encoding require satisfying a constraint for each variable, $\sum_{j=1,M}x_{ij}-1=0, \ \forall i=1, \ldots, N$, which ensures each node is uniquely assigned to a community. These constraints are typically relaxed by introducing Lagrange multipliers in the QUBO formulation. However, determining appropriate values for these multipliers often requires manual tuning or guessing, a process that becomes impractical for weighted networks, where even slight variations can alter the optimal value. While constrained optimization methods can be effective \cite{Ushijima-Mwesigwa2017,Negre2020}, they add a layer of complexity. DQM, a popular hybrid model of DWAVE, can solve discrete cases~\cite{Wierzbinski2023}, but incurs increased computational requirements (Table S1), sacrifices transparency due to undisclosed classical–quantum allocations, and ultimately exhibits behavior similar to our fully open approach.

While the results presented here were obtained using D-Wave hardware, the hierarchical strategy is broadly applicable to any solver that encodes communities via binary variables. In particular, the method can be implemented on quantum-inspired devices such as the Fujitsu Digital Annealer~\cite{jiang_benchmarks_2024} or on emerging quantum annealing systems such as those developed by the Nippon Electric Company (NEC) corporation~\cite{yamaji_development_2022}. In the case of gate-based devices, our solution can be used with the Quantum Approximate Optimization Algorithm (QAOA) algorithm~\cite{farhi_quantum_2022}. However, practical experiments in this direction are limited by additional issues, especially the high noise level of those devices and the need to set additional variational hyperparameters.

The applicability of our approach to quantum-inspired algorithms depends on the encoding requirements. For instance, the  QIEA-net and iQIEA-net algorithms~\cite{yuanyuan_quantum_2018} use integer-type encodings to assign communities to nodes. Consequently, our solution is not directly applicable in this context. Being classical, these algorithms are not bound by the binary encoding constraints of quantum devices. In practice, they achieved lower modularities for the Karate network (Table~\ref{tab:karate}). In contrast, the quantum-inspired simulated bifurcation algorithm~\cite{li_community_2025}, which correctly resolved the Karate club system, adopts a QUBO formulation with one-hot encoding constraints and Lagrange multipliers. While this approach conflicts with our goal of avoiding explicit constraints and one-hot encodings, it could benefit from recursively encoding binary variables, albeit not being affected by the practical aspects of using penalty weights with real annealers.

Finally, although we employed a recursive process, other combinations of multiple binary divisions could be designed to tackle this and other optimization problems \cite{Nembrini2022}, potentially leading to quantum-classical algorithms similar to the Louvain method. For example, in \cite{rathore_load_2025}, a similar hierarchical approach is used to divide the graph of data points into $2^n$ equal parts to achieve load balancing for parallel processing.

\subsection{Hardware constraints and scalability}
The dynamic range of values in the Ising Hamiltonian and the corresponding QUBO coefficients for optimization problems solved by D-Wave annealing devices may be limited. This arises from the need to represent these values in the D-Wave QPU, which uses an analog process and can result in limited precision for coefficient representation\footnote{For details, see the DWAVE documentation on integrated control errors (ICE).}. Therefore, if the ratio between the largest and smallest values in QUBO is relatively small, the problem is easier to solve by QPU (Supplementary Material S5; see also Fig. S21). Importantly, the range of the weights considered throughout this work did not compromise performance. Nonetheless, research should focus on mitigating the detrimental impact derived from these misrepresentations. Currently, there is ongoing research to overcome this issue~\cite{mucke2025}, but no unique or recommended solution exists.

Noise also affects quantum annealers (Supplementary Material S6). An indirect way to assess its presence is through the breaking of chains of physical qubits~\cite{Jeong2025}, where each chain encodes a single logical qubit. Chains and their breaking, therefore, indicate that the annealer is effectively solving a slightly perturbed version of the original QUBO problem~\cite{Grant2022}. Although both noise and chain breaks tended to increase with the number of nodes, they were not associated with a reduced ability to reach good solutions (Figs. S22 and S23). Instead, noise appeared to promote exploration of a broader set of alternatives, some better and some worse than the mean. The modularity $\hat{Q}$ is the best possible solution, not the average, and appropriately setting the chain strength guarantees a competitive performance.

Another key question is whether our QA approach can scale beyond the 166‑node networks tested. Pegasus connectivity limits embeddings to 169 variables on Advantage 5.4 (177 on Advantage 4.1). While heuristic embeddings could occasionally handle slightly larger graphs, our cached clique embeddings proved faster and more reliable, since they are computed once at the start. Consequently, scaling to larger problems necessitates D‑Wave’s hybrid solvers \cite{Wierzbinski2023}, but these incur long runtimes and opaque “black‑box” workflows that limit user control. Local compute resources can ease runtime issues, but transparency and flexibility remain unresolved.

In Algorithm \ref{algo:HAnnealing}, while the modularity matrix requires $\mathcal{O}(n^2)$, the generalized modularity matrix \textbf{B}\textsuperscript{$k$} is computed in linear time, $\mathcal{O}(n)$, since it is a subset of \textbf{B}. Therefore, a realistic estimate of the complexity of the recursion tree is \(\mathcal{O}(n\log n)\) corresponding to the average depth of the tree \cite{Newman2006}. The Akra-Bazzi theorem allows for a more formal proof. Highly unbalanced trees have linear complexity, but they are very unlikely. Thus, the overall complexity of the hierarchical procedure is $\mathcal{O}(n^2 + n\log n+\mathcal{K}_n)$, where $\mathcal{K}_n$ is the complexity attributed to the repeated binary QA steps. Hence, the computation of \textbf{B} dominates over the recursion, regardless of the intermediate quantum-based steps.

The Louvain and Leiden algorithms scale between \(\mathcal{O}(n\log n)\) and \(\mathcal{O}(n^2)\)~\cite{Lancichinetti2009}, placing the classical component of our approach within the range of established methods~\cite{Girvan2002,Duch2005,Newman2004}. Annealing time scales inversely with the square of the minimum energy gap, which depends on the Hamiltonian~\cite{Farhi2001,Albash2018}, yielding \(\mathcal{K}_n\) complexities from polynomial to exponential. Empirically, QPU access time followed a logarithmic trend, \(\mathcal{K}_n \sim \mathcal{O}(\log n)\). Although extrapolation to larger networks was not feasible, our results suggest a mixed linear--logarithmic dependence of QPU time on problem size (Table~\ref{tab:r-squared}). QPU access times remained below 1~s even for the largest tested networks, with a strong dependence on network topology. However, communication with the DWAVE servers, combined with internal processing and queueing, consistently exceeded QPU computation times. This overhead may hinder scaling to larger networks, whereas local QPU access (e.g., in supercomputing centers) or user-specific sessions could substantially improve the availability and responsiveness of quantum resources.

The advantage of Algorithm \ref{algo:HAnnealing} is most evident for discrete problems, where the one-hot vector dimension $M\leq n$ must accommodate the correct solution. Setting \(M \propto n\) increases the annealing complexity to \(\mathcal{O}(\mathcal{K}_{n^2})\), likely worsened by the need to fit Lagrange multipliers. A more practical estimate for \(M\) follows from the resolution limit~\cite{Fortunato2007,Kumpula2007}, where the maximum number of detectable communities scales as \(M \sim \sqrt{n}\), resulting in \(\mathcal{O}(\mathcal{K}_{n\sqrt{n}})\). In both cases, nonetheless, recursion is expected to provide a notable speedup while avoiding Lagrange multiplier fitting (Table~\ref{tab:complexity_onehot}). Consequently, as QPU connectivity and qubit counts improve, hybrid and quantum methods like ours could scale competitively against classical heuristics without compromising the results.

\begin{table}[h] 
    \centering
    \resizebox{\columnwidth}{!}{%
        \begin{tabular}{l l||c|c}
        & $f(n)$ & Recursion complexity & One-hot encoding complexity \\
        \hline
        I. & $c$ & $O(n\log n)$ & $\mathcal{O}(1)$ \\
        II. & $\log n$ & $O(n\log n)$ & $\mathcal{O}(\log n)$ \\
        III. & $n$ & $\mathcal{O}(n \log n)$ & $\mathcal{O}(n\sqrt{n})$\\
        IV. & $n^s, \ s > 1$ & $\mathcal{O}(n^s)$ & $\mathcal{O}(n^sn^{s/2})$\\
        V. & $e^{n^s}, \ s \in \mathbb{R}^+$ & $\mathcal{O}(n e^{n^s})$ & $\mathcal{O}(e^{n^sn^{s/2}})$\\
        \end{tabular}
    }
    \caption{\textbf{Computational complexities of the recursion in Algorithm~\ref{algo:HAnnealing} and one-hot encoding for different $f(n)$ forms.} In the $n \to \infty$ limit, cases~III to V favor recursion, whereas cases~I and II favor one-hot encoding. The parameter $c$ is any positive and finite constant. The total QA $\mathcal{K}_n$ and recursion complexities depend on $f$ via Eq.~\ref{eq:complexity_quantum}. The corresponding limits and calculations are detailed in the Supplementary Materials (Section S4.3).}
    \label{tab:complexity_onehot}
\end{table}

\subsubsection*{Limitations and future work}
We did not study signed networks. Although these networks allow the same spin-glass description used throughout this work \cite{Traag2009}, some definitions are tailored to each scientific discipline \cite{fornito2016fundamentals}. This calls for specific rather than generic descriptions of the function to optimize, something that we wanted to avoid here. However, it is necessary to assess the performance of QA in these types of systems where non-trivial interactions can emerge \cite{kirkley2019balance}.

\section{Conclusions}

We use the D-Wave Advantage annealer to maximize modularity via a hierarchical scheme that solves binary subproblems, avoids one-hot encoding, and automatically determines the number of communities. We show analytically that the objective is independent of the resolution parameter and network type. Experiments demonstrate robustness across topologies, with strong performance on scale-free networks, while the hierarchical framework could support multiresolution analysis to yield interpretable community structures.

\section*{Acknowledgments}
The publication was created within the project of the Minister of Science and Higher Education "Support for the activity of Centers of Excellence established in Poland under Horizon 2020" on the basis of the contract number MEiN/2023/DIR/3796. This project has received funding from the European Union’s Horizon 2020 research and innovation programme under grant agreement No 857533. This publication is supported by the Sano project carried out within the International Research Agendas programme of the Foundation for Polish Science, co-financed by the European Union under the European Regional Development Fund (J.F.-R., B.W., K.C., and K.R.). The research presented in this paper received support from the funds assigned by the Polish Ministry of Science and Technology to AGH University (K.J., K.C., and K.R.). We gratefully acknowledge the Polish high-performance computing infrastructure PLGrid (HPC Center: ACK Cyfronet AGH) for providing computer facilities and support within the computational grant no. PLG/2025/018857 (J.F.-R., K.J., B.W., and K.R.), PLG/2024/017319 (J.F.-R.), and PLG/2024/017418 (B.W.).

\FloatBarrier
\bibliography{bibliography}

\clearpage

\setcounter{figure}{0}
\renewcommand{\thefigure}{S\arabic{figure}}
\setcounter{equation}{0}
\renewcommand{\theequation}{S1.\arabic{equation}}
\setcounter{section}{0}
\renewcommand{\thesection}{S\arabic{section}}
\setcounter{table}{0}
\renewcommand{\thetable}{S\arabic{table}}
\clearpage
\vfill
\begin{center}
    {\Large\bfseries Supplementary Material}
    
    \vspace{1em}
\end{center}
\vfill

\section{Weighted and undirected networks} \label{app:A}
For an undirected network, Newman's quality function for a given community structure consisting of only two modules is written as \cite{Wierzbinski2023}:
\begin{equation} \label{eq_A:ising}
    \begin{split}
        Q & = \sum_{c=1}^{2}\left[ \frac{L_c}{m} - \gamma \left(\frac{g_c}{2m}\right)^2\right] \\
          & = \frac{1}{2m} \sum_{ij\in G}\left( A_{ij} - \gamma \frac{g_i g_j}{2m} \right) \left(\frac{s_i s_j + 1}{2}\right)\\
          & = \frac{1}{4m} \sum_{ij\in G} B_{ij} s_i s_j + \frac{(1-\gamma)}{2},
    \end{split}
\end{equation} Quantum annealing (QA) can also work in binary variables $x_i=\{0,1\}$ instead of spin-like ones. We perform a simple transformation $s_i = 2x_i - 1$ for that. With this in mind, the complete quality function is as follows:
\begin{equation} \label{eq_A:QUBO}
    Q = \frac{1}{m}\sum_{ij\in G} B_{ij} x_i x_j - \frac{(1-\gamma)}{m}\sum_{i\in G} g_{i} x_i + 1-\gamma.
\end{equation}

Since QA minimizes a function up to a constant we can rewrite the function to optimize as follows:

\begin{equation} \label{eq_A:QUBO_optim}
    \begin{split}
        \Tilde{Q} & \doteq - m \left(Q - 1 + \gamma \right) \\
                  & = -\sum_{ij\in G} B_{ij} x_i x_j + (1-\gamma)\sum_{i\in G} g_{i} x_i \\
                  & = -\sum_{ij\in G} \left( A_{ij} - \gamma \frac{g_i g_j}{2m} \right) x_i x_j + (1-\gamma)\sum_{i\in G} g_{i} x_i,
    \end{split}
\end{equation} which can be optimized via QA to obtain high-quality divisions of a network into two communities \cite{Ushijima-Mwesigwa2017,Negre2020,Nembrini2022,Wierzbinski2023}. Thus, we base our method on the fact that QA works well in simple QUBO problems without one-hot encoding.

\subsection{Properties of the symmetric modularity matrix}
The sum of the elements in a single column (rows):
\begin{equation}\label{eq_A:sym_und_rows}
    \sum_{j \in G}B_{ij} = \sum_{j \in G}\left( A_{ij} - \gamma \frac{g_i g_j}{2m}\right) = (1 - \gamma)g_i.
\end{equation}
The sum of the elements in a single row (columns):
\begin{equation}\label{eq_A:sym_und_cols}
    \sum_{i \in G}B_{ij} = \sum_{i \in G}\left( A_{ij} - \gamma \frac{g_i g_j}{2m}\right) = (1 - \gamma)g_j.
\end{equation}
Given that the matrix is symmetric, $\forall i,j \in G$ the following also holds: $g_i=g_j$. Finally, the sum of all elements,
\begin{equation} \label{eq_A:sym_und_elements}
    \sum_{ij \in G}B_{ij} = 2m(1-\gamma).
\end{equation} With these, we can rewrite the modularity $Q$ using both Ising and binary variables (i.e., Eqs. \ref{eq_A:ising} and \ref{eq_A:QUBO}).

\subsection{Properties of the symmetric generalized modularity matrix}

\textit{The entire network forms a single community }($g=G$):

The generalized modularity matrix \textbf{B}\textsuperscript{g} can be simplified to
\begin{equation} \label{eq_A:gen_mod_full_G}
    \begin{split}
    B^{\text{g}}_{ij} & = B_{ij} - \delta_{ij}\sum_{k \in G} B_{ik} \\
                      & = B_{ij} - (1-\gamma)g_i\delta_{ij}, \\
    \end{split}
\end{equation} where the second equality follows from Eqs. \ref{eq_A:sym_und_rows}, \ref{eq_A:sym_und_cols} and, as stated in the main text, $\delta_{ij}$ is the Kronecker delta function. The following is thus straightforward, but for completeness, we explicitly sum the elements of rows and columns. The sum of the elements in a single column (rows):
\begin{equation}\label{eq_A:gen_sym_und_rows_full_g}
    \begin{split}
    \sum_{j \in G} B^{\text{g}}_{ij} & = \sum_{j \in G} \left[ B_{ij} - (1-\gamma)g_i\delta_{ij} \right ]\\
                                     & = (1-\gamma)g_i - (1-\gamma)g_i \sum_{j \in G} \delta_{ij} \\
                                     & = 0 \ \ \forall \gamma,
    \end{split}
\end{equation} where the third equality stems from $\sum_{i \in G} \delta_{ij}=1$. The sum of the elements in a single row (columns): 
\begin{equation}\label{eq_A:gen_sym_und_cols_full_g}
    \begin{split}
    \sum_{i \in G} B^{\text{g}}_{ij} & = \sum_{i \in G} \left[ B_{ij} - (1-\gamma)g_i\delta_{ij} \right ] \\
                                     & = (1-\gamma)g_j - (1-\gamma)g_j \sum_{i \in G} \delta_{ij} \\
                                     & = 0 \ \ \forall \gamma,
    \end{split}
\end{equation} where, once again, the third equality is true given that $g=G$. Contrary to the modularity matrix \textbf{B}, the sum of all elements of the generalized modularity matrix \textbf{B}\textsuperscript{g} is equal to zero, that is,  $\forall \gamma$,
\begin{equation} \label{eq_A:gen_sym_und_elements_full_g}
    \sum_{ij \in G}B^{\text{g}}_{ij}=0.
\end{equation} Noteworthy, the two matrices are identical if the resolution parameter is equal to 1 \cite{Newman2006}.

\textit{The entire network has already been divided into an arbitrary number of communities and we wish to further subdivide one of them} ($g \subset G$):

The sum of the elements in a single column (rows):
\begin{equation}\label{eq_A:gen_sym_und_cols_g}
    \begin{split}
    \sum_{j \in g} B^{\text{g}}_{ij} & = \sum_{j \in g} \left[ B_{ij} - \delta_{ij}\sum_{k \in g} B_{ik} \right ] \\
    & = \sum_{j \in g} \left[ A_{ij} - \gamma \frac{g_i g_j}{2m} - \delta_{ij}\sum_{k \in g} \left( A_{ik} - \gamma \frac{g_i g_k}{2m}\right) \right ] \\
    & = \sum_{j \in g} A_{ij} - \frac{\gamma g_i}{2m} \sum_{j \in g}g_j - \sum_{j \in g} \delta_{ij}\sum_{k \in g} A_{ik} + \frac{\gamma g_i}{2m}\sum_{j \in g} \delta_{ij} \sum_{k \in g} g_k \\
    &= \sum_{j \in g} A_{ij} - \sum_{j \in g} \delta_{ij}\sum_{k \in g} A_{ik} - \frac{\gamma g_i}{2m} \left[ \sum_{j \in g}g_j - \sum_{j \in g} \delta_{ij} \sum_{k \in g} g_k \right] \\
    &= \sum_{j \in g} A_{ij} - \sum_{k \in g} A_{ik} - \frac{\gamma g_i}{2m} \left[ \sum_{j \in g}g_j - \sum_{k \in g} g_k \right] \\
    & = 0 \ \ \forall \gamma,
    \end{split}
\end{equation} where we have used $\sum_{j \in g}\delta_{ij}=1$. A similar derivation could be done for the sum of the elements in a single column (rows); however, we can use make use of the fact that the modularity matrix is symmetric for this case. That is,
\begin{equation}\label{eq_A:gen_sym_und_rows_g}
    \sum_{i \in g} B^{\text{g}}_{ij} = \sum_{i \in g} B^{\text{g}}_{ji} = 0 \ \ \forall \gamma.
\end{equation} Therefore, as opposed to the sum of the elements of the modularity matrix, the sum of the elements in $g$ of the generalized modularity matrix add up to zero for an arbitrary resolution parameter; that is, $\forall \gamma$,
\begin{equation} \label{eq_A:gen_sym_und_elements_g}
    \sum_{ij \in g}B^{\text{g}}_{ij}=0.
\end{equation} 

\setcounter{equation}{0}
\renewcommand{\theequation}{S2.\arabic{equation}}
\section{Weighted and directed networks} \label{app:B}
For directed networks, the quality function is slightly different but the procedure is similar to the one outlined before.
\begin{equation} \label{eq_B:ising_dir}
    \begin{split}
        Q & = \sum_{c=1}^{2}\left[ \frac{L_c}{m} - \gamma \left(\frac{g_c^{in}g_c^{out}}{m^2}\right)\right] \\
          & = \frac{1}{m} \sum_{ij\in G}\left( A_{ij} - \gamma \frac{g_i^{in} g_j^{out}}{m} \right) \left(\frac{s_i s_j + 1}{2}\right)\\
          & = \frac{1}{2m} \sum_{ij\in G} B_{ij}^{d} s_i s_j + \frac{(1-\gamma)}{2},
    \end{split}
\end{equation} where $B_{ij}^d = A_{ij} - \gamma \frac{g_i^{in} g_j^{out}}{m}$ is the now the \textit{directed} modularity matrix, $m=\sum_{ij}A_{ij}$ is the total edge-weight, $g_i^{in}=\sum_j A_{ij}$ is the node weighted in-degree, $g_j^{out}=\sum_i A_{ij}$ is the node weighted out-degree, and $s_i = \{-1,1\} \ \forall i \in G$ are spin-like variables. Once again, we apply the same transformation $s_i=2x_i-1$ to obtain the quality function in terms of binary variables.

\begin{equation} \label{eq_B:QUBO_dir}
    Q = \frac{2}{m}\sum_{ij\in G} B_{ij}^d x_i x_j - \frac{(1-\gamma)}{m}\left[\sum_{i\in G} g_{i}^{in} x_i + \sum_{j\in G} g_{j}^{out} x_j\right] + 1-\gamma.
\end{equation}

Once again, we can optimize the previous function up to a constant. Then, we proceed in an identical manner:

\begin{equation} \label{eq_B:QUBO_dir_optim}
    \begin{split}
        \Tilde{Q} & \doteq - \frac{m}{2} \left(Q - 1 + \gamma \right) \\
                  & = -\sum_{ij\in G} B_{ij}^d x_i x_j + \frac{(1-\gamma)}{2}\left[\sum_{i\in G} g_{i}^{in} x_i + \sum_{j\in G} g_{j}^{out} x_j \right]\\
                  & = -\sum_{ij\in G} \left( A_{ij} - \gamma \frac{g_i^{in} g_j^{out}}{m} \right) x_i x_j + \frac{(1-\gamma)}{2}\left[\sum_{i\in G} g_{i}^{in} x_i + \sum_{j\in G} g_{j}^{out} x_j \right].
    \end{split}
\end{equation} which can be optimized via QA to obtain high-quality divisions of a directed network into two communities. Noteworthy, the undirected and directed cases are equivalent only if the resolution is equal to 1. 

\subsection{Properties of the asymmetric modularity matrix}
The sum of the elements in a single column (rows):
\begin{equation}\label{eq_B:dir_rows}
    \sum_{j \in G}B_{ij}^d = \sum_{j \in G}\left( A_{ij} - \gamma \frac{g_{i}^{in} g_{j}^{out}}{m}\right) = (1 - \gamma)g_i^{in}.
\end{equation}
The sum of the elements in a single row (columns):
\begin{equation}\label{eq_B:dir_cols}
    \sum_{i \in G}B_{ij}^d = \sum_{i \in G}\left( A_{ij} - \gamma \frac{g_{i}^{in} g_{j}^{out}}{m}\right) = (1 - \gamma)g_j^{out}.
\end{equation}
Finally, the sum of all elements,
\begin{equation} \label{eq_B:dir_elements}
    \sum_{ij \in G}B_{ij}^d = m(1-\gamma).
\end{equation} With these, we can rewrite the modularity $Q$ using both Ising and binary variables (i.e., Eqs. \ref{eq_B:ising_dir} and \ref{eq_B:QUBO_dir}).

\subsection{Properties of the asymmetric generalized modularity matrix}

\textit{The entire network forms a single community }($g=G$):

The directed generalized modularity matrix \textbf{B}\textsuperscript{d,g} can be simplified to
\begin{equation} \label{eq_B:dire_gen_mod_full_G}
    \begin{split}
    B^{d,\text{g}}_{ij} & = B_{ij}^d - \delta_{ij}\sum_{k \in G} B_{ik}^d \\
                      & = B_{ij}^d - (1-\gamma)g_{i}^{in}\delta_{ij}, \\
    \end{split}
\end{equation} where the second equality follows from Eqs. \ref{eq_B:dir_rows}, \ref{eq_B:dir_cols} and, as stated in the main text, $\delta_{ij}$ is the Kronecker delta function. The following is thus straightforward, but for completeness, we explicitly sum the elements of rows and columns. The sum of the elements in a single column (rows):
\begin{equation}\label{eq_B:gen_dir_rows_full_g}
    \begin{split}
    \sum_{j \in G} B^{d,\text{g}}_{ij} & = \sum_{j \in G} \left[ B_{ij}^d - (1-\gamma)g_{i}^{in}\delta_{ij} \right ]\\
                                     & = (1 - \gamma)g_i^{in} - (1 - \gamma)g_i^{in} \sum_{j \in G} \delta_{ij} \\
                                     & = 0 \ \ \forall \gamma,
    \end{split}
\end{equation} where the third equality stems from $\sum_{i \in G} \delta_{ij}=1$. The sum of the elements in a single row (columns): 
\begin{equation}\label{eq_B:gen_dir_cols_full_g}
    \begin{split}
    \sum_{i \in G} B^{d,\text{g}}_{ij} & = \sum_{i \in G} \left[ B_{ij}^d - (1-\gamma)g_{i}^{in}\delta_{ij} \right ] \\
                                     & = (1-\gamma)g_j^{out} - (1-\gamma) \sum_{i \in G} g_i^{in}\delta_{ij} \\
                                     & = (1-\gamma)(g_j^{out}-g_j^{in}) \ \ \forall \gamma.
    \end{split}
\end{equation} Contrary to the directed modularity matrix \textbf{B}\textsuperscript{d}, the sum of all elements of the directed generalized modularity matrix \textbf{B}\textsuperscript{d,g} is equal to zero, that is,  $\forall \gamma$,
\begin{equation} \label{eq_B:gen_dir_elements_full_g}
    \sum_{ij \in G}B^{\text{g}}_{ij}=0.
\end{equation} This is due to the fact that summing the in- and out-degrees is equivalent. Noteworthy, the two matrices are identical if the resolution parameter is equal to 1 \cite{Leicht2008}.

\textit{The entire network has already been divided into an arbitrary number of communities and we wish to further subdivide one of them} ($g \subset G$):

The sum of the elements in a single column (rows):
\begin{equation}\label{eq_B:gen_dir_cols_g}
    \begin{split}
    \sum_{j \in g} B^{d,\text{g}}_{ij} & = \sum_{j \in g} \left[ B^{d}_{ij} - \delta_{ij}\sum_{k \in g} B^{d}_{ik} \right ] \\
    & = \sum_{j \in g} \left[ A_{ij} - \gamma \frac{g^{in}_i g^{out}_j}{2m} - \delta_{ij}\sum_{k \in g} \left( A_{ik} - \gamma \frac{g^{in}_i g^{out}_k}{2m}\right) \right ] \\
    & = \sum_{j \in g} A_{ij} - \frac{\gamma g^{in}_i}{2m} \sum_{j \in g}g^{out}_j - \sum_{j \in g} \delta_{ij}\sum_{k \in g} A_{ik} + \frac{\gamma g^{in}_i}{2m}\sum_{j \in g} \delta_{ij} \sum_{k \in g} g^{out}_k \\
    &= \sum_{j \in g} A_{ij} - \sum_{j \in g} \delta_{ij}\sum_{k \in g} A_{ik} - \frac{\gamma g^{in}_i}{2m} \left[ \sum_{j \in g}g^{out}_j - \sum_{j \in g} \delta_{ij} \sum_{k \in g} g^{out}_k \right] \\
    &= \sum_{j \in g} A_{ij} - \sum_{k \in g} A_{ik} - \frac{\gamma g^{in}_i}{2m} \left[ \sum_{j \in g}g^{out}_j - \sum_{k \in g} g^{out}_k \right] \\
    & = 0 \ \ \forall \gamma,
    \end{split}
\end{equation}

Contrary to the symmetric case, the sum of the elements in a single rows (columns) is different than zero. To show this, we proceed similarly as above.
\begin{equation}\label{eq_B:gen_dir_rows_g}
    \begin{split}
    \sum_{i \in g} B^{d,\text{g}}_{ij} & = \sum_{j \in g} \left[ B^{d}_{ij} - \delta_{ij}\sum_{k \in g} B^{d}_{ik} \right ] \\
    & = \sum_{i \in g} B^{d}_{ij} - \left( \sum_{i \in g} \delta_{ij}\sum_{k \in g} B^{d}_{ik} \right ) \\
    & = \sum_{i \in g} B^{d}_{ij} - \sum_{k \in g} B^{d}_{jk}\\
    & = \sum_{i \in g} \left ( B^{d}_{ij} - B^{d}_{ji} \right ) \ \ \forall \gamma.
    \end{split}
\end{equation} In general, the above results are different zero given that the matrix is not symmetric (i.e., $B^{d}_{ij} \neq B^{d}_{ji}$). Straightforwardly, the sum of the elements in $g$ of the directed generalized modularity matrix also adds up to zero.
\begin{equation} \label{eq_B:gen_dir_elements_g}
    \sum_{ij \in g}B^{d,\text{g}}_{ij}=0.
\end{equation}

\subsection{Computation of the \textit{k}-th element of the process without symmetrizing the network}
Proceeding similarly to the undirected case, the modularity of an undirected network of $|C^{k-1}|$ different communities can be decomposed as 
$$ Q(C^{k}) = \sum_{c \in \Tilde{C}^{k-1}} \left[ \frac{L_c}{m} - \gamma \left(\frac{g^{in}_c g^{out}_c}{m^2}\right)\right] + \frac{L_k}{m} - \gamma \left(\frac{g^{in}_k g^{out}_k}{m^2}\right). $$ Then, the modularity index of the same network after splitting the $k$-th community in two can also be decomposed as
$$ Q(C^{k+1}) = \sum_{c \in \Tilde{C}^{k-1}} \left[ \frac{L_c}{m} - \gamma \left(\frac{g^{in}_c g^{out}_c}{m^2}\right)\right] + \sum_{r=1}^{2} \left[ \frac{L_{k_{r}}}{m} - \gamma \left(\frac{g^{in}_{k_{r}} g^{out}_{k_{r}}}{m^2}\right) \right]$$ where $k_1 \cup k_2 = k$. Subtracting both expressions and using Eq. \ref{eq_B:ising_dir} we obtain a compact expression for the elements of $\mathcal{P}$ in terms of the directed modularity matrix,
\begin{equation} \label{eq_B:pk_dir}
    \begin{split}
        p_k &= \frac{-2}{m} \sum_{ij \in k} \left( B_{ij}^ds_i s_j - B^d_{ij}\right) - \frac{1-\gamma}{2} \\
        &= \frac{-2}{m} \sum_{ij \in k} \left( B^d_{ij} - \delta_{ij} \sum_{r\in k} B^d_{ir}\right)s_i s_j = \frac{-2}{m} \sum_{ij \in k} B_{ij}^{d,\text{k}} s_i s_j \\
        &= \frac{-2}{m} \left[ 4 \sum_{ij \in k} B_{ij}^{d,\text{k}} x_i x_j -2 \sum_{ij \in k} B_{ij}^{d,\text{k}} x_i + -2 \sum_{ij \in k} B_{ij}^{d,\text{k}} x_j + \sum_{ij \in k} B_{ij}^{d,\text{k}} \right]\\
        &= \frac{-8}{m} \left[ \sum_{ij \in k} B_{ij}^{d,\text{k}} x_i x_j  - \frac{1}{2} \sum_{j\in k} x_j \sum_{i\in k} \left(B_{ij}^d - B_{ji}^d \right)\right],
    \end{split}
\end{equation} where $\delta_{ij}$ is the Kronecker delta function, we used $s_i^2=1$, and 
\begin{equation} \label{eq_B:dir_gen_mod_matrix}
    B_{ij}^{d,\text{k}} = B_{ij}^d - \delta_{ij} \sum_{r\in k} B_{ir}^d \ \ \forall i,j \in k,
\end{equation}is the directed generalized modularity matrix \textbf{B}\textsuperscript{$d$,k} for the $k$-th community \cite{Newman2006}. In the second row, we have applied a linear mapping from the Ising variables to a set of binary ones $s_i=2x_i-1$ to obtain the QUBO formulation. The last row follows from some of the properties of the generalized modularity matrix outlined in the previous section in the Supplements.

Finally, for $k=0$, $$p_0 = -4 \left[ \frac{2}{m} \sum_{ij\in G} B_{ij}^d x_i x_j - \frac{2(1-\gamma)}{m}\sum_{i \in G} g^{in}_i x_i - \frac{1}{2m} \sum_{i\in k} x_i \left(B_{ij}^d - B_{ji}^d \right)\right],$$ which does not resemble Eqs. \ref{eq_B:QUBO_dir} or \ref{eq_B:QUBO_dir_optim}.

\setcounter{equation}{0}
\renewcommand{\theequation}{S3.\arabic{equation}}
\section{Sampling the Hierarchical Annealing procedure a reasonable number of times}
Because our algorithm has inherent stochasticity, the number of times we observe the maximum modularity value across repeated runs can be modeled as a discrete random variable. Let a \textit{success} denote the event that a single run of the hierarchical annealing procedure achieves the maximum modularity value $\max Q$ (the global optimum for the problem under study). If we repeat the experiment independently $n$ times (i.e., $N_{runs}$), and if $p$ denotes the probability of observing this optimal value in a single run, then the number of successes $X$ follows a binomial distribution,
\begin{equation}
    X \sim \mathrm{Binomial}(n,p).
\end{equation}

The probability of observing exactly $k$ successes is
\begin{equation}
    \begin{split}
        P(X = k) &= \binom{n}{k} p^{k}(1-p)^{n-k} \\
        &= \frac{n!}{k!(n-k)!}p^{k}(1-p)^{n-k},
    \end{split}
\end{equation}
for $k = 0,1,\dots,n$. We are primarily interested in the probability of observing at least one success: achieving $\max Q$ at least once across the $n$ independent runs:
\begin{equation} \label{eq:atleast_one_success}
    \begin{split}
        P(X \geq 1) &= \sum_{r=1}^{n} P(X = r) \\
        &= 1 - P(X = 0) \\
        &= 1 - (1-p)^n.
    \end{split}
\end{equation}

The challenge is that the success probability $p$ is unknown in practice, and it may vary substantially depending on the structure and modularity landscape of the complex network under investigation. Consequently, before determining an appropriate number of repetitions $n$ for the hierarchical annealing procedure, we must form a reasonable estimate of $p$. Ideally, one could run the experiment an arbitrarily large number of times. However, each run is costly; for example, executing a single experiment on DWAVE consumes a non-negligible number of computing minutes. Although repeatedly running the experiment guarantees that we will eventually observe $\max Q$ even when $p$ is small (Fig.~\ref{fig:sampling_runs}a; see also Eq. \ref{eq:atleast_one_success}), doing so is highly inefficient for most plausible values of $p$. In practice, once we are only concerned with observing at least one success ($k \geq 1$), excessively repeating the experiment yields diminishing returns and substantial waste of computational resources.

\subsection{Estimating the probability of success}
Suppose we repeat the hierarchical annealing procedure a total of \texttt{num\_runs} times. Using Laplace’s principle, we can estimate the probability of observing a success by the empirical frequency with which the algorithm achieves $\max Q$:
\begin{equation} \label{eq:p_laplace}
    p \approx \frac{\text{\# times where } Q = \max Q}{N_{runs}}.
\end{equation} However, this estimate can be highly variable when the number of repetitions is small. In our case, with only $\texttt{num\_runs} = 20$, the resulting estimate in Eq. \ref{eq:p_laplace} may be too noisy to rely on directly. 

To obtain a more robust estimate of the probability of observing the maximum modularity value, we used a nonparametric bootstrap procedure. For each network, we treated the set of modularity values obtained from the original \texttt{num\_runs} repetitions as the empirical distribution. We then generated bootstrap samples by resampling these values with replacement. Each bootstrap sample consisted of $N_{\mathrm{resamples}}$ draws, ranging from 10 to 1000 in increments of 10, to study the convergence of the boostrapped estimate of $p$. For all bootstrap samples, we computed the relative frequency with which the maximum modularity $\max Q$ was observed. Repeating this procedure $N_{\mathrm{experiments}}=500$ times for each value of $N_{\mathrm{resamples}}$ yielded a collection of estimates of the success probability. We recorded the mean of these bootstrap estimates to characterize the expected likelihood of observing a success $\mathcal{L}(\max Q) \to p$. This provided a non-parametric approximation of the underlying probability in Eq. \ref{eq:p_laplace}.

Once we obtained a stable estimate of $p$ from the bootstrap procedure with $N_{\mathrm{resamples}}=1000$, we computed the theoretical probability of observing the maximum modularity at least once (Eq.~\ref{eq:atleast_one_success}). Across different networks, this probability remained above 50\% for most cases and dropped to 40\% in only two instances (Fig.~\ref{fig:sampling_runs}b). These results confirm that even if the likelihood of observing $\max Q$ is relatively small (Fig.~\ref{fig:sampling_runs}c), running the hierarchical annealing procedure 20 times provides a reasonable chance of achieving very high modularity values once.

Furthermore, to provide a transparent account of the behavior of the hierarchical annealing procedure, we report both $\min Q$ and $\max Q$ along with the corresponding bootstrapped likelihood estimates (Tables~\ref{tab:powerlaw}-\ref{tab:directedscalefree}). As expected, the frequency of observing $\max Q$ decreased as the number of nodes increased, but the bootstrap estimate of $p$ remained stable as $N_{\mathrm{resamples}}$ increased. Notably, the most likely value of $Q$ did not differ substantially from $\max Q$, reflecting both the high degeneracy of the modularity function and the very competitive performance of the recursive sampling performed by the DWAVE Advantage annealer.

\setcounter{equation}{0}
\renewcommand{\theequation}{S4.\arabic{equation}}
\section{Computational complexity}
The computational complexity of the recursive part of Algorithm 1 (see Main text) can be estimated using big-$\mathcal{O}$ notation, where $n$ indicates the size of the problem (i.e., number of nodes), $k$ indicates a given step, and $T(n_k)$ depicts the time needed to perform the computation for $n_k$. %Also, $n_0 \equiv n$.

\subsection{Generalized modularity matrix}
The modularity matrix \textbf{B} requires $\mathcal{O}(n^2)$ computations, but it is calculated and stored only once at the start of the recursion. Then, it is read to obtain the generalized modularity matrix \textbf{B}\textsuperscript{k} in $\mathcal{O}(n)$ linear time, since a single summation over the nodes within a given community needs to be performed. Thus, in each level of the recursion, the computations involve linear operations and the realization of a quantum annealing (QA) step.

\subsection{Recursion tree}
Several cases must be carefully considered. Moreover, the general case is better understood after analyzing the complexity of simpler scenarios. For this reason, we start with the worst-case scenario and build up to the most realistic approximation of the problem.

Additionally, as stated in the main text, the running time of the quantum annealing (QA) step can vary from polynomial \cite{Farhi2001} to exponential times \cite{Albash2018}. However, this is problem-specific and difficult to address analytically. For this reason, here we denote the complexity of a single quantum split as $f(n)$. Crucially, the final complexity and scalability of the algorithm will be modulated by the total QA running time, $\mathcal{K}_n$, but the functional dependence of the classical counterpart will not be affected by it.

\subsubsection*{Quadratic complexity} Although very unlikely, some networks could be recursively divided into two branches where one of the two can no longer be split in the next recursion steps. That is, $n_k = n_{k+1} + a \ \ \forall k$ and $a\in (0,n)$. In this case, the total time to divide the entire network into two communities can be written as $$T(n) = T(n-a) + T(a) + n + f(n) + 1,$$ and, subsequently, $$T(n-a) = T(n-2a) + n-a + f(n-a) + 1,$$ $$\ldots$$ Crucially, the "right" communities, always containing $a$ nodes, are not split again, thus no longer contributing to the tree. Then, the general running time for the $k$-th step is easily calculated by recursively substituting the different running times. 
\begin{equation} \label{eq:quadratic_time}
    \begin{split}
        T(n) &= T(n-ak) + \sum_{r=0}^{k-1}\left[1+n-ar+f(n-ar)\right] \\
        &= T(n-ak) + k + nk - \frac{a}{2}(k-1)k + \sum_{r=0}^{k-1}f(n-ar)
    \end{split}    
\end{equation} Assuming $n_k=n-ak\leq a$ as a stopping criterion yields $k=\frac{n}{a}-1>0$ number of steps. Therefore, 
\begin{equation}
    \begin{split}
    T(n) &= T(a) + \frac{n}{a}-1 + n\left(\frac{n}{a}-1\right) + \frac{a}{2}\left(\frac{n}{a}-2\right)\left(\frac{n}{a}-1\right) + \sum_{r=0}^{n/a-2}f(n-ar),  \\
    &\underset{n \gg 1}{\approx} \frac{3n^2}{2a} + \sum_{r=0}^{n/a}f(n-ar) \longrightarrow \mathcal{O}(n^2 + \mathcal{K}_n)
    \end{split}
\end{equation} where $T(a)$ is a constant, and $\mathcal{K}_n$ is the overall QA running time.

\subsubsection*{Unbalanced recursive trees}
The quadratic instance of the recursion represents the worst-case scenario. In practice, however, each community found at step $k$ is typically subdivided in step $k+1$, producing two new communities of unequal size, \begin{equation} \label{eq:unbalanced_sizes}
    n_{k}\frac{1}{a} \ \text{   and   } \ n_{k}\left(1-\frac{1}{a}\right),
\end{equation} with $a \in (1, M)$ for some finite $M$. Notably, the stopping criterion is determined by the properties of the generalized modularity matrix, which differs for all discovered communities. Fortunately, due to the resolution limit, communities with fewer than $\sim\sqrt{m}$ nodes (modulated by $\gamma^{1/2}$ for arbitrary resolution parameters), where $m$ is the total number of edges, cannot be resolved~\cite{Fortunato2007,Kumpula2007}. This implies that the recursion will halt no later than at some minimum effective size $n_0 \in [1, n - a]$. Then, considering the shrinking in Eq. \ref{eq:unbalanced_sizes} and setting $T(n_0) = \mathcal{O}(1)$, the Akra-Bazzi theorem can be used to derive a tight bound on the algorithm's complexity. Briefly, the theorem states that recurrences of the form
\begin{equation} \label{eq:recursion}
T(n) = \sum_{i=1}^s a_i T(b_i n + h_i(n)) + g(n),
\end{equation} have an asymptotic solution 
\begin{equation}\label{eq:akra-bazzi_S}
    T(n) \in \Theta\left(n^p \left(1 + \int_{n_0}^n \frac{g(u)}{u^{p+1}} du \right)\right),
\end{equation} where $p$ satisfies the characteristic equation
\begin{equation} \label{eq:characteristic}
    \sum_{i=1}^s a_i b_i^p - 1 = 0,
\end{equation} with $a_i > 0$, $0 < b_i < 1$ for all $i = 1, \ldots, s$, and $g(n)$ is a continuous, positive, and regular function for all sufficiently large $n$. Crucially, the QA complexity $f(n)$ seems exponential at worst \cite{Albash2018}, which still satisfies the conditions of regularity. The functions $h_i(n)$ represent small perturbations in the arguments of $T$, which, in our case, arise because they are integers rather than real numbers. In this case, $h_i(n)\in \mathcal{O}(1)$, since the rounding introduces a constant. 

Applying the theorem to our case with $s=2$, $a_1=a_2=1$, $b_1=\frac{1}{a}$, and $b_2=1-\frac{1}{a}$ yields the initial recursion $$T(n)=T\left(\left[\frac{n}{a}\right]+c_1\right)+T\left(\left[n\left(1-\frac{1}{a}\right)\right]+c_2\right)+n+f(n),$$ where $\left[x\right]$ is the integer function, and $c_1, \ c_2$ are some unknown constants acting as the perturbations. Then, the corresponding characteristic equation $$\left(\frac{1}{a}\right)^p+\left(1-\frac{1}{a}\right)^p-1=0$$ has a unique solution $p=1, \ \forall a \in (1, M)$. This follows because the left-hand side defines a continuous, strictly decreasing function of $p$, with limits $(+\infty,-1)$ as $p\to\mp\infty$. By the intermediate value theorem, there is exactly one real solution, and $p=1$ satisfies the equation.

The asymptotic solution for Eq. \ref{eq:akra-bazzi_S} is then 
\begin{equation} \label{eq:unbalance_O}
    T(n)\in\mathcal{O}\left(n+n\log n + n\int_{n_0}^n\frac{f(u)}{u^2}du\right)\sim\mathcal{O}\left(n\log n +  \mathcal{K}_n\right),
\end{equation} being
\begin{equation} \label{eq:QA_rec_complex}
    \mathcal{K}_n \doteq n\int_{n_0}^n\frac{f(u)}{u^2}du
\end{equation} the total complexity derived from the repeated QA instances.

\subsection*{Unbalanced and dynamic recursive trees}
There is another important consideration. The most realistic scenario corresponds to the following:
\begin{equation} \label{eq:unbalanced_dynamic_sizes}
    n_{k}\frac{1}{a_k} \ \text{   and   } \ n_{k}\left(1-\frac{1}{a_k}\right),
\end{equation} that is, the fraction of nodes assigned to each community changes along the hierarchy. However, this does not change the final result in Eqs. \ref{eq:unbalance_O} and \ref{eq:QA_rec_complex}. 

To understand this, we analyze the dynamics of the sequence $\{a_k\}_{k=0,...}$ representing the fraction of nodes assigned to each community at the $k$-th step. The elements $a_k$ can be modeled as uniform perturbations
$$a_k = a+\delta_k+O(\delta_k^2)$$ where $a \in (1, M]$ and $\delta_k\sim U(-a,M)$ for some finite $M$ that depends on the average split size along the tree. The community sizes can then be expressed as,
\begin{equation}
    \begin{split}
        \frac{n_k}{a_k} &= \frac{n_k}{a}\left(\frac{1}{1+\delta_k/a}\right)=\frac{n_k}{a}\left(1-\frac{\delta_k}{a}+O(\delta_k^2)\right) = \frac{n_k}{a}+h_1(n_k)\\
        n_k\left(1-\frac{1}{a_k}\right) &= n_k\left(1-\frac{1}{a}\frac{1}{(1+\delta_k/a)}\right)=n_k\left(1-\frac{1}{a}+\frac{\delta_k}{a^2}+O(\delta_k^2)\right)= n_k\left(1-\frac{1}{a}\right)+h_2(n_k)
    \end{split},
\end{equation} where $h_1(n_k)$ and $h_2(n_k)$ represent the perturbations in the sizes of the two communities, and both are of $\mathcal{O}(n_k)$. These perturbations are bounded and decrease along the recursion. Therefore, even in the most realistic case of dynamic and unbalanced trees, the classical part of Algorithm 1 remains $\mathcal{O}(n \log n)$, and the quantum annealing complexity is still $\mathcal{K}_n$.

\subsection{Quantum annealing}
We now evaluate the total complexity contributed by the quantum annealing steps, $\mathcal{K}_n$, under different assumptions for the function $f(n)$, which accounts for the cost of a single QA subproblem:
\begin{enumerate}[label=\Roman*.]
    \item If $f(n)=c \ \text{with} \ (0<c<M)$, then $\mathcal{K}_n=O(1)$ and $T(n)\in\mathcal{O}(n\log n)$;
    \item if $f(n) \sim \log n$, then $\mathcal{K}_n=O(\log n)$ and $T(n)\in\mathcal{O}(n \log n + \log n) \sim \mathcal{O}(n \log n)$; 
    \item if $f(n)\sim n$, then $\mathcal{K}_n=O(n\log n)$ and $T(n)\in\mathcal{O}(n\log n)$;
    \item if  $f(n)\sim n^s, \ s>1$, then $\mathcal{K}_n=O(n^s)$ and $T(n)\in\mathcal{O}(n\log n+n^s)\sim\mathcal{O}(n^s)$; 
    \item and, if $f(n)\sim e^{n^s}\ \text{with} \ s\in\mathbb{R}^+$, then $\mathcal{K}_n$ has no closed-form expression, but the exponential complexity dominates, and $T(n)\sim\mathcal{O}(ne^{n^s})$.
\end{enumerate}

\subsubsection*{One-hot encoding constraints}
The community detection problem naturally lies in the discrete domain, where each node is assigned to one of the $|C|$ communities discovered. As discussed in the main text, QA requires the computational variables (i.e., the nodes) to be encoded in the binary domain (i.e., $x_i=\{0,1\} \ \forall i=1,\ldots,n$). A simple solution to this inconsistency is to use one-hot encoding, where each node's community assignment is represented by a vector with a single 1 at the index of its assigned community and 0s elsewhere. That is, $$\mathbf{x}_i=(0,\ldots,1,\ldots,0), \ \forall i=1,\ldots,n.$$ Since each node belongs to one and only one community, these one-hot vectors must satisfy the following property, $$\sum_{i=1}^{n}\left(\sum_{j=1}^{M}x_{ij}-1\right)^2=0,$$ where $M$ is the dimension of the vectors. The constraint is imposed on the QUBO function via a Lagrange multiplier $\lambda$, whose optimal value is not known a priori. There is ongoing research on this topic~\cite{roch_2021,mayowa_2022}. However, in practice, $\lambda$ is often set through trial-and-error or grid search. Additionally, $M$ defines an upper bound on the number of detectable communities. Thus, even with an optimal $\lambda$, the choice of $M$ remains an open issue. We aim to understand how $M$ should scale with the number of nodes $n$. That is, if $M$ is appropriate for a network of size $n$, will it remain suitable for a network of size $2n$? And how will the computational complexity of the recursion be affected by this? 

While in the extreme case $\gamma \to \infty$ modularity maximization would favor each node forming its own community (i.e., $M = n$), this limit is not realistic in practice~\cite{Kumpula2007}. The resolution limit inherent to the modularity function $Q$ implies that the number of communities that can be reliably detected is bounded above, with the exact maximum depending on network topology. This upper bound tends to scale between $\sqrt{n}$ and $n$ in sparsely and densely connected networks, respectively ~\cite{Fortunato2007,Kumpula2007}. With this in mind, revisiting the complexities of a QA step $f(n)$, it's trivial to see that the only case where using a one-hot encoding scheme clearly favors the scalability is if the running time remains constant w.r.t. the size of the network (i.e., case \rom{1}). Additionally, although using recursion theoretically degrades scalability when the annealing complexity is logarithmic, i.e., case \rom{2}, the need to incorporate Lagrange multipliers in the one-hot encoding scheme may counteract this improvement. Even if we set $M=\sqrt{n}$, 

\begin{enumerate}[label=\Roman*.]
    \item If $f(n)=c \ \text{with} \ (0<c<M)$, the recursion is $\mathcal{O}(n\log n)$ while the one-hot encoding remains constant;
    \item if $f(n)\sim \log n$, the recursion is $\mathcal{O}(n\log n)$ while the one-hot encoding is $\mathcal{O}(\log n\sqrt{n})\sim \mathcal{O}(\log n)$
    \item if $f(n)\sim n$, the recursion is $\mathcal{O}(n\log n)$ while the one-encoding is $\mathcal{O}(n\sqrt{n})$;
    \item if  $f(n)\sim n^s \ \text{with} \ s>1$, the recursion is $\mathcal{O}(n^s)$ while the one-hot encoding is $\mathcal{O}(n^sn^{s/2})$; 
    \item and, if $f(n)\sim e^{n^s}\ \text{with} \ s\in\mathbb{R}^+$, the recursion is $\mathcal{O}(ne^{n^s})$ while the one-hot encoding is $\mathcal{O}(e^{n^sn^{s/2}})$.
\end{enumerate} 
Examples III to V might benefit from the explicit calculation when $n \to \infty$. Thus, $$\lim_{n\to\infty} \frac{n\log n}{n\sqrt{n}}=\lim_{n\to\infty} \log \left(n^{n^{-1/2}}\right)=0 \ \text{  (also derived using l'H\^opital's rule)}$$ $$\lim_{n\to\infty} \frac{n^s}{n^sn^{s/2}}=\lim_{n\to\infty} n^{-s/2}=0$$ $$\lim_{n\to\infty}\frac{ne^{n^s}}{e^{n^sn^{s/2}}}=\lim_{n\to\infty} ne^{-n^s(n^{s/2}-1)}=0$$ favouring, in both cases, the recursion over an explicit one-hot encoding.

In the scenario where $M=n$, the complexity derived from $f(n^2)$ scales even worse. In this aspect, using recursion and binary QA significantly improves the scalability of the algorithm, since it is a process that takes on average $n\log n$ time to complete. However, the practical speed-up stemming from the usage of quantum resources might largely depend on the running times of the QA processes themselves. 

\setcounter{equation}{0}
\renewcommand{\theequation}{S5.\arabic{equation}}
\section{QUBO and the dynamical range of the modularity}
The $k$-th element $p_k$ of the process to optimize using QA is given by the elements of the generalized modularity matrix \textbf{B}\textsuperscript{k} (see Eqs. 10 and 11 in Methods),
\begin{equation} \label{eq:qubo_mod}
    p_k = \frac{-1}{m} \sum_{ij \in k} B_{ij}^{\text{k}} x_i x_j.
\end{equation} The constant $m$, corresponding to the number of edges in the network, plays no role in the optimization, but it is kept here for completeness. To analyze the information (measured in \textit{bits}) needed to encode the entries of the modularity matrix, we can compute the dynamic range~\cite{mucke_optimum-preserving_2024},
\begin{equation} \label{eq:dynamic_range}
    \mathcal{DR}\doteq\ \log_2\left(\frac{\max D}{\min D}\right),
\end{equation} where $D$ is the set of absolute differences between all the elements of the (generalized) modularity matrix, $$D(\mathbf{B}^{\text{k}})\doteq\{ |B_{ij}^{\text{k}}-B_{rs}^{\text{k}}|: \quad B_{ij}^{\text{k}},B_{rs}^{\text{k}} \in \mathbf{B}^{\text{k}}, \quad |B_{ij}^{\text{k}}- B_{rs}^{\text{k}}|\geq \epsilon\}.$$ The parameter $\epsilon$ controls for the numerical precision, and it was taken as $10^{-6}$. In principle, at each step of the recursion, the information decreases as the size of the communities diminishes and the ground state of the QUBO approaches the trivial solution. Because of this, we can treat the $k=0$ element as an upper bound, simplifying the numerical study, i.e., $D(\mathbf{B}^{\text{k}=0})=D(\mathbf{B})$. 

To evaluate the sensitivity of Advantage to the $\mathcal{DR}$ and to create unrealistically harsh conditions, we simulated two unrealistic experiments. Initially, we generated chains of 3-cliques with 2, 6, and 20 3-cliques (i.e., 6,18, and 60 nodes, respectively). Then, two randomly selected edges were monotonously changed to (1,1), (10,.1), (50,.05), (100,.001), and the modularity was maximized with both Louvain and Hierarchical annealing (see Algorithm 1 in the Methods). In the second experiment, we generated 3 power-law clustered networks~\cite{Holme2002} of sizes $N=20,50,100$ with parameters $m=1$ and $p=.1$. Similarly, one randomly selected edge was monotonously changed to 1, 10, 50, 100, 500, 1000, 5000, and 10000. The performance of the Hierarchical annealing algorithm was compared to the Louvain in the same manner (Fig. \ref{fig:dynamic_range}).

\setcounter{equation}{0}
\renewcommand{\theequation}{S6.\arabic{equation}}
\section{Embedding strategy, chain strength, and noise}
\subsection{Setting the chain strength in the embedding}
The modularity function $Q$ can be mapped onto a QUBO problem whose topology is determined by the entries of the modularity matrix $\mathbf{B}$. The solution is encoded in the maximization of the modularity,
\begin{equation}
    \hat{Q} \doteq \max_{\forall C \in \mathcal{C}} Q(C)
= - \min_{\forall C \in \mathcal{C}} \bigl[-Q(C)\bigr].
\end{equation}
In contrast, the quantum annealer (i.e., the QPU) evolves toward the minimum-energy configuration of the Hamiltonian of a distinct physical system, $H(\mathbf{J})$, which depends on the physical couplings $\mathbf{J}$ between qubits. The mapping from the QUBO formulation to the D-Wave annealer is achieved via a \emph{minor embedding} procedure~\cite{Choi2008,choi_minor-embedding_2011}. This embedding defines the physical couplings between qubits $a$ and $b$ as a function of the logical couplings between nodes $i$ and $j$ in the network, such that $\mathbf{J}=f(\mathbf{B},k)$. Except in special cases where the logical and physical topologies are nearly identical, this mapping requires the introduction of chains of physical qubits to represent single logical variables. Within each chain, physical qubits are coupled by a \emph{chain strength} $k$, which must be chosen to ensure a faithful representation of the logical problem while avoiding excessive distortion of the physical energy landscape~\cite{Grant2022}. In practice, the annealer seeks the ground state of $\mathcal{H}$ under the constraint that all qubits within a chain adopt the same state. When this constraint is violated, broken chains are resolved using a majority voting strategy, whereby the logical value is assigned according to the most common spin state along the chain. 

To set the chain strength, we adopted D-Wave’s default strategy, which aims to mitigate inhomogeneities in the effective coupling strengths induced by the specific minor embedding~\cite{raymond2020improving}. This approach defines the chain strength as a function of both the logical problem and the topology of the physical connectivity graph:
\begin{equation}
    k^{*} = k_0 \sqrt{\sigma^2 N},
    \label{eq:chain_strength}
\end{equation} where $\sigma^2 = \frac{2}{N(N-1)} \sum_{a<b} J_{ab}^2$ denotes the variance of the physical couplings, $N$ is the number of logical variables, and $k_0=\sqrt{2}$ was determined empirically~\cite{raymond2020improving}. Alternatively, the chain strength $k$ can be manually tuned by evaluating multiple values and selecting the one that yields the best empirical performance. 

\subsection{Relationship between the chain strength and noise}
Quantum annealers are intrinsically sensitive to noise, which in practice means that the device evolves toward the ground state of a perturbed Hamiltonian,
\begin{equation}
    H^{\delta} = H(\mathbf{J} + \delta \mathbf{J}),
\end{equation}
where $\delta \mathbf{J}$ models deviations between the programmed and realized couplings. Under the integrated control error (ICE) model, these deviations are described as independent, zero-mean Gaussian perturbations with finite variance~\cite{yarkoni2022quantum}. Within this framework, a key observation is that the probability of chain breaks is related to the magnitude of the noise~\cite{Jeong2025}, understood as the variance of $\delta \mathbf{J}$. Other factors at play include embedding-specific factors and energy scale separations (e.g., annealing temperature). Importantly, noise is itself a statistical abstraction: it provides an effective description of unknown and uncontrolled physical processes that give rise to discrepancies between the ideal results and empirical outcomes.

Consequently, an indirect strategy to assess the noise level of the QPU is to measure the fraction of broken chains, defined as the ratio between the number of chains that violate the embedding constraint and the total number of chains in a given embedded problem. This information can be directly extracted from the D-Wave samplers. For convenience, we define the \emph{chain break probability} as the percentage of broken chains, $CBP = 100 \cdot CBF$, where $CBF$ denotes the chain break fraction. While this definition is operational, it is worth noting that $CBP$ and $CBF$ can be formally related to probabilistic measures through expectations under specific (and not always strictly satisfied) modeling assumptions~\cite{yamaji_development_2022}.

\subsection{Experimental measurements of chain breaks and their relationship to annealing performance}
To assess the performance of the annealing strategy proposed in this work and to evaluate whether noise constitutes a limiting factor, we systematically analyzed chain break behavior as a function of network size and the dispersion of outcomes across independent runs. Performance was quantified by the distance between individual solutions and the best result obtained, defined through the relative decrease in modularity as
\begin{equation}
    \text{Relative decrease}(N) \doteq 
    \frac{Q(N)}{\max_{r \in N_{\mathrm{runs}}} Q(N)},
\end{equation} the relative distance was defined as $$100\cdot(1-\text{Relative decrease}).$$ In addition, we measured the chain break fraction as a function of network size and computed statistical descriptors of the relative decrease in modularity across runs. Finally, to provide a quantitative assessment of the effectiveness of the default chain strength strategy, we manually tuned the chain strength $k$ for the largest network embeddable on the Pegasus topology ($N=169$) and compared the resulting performance against the default setting ($k=k^{*}$ in Eq.~\ref{eq:chain_strength}). The chain break fraction was defined as the maximum $CBF$ observed along the hierarchical tree, serving as a proxy for the most challenging subproblem associated with each instance. In practice, in almost all cases, this corresponded to the values measured at the highest level of the hierarchies (Fig.~\ref{fig:noise_hierarchy}). For these experiments, we considered only Erd\H os--Renyi and power-law networks, as they represent two extremes of the same problem. The former corresponds to the canonical academic example of a random QUBO topology, whereas the latter represents a highly structured topology.

The relative decrease in modularity exhibited a clear degradation as a function of network size, accompanied by an increase in chain break fractions, suggesting an apparent global relationship between noise-related effects and solution quality (Fig.~\ref{fig:chain_strength}a). However, this trend did not persist when conditioning on individual network sizes (Fig.~\ref{fig:chain_strength}b). Specifically, for a fixed problem size, variations in chain break fractions across independent runs were not significantly associated with variations in solution quality (Pearson and Spearman $\rho$, $p$-values$>0.05$, exact tests). This indicates that the observed negative trends couldn't be directly attributed to chain breaks at the level of individual instances. Rather, both increased chain break fractions and reduced solution stability emerged as concomitant consequences of scaling the logical problem size and its associated embedding complexity. In this sense, $CBF$s acted as a marker of problem difficulty under scaling of logical variables, rather than as a causal determinant of solution quality.

Indeed, the proportion of broken chains increased with the number of logical qubits (Fig.~\ref{fig:chain_strength}c), although it remained below 10\% in the worst cases. Interpreting chain breaks as an indirect measure of noise, this result suggests that noise increased with problem size, as expected, but remained within acceptable levels throughout. Interestingly, this increase was significantly more pronounced for unstructured than for structured topologies (i.e., Erd\H os--Renyi and Powerlaw networks, respectively). 

Concurrently, the chain strength increased with problem size (Fig.~\ref{fig:chain_strength}d), following a power-law scaling with an exponent of approximately $0.5$ for Erd\H os-Renyi networks and $\sim 0.06$ for Powerlaw networks. In the former case, this behavior closely matched the analytical prediction in Eq.~\ref{eq:chain_strength}, as the variance of physical couplings in random Gaussian systems is expected to remain constant~\cite{raymond2020improving}. In contrast, for Powerlaw networks, the variance of the couplings played a dominant role in determining the chain strength, with only a weak contribution arising from the increasing number of logical qubits. Lower chain strengths are expected to induce larger chain break fractions and, consequently, reduced resilience to noise. However, this behavior was not observed (Fig.~\ref{fig:chain_strength}c). This indicates that not only the number of logical qubits, but also the underlying topology of the problem, plays a critical role in determining appropriate chain strengths. As a result, the design of ad hoc or in-house strategies for setting the chain strength becomes challenging, and it is therefore preferable, in principle, to rely on the default and well-characterized strategy proposed by D-Wave~\cite{raymond2020improving}. 

These topological differences, as well as the different chain strength values, were reflected in the distribution of the solutions (Fig.~\ref{fig:chain_strength}e). In Erd\H os--Renyi networks, the standard deviation of the modularity values exhibited a discrete jump from zero to a relatively constant level, remaining stable as network size increased. In contrast, for Powerlaw networks, the variability of the solutions progressively increased with network size. These distinct patterns of variability were also mirrored in the average quality of the solutions. For Erd\H os--Renyi networks, the average distance to the best solution showed an abrupt increase and then stabilized at approximately 3\% from the optimum. Conversely, in Powerlaw networks, the average solution quality deteriorated gradually to similar levels. Thus, the degradation in the average performance was driven by the increasing variability of the solutions, which broadly tracked the rise in chain break fractions, rather than purely by chain breaks.

Crucially, our primary interest was not the average modularity, but the highest. Accordingly, a more informative characterization of the solution landscape can be achieved by examining the upper percentiles of the modularity distribution. Since the best solution corresponds to the 100th percentile, the relevant question is how the high percentiles evolve as the variability of the solutions increases; i.e., how far do they move the best result. These percentiles remained largely constant for all problem sizes once the variability became non-zero (Fig.~\ref{fig:chain_strength}f). It is also worth noting that for the largest problem size, $N=169$, which corresponds to the practical limits of the current hardware, a "second jump" was observed, analogous to the one occurring at $N=60$. This behavior suggests that the annealer retained its ability to explore high-quality solutions largely independent of the noise and concurrent chain breaks. The degradation in average modularity could instead be attributed to increased noise levels, which led to higher chain break fractions and, consequently, larger standard deviations in the distribution of solutions.

Lastly, we assessed the robustness of the default strategy for setting the chain strength by manually varying the chain strength $k$ for the largest network that could be embedded on Advantage ($N=169$). For simple networks, such as the Karate club graph, the solution was effectively independent of the chain strength (Fig.~\ref{fig:chain_strength}g). In contrast, for larger networks, the sensitivity of the solutions to the value of $k$ was evident (Fig.~\ref{fig:chain_strength}h). The default value $k^{*}$ consistently stayed within a range that ensured optimal performance, without compromising the annealer’s ability to reach the best possible solutions. This observation justifies the use of the default strategy. However, in scenarios where sufficient computational resources are available and maximal precision is required, exploring a neighborhood around the default value may result in marginal but relevant improvements.

In conclusion, although noise increased with problem size, the default chain strength strategy, accounting for both the topology and size of the QUBO, was effective in keeping its impact under control. Importantly, the presence of noise did not hinder the annealer’s ability to explore the solution landscape, allowing it to sample both lower- and higher-quality than average solutions. As a result, the annealer consistently achieved highly competitive performance when compared with state-of-the-art algorithms for modularity maximization. Using the default strategy to set the chain strength yields noise-resilient annealing strategies whose performance is ultimately governed by the problem size and the topology of the embedding.

\section{Software specifics: \textit{Qommunity}}

\subsection{Architecture}
\begin{figure}[h!]
    \centering
    \includegraphics[width=1\linewidth]{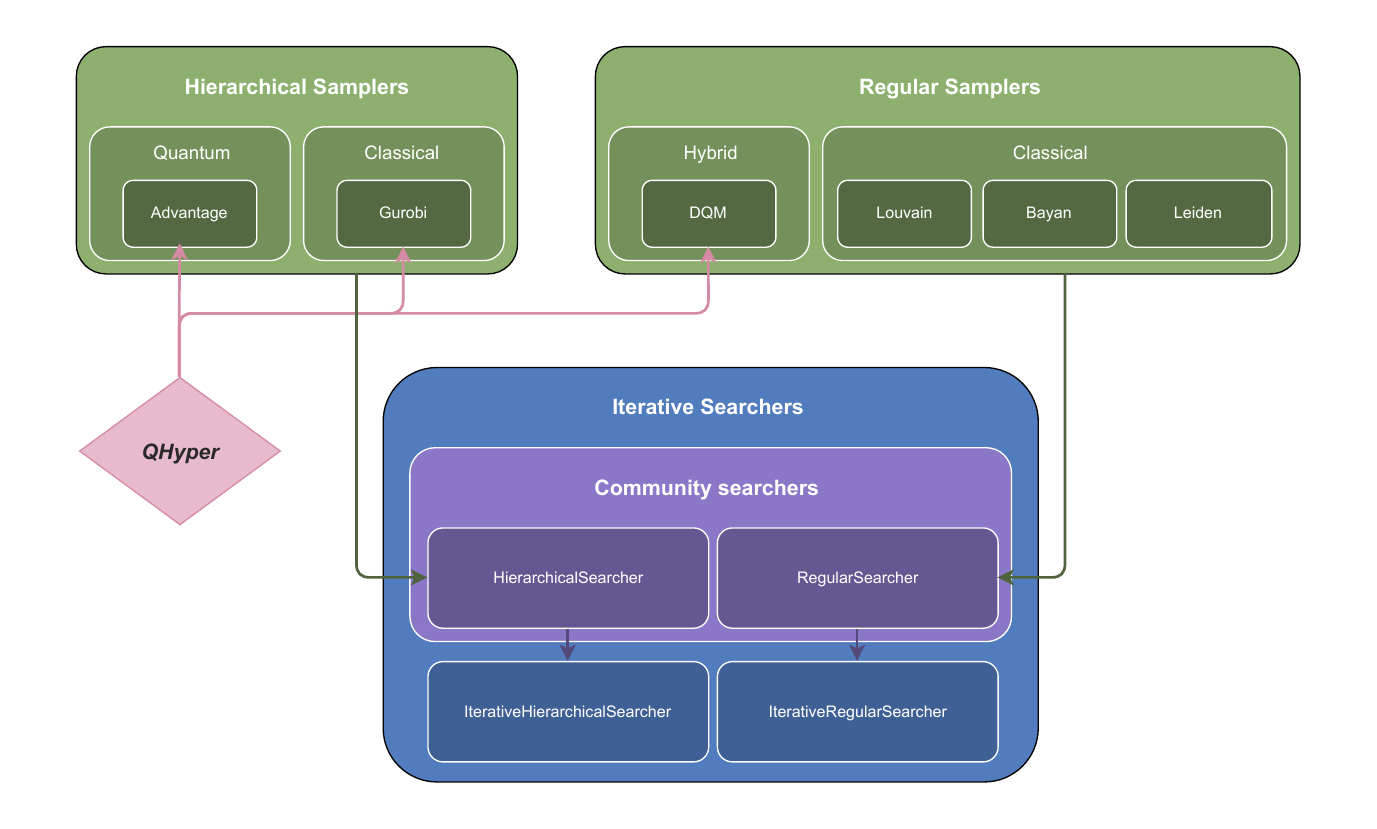}
    \caption{\textbf{Architecture of the \textit{Qommunity} Python library}. The package utilizes the quantum computing library QHyper \cite{lamza_qhyper_2024} to communicate with the quantum annealer by employing multiple instances of a quadratic binary optimization problem. The Qommunity library also allows the user to choose from several other algorithms and solvers described in the main text. Lastly, modularity maximization occurs at the {\tt IterativeSearcher} class, where any solver and algorithm can be explicitly and automatically run an arbitrary number of times to produce a sufficiently large number of solutions to choose from.}
    \label{fig:qommunity_architecture}
\end{figure}

\begin{figure}[h!]
    \centering
    \includegraphics[width=.95\linewidth]{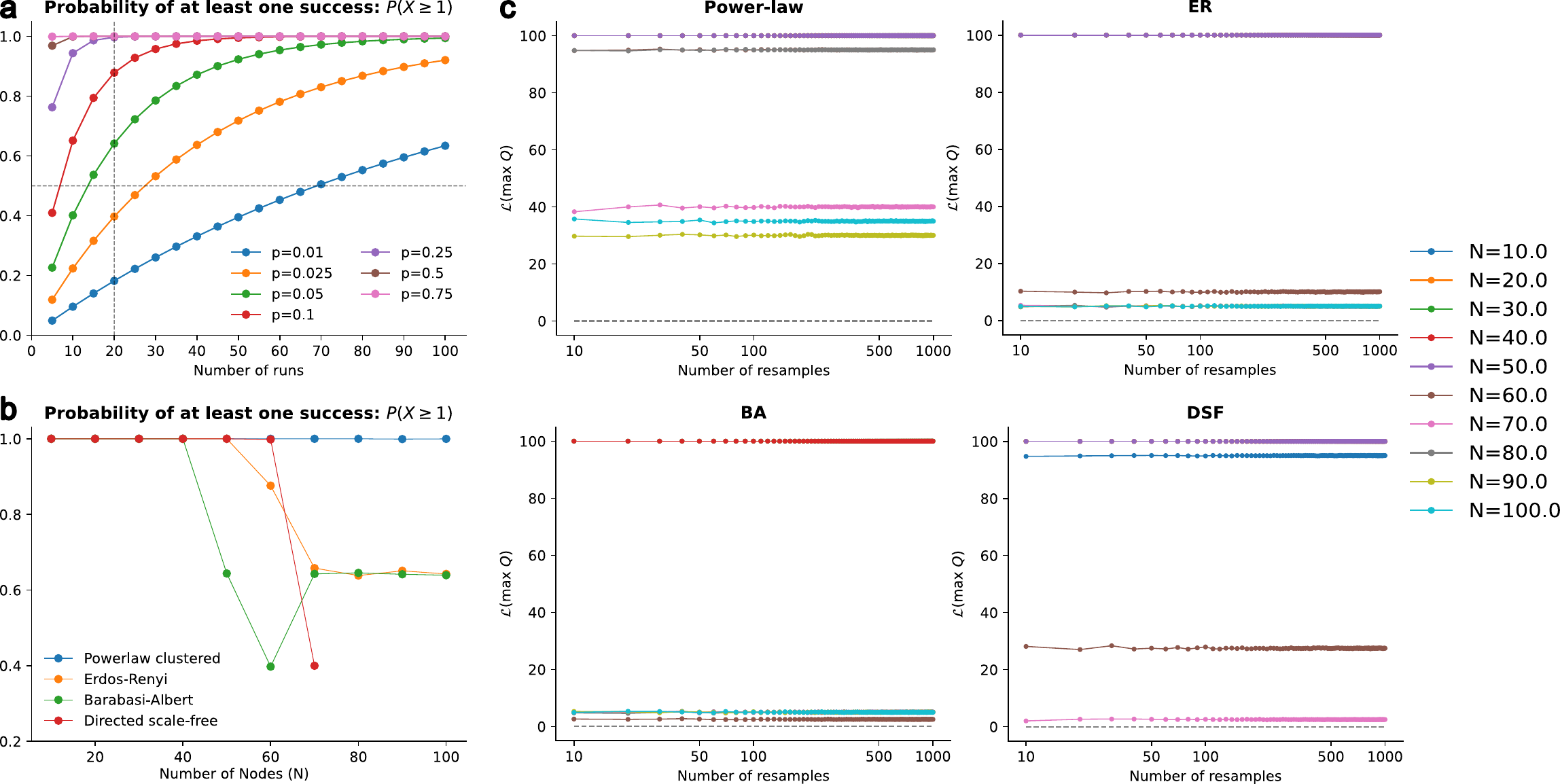}
    \caption{\textbf{Estimated probabilities of observing the maximum modularity for different networks.} \textbf{a} Theoretical probabilities of observing at least one successful event of a Binomial distribution as a function of the times that a given experiment can be run. \textbf{b} Estimated probabilities of observing $\max Q$ in the networks studied as a function of the number of nodes. \textbf{c} Convergence of the bootstrap Laplace estimates of the probability of success $p$ in Eq. \ref{eq:p_laplace} for different networks and numbers of nodes. All the bootstrap samples were obtained from the empirical distribution using $N_{runs} = 20$.}    
    \label{fig:sampling_runs}
\end{figure}

The \textit{Qommunity} library was designed to integrate several methods for detecting communities in graphs in one place. The implemented methods rely on classical, quantum, and hybrid solutions. The idea was to simplify and standardize the interface for conducting experiments. Several components compartmentalize the architecture, offering user-friendly features: Samplers, Searchers, and IterativeSearchers (Fig. \ref{fig:qommunity_architecture}).

\textit{\textbf{Samplers:}} Through Samplers we specify the method by which we want to detect communities, providing the necessary parameters. Currently, several classical solutions are implemented, such as Gurobi, Louvain \cite{Blondel2008}, Bayan\cite{aref2023}, Leiden \cite{Traag2019}, Infomap~\cite{Rosvall2008}, SBM Inference~\cite{Peixoto2014}, Spectral clustering, hybrid -- using DQM \cite{Wierzbinski2023}, and purely quantum -- using Advantage \cite{Johnson2011,lamza_qhyper_2024}. We divided the samplers based on how the divisions are done;
\begin{itemize} 
\item regular: the dividing into communities is done according to the algorithms specified inside the Sampler.\item hierarchical: perform binary divisions using classical methods or quantum annealing, depending on the Sampler specified. If no division is optimal, the Sampler returns a single community. Alternatively, each binary division can be further continued in the course of recursion logic provided by the HierarchicalSearcher.
 %binary splits are done, that can be further continued in the course of recursion, if used with the help of a HierarchicalSearcher,
%\item \textcolor{blue}{hierarchical: perform binary splits and are capable of performing optional binary splits, i.e. split only in favour of an optimal solution. The splitting happens as a result of objective function opitimisation or in the course of sampling, depending on the algorithm specified inside the Sampler class},
\end{itemize}

Although Samplers provide community detection methods and can perform community divisions themselves, the Seacher and IterativeSearcher are dedicated tools that enable the user to perform the community detection process comprehensively. 
%To perform community detection with the appropriate Sampler, an instance of it needs to be passed to either the Searcher or the IterativeSearcher.

\textit{\textbf{Searchers:}} Searchers are responsible for executing community detection and for the appropriate postprocessing of its results.  There are two types of Searchers - hierarchical, which accept Samplers that perform hierarchical divisions of communities, and regular, which perform a non-hierarchical, regular community detection method. Each option is defined by the algorithm of choice 
%within the Sampler 
as follows: 
%However, the Searchers full scope of responsibility and action depends on their type:
\begin{itemize}
\item regular: they serve as wrappers for the RegularSampler, facilitating the regular community detection.
\item hierarchical: they extend the scope of action far beyond the RegularSearchers, by implementing the logic of recursion in the hierarchical search.  They build the construct of recursive calls. Its execution results in a tree of binary splits. Similar to RegularSearchers, they also wrap the appropriate Sampler, but with a different purpose - to call the community detection method on HierarchicalSampler on each recursion level and, depending on whether a binary split has occurred as its result, decide whether to proceed with recursion. Hence, they build and oversee the process of hierarchical community detection search, described in Algorithm 1 in the Main text.

The hierarchical searchers track the trace of the hierarchical divisions and, if specified by the user, can return it as a division tree, alongside division modularities, corresponding to each hierarchy division level.
\end{itemize}

\textit{\textbf{Iterative Searchers:}}  Iterative Searchers serve as wrappers for community Searchers, responsible for calling their corresponding (i.e. regular or hierarchical) community search method a given number of times. These multiple calls are useful due to the stochastic and probabilistic natures of the algorithms \textit{Qommunity} provides.
As a result, for each iteration, we get a graph divided into communities, its modularity measure, and the execution time. The result may also contain the division trees and the division modularities, in the case of hierarchical community search. Iterative Searchers are used to simplify the execution of experiments aimed at maximizing the modularity of a given network (see Algorithm 2 in the Main text).

\subsection{High-level Python code examples to use the quantum resources for modularity maximization and interpretation}
\begin{lstlisting}[language=Python, caption=Modularity maximization and community detection using \textit{Qommunity}. After importing the relevant Python modules maximizing the modularity of the network using quantum computing would be trivial., label=python1]
import networkx as nx # Hagberg, et al. 2008
import numpy as np

## ORIGINAL FROM THIS WORK
from Qommunity.samplers.hierarchical.advantage_sampler import AdvantageSampler # H. Annealing
from Qommunity.samplers.hierarchical.gurobi_sampler import GurobiSampler # H. Gurobi

## HYBRID QUANTUM SOLVER (Wierzbinski, et al. 2023)
from Qommunity.samplers.regular.dqm_sampler import DQMSampler 

## ALETERNATIVE BENCHAMRKS
from Qommunity.samplers.regular.bayan_sampler import BayanSampler 
from Qommunity.samplers.regular.leiden_sampler import LeidenSampler
from Qommunity.samplers.regular.louvain_sampler import LouvainSampler

## MODULARITY MAXIMIZATION
from iterative_searcher.iterative_searcher import IterativeSearcher

## NETWORK TO ANALYZE
Graph = nx.karate_club_graph()
# Graph = nx.from_numpy_array(np.genfromtxt("A.csv", delimiter=','))

## PARAMETERS
num_runs = 20
resolution = 1 #

## Initiate the quantum processor and generate the QUBO
adv_sampler = AdvantageSampler(
    Graph, 
    resolution=resolution, 
    num_reads=100, 
    use_clique_embedding=True
)
adv_iterative= IterativeSearcher(adv_sampler)
communities, modularities, times = adv_iterative.run(num_runs=num_runs, save_results=False)

# Only the community with the highest modularity
community_structure = communities[modularities.argmax()]
modularity = modularities.max()
elapsed_time = times[modularities.argmax()]

\end{lstlisting}

\begin{lstlisting}[language=Python, caption=Example to obtain the hierarchical structure of a given network using modularity maximization., label=python2]
import networkx as nx
import numpy as np
import matplotlib.pylab as plt

## IMPORT THE RELEVANT MODULES
from Qommunity.samplers.hierarchical.advantage_sampler import AdvantageSampler
from iterative_searcher.iterative_searcher import IterativeSearcher
from dendro import Dendrogram

## PARAMETERS
num_runs = 20
resolution = 1

## NETWORK
G = nx.karate_club_graph()
# Graph = nx.from_numpy_array(np.genfromtxt("A.csv", delimiter=','))

# COMMUNITY DETECTION USING RECUSRIVE QUANTUM ANNEALING
searcher = IterativeSearcher(
    AdvantageSampler(
        G, 
        resolution=resolution, 
        num_reads=100, 
        use_clique_embedding=True
    )
)
results = searcher.run_with_sampleset_info(num_runs=num_runs, save_results=False, saving_path=None, iterative_verbosity=0)

## MODULARITY MAXIMIZATION (Algorithm 2 in the main Methods)
mods_Adv = results.modularity
communities = results.communities[mods_Adv.argmax()]
division_tree = results.division_tree[mods_Adv.argmax()]
division_modularities = results.division_modularities[mods_Adv.argmax()]
time = results.time[mods_Adv.argmax()]

## PLOTTING AND CUSTOMIZING THE HIERARCHY PLOT
dendro = Dendrogram(G, communities, division_modularities, division_tree)
fig, ax = plt.subplots(1,1,figsize=(13,6))
dendro.draw(
    display_leafs=False,
    yaxis_abs_log=True,
    ax=ax,
    fig=fig,
    communities_labels=["Group 1", "Group 2", "Group 3", "Group 4", "Group 5", "Group 6", "Group 7", "Group 8"],
    fig_saving_path="./Karate/dendrogram.svg",
    title='Hierarchicies inside the Karate network'
)
\end{lstlisting}

\clearpage
\section*{Supplementary figures}

\begin{figure}[h!]
    \centering
    \includegraphics[width=\linewidth]{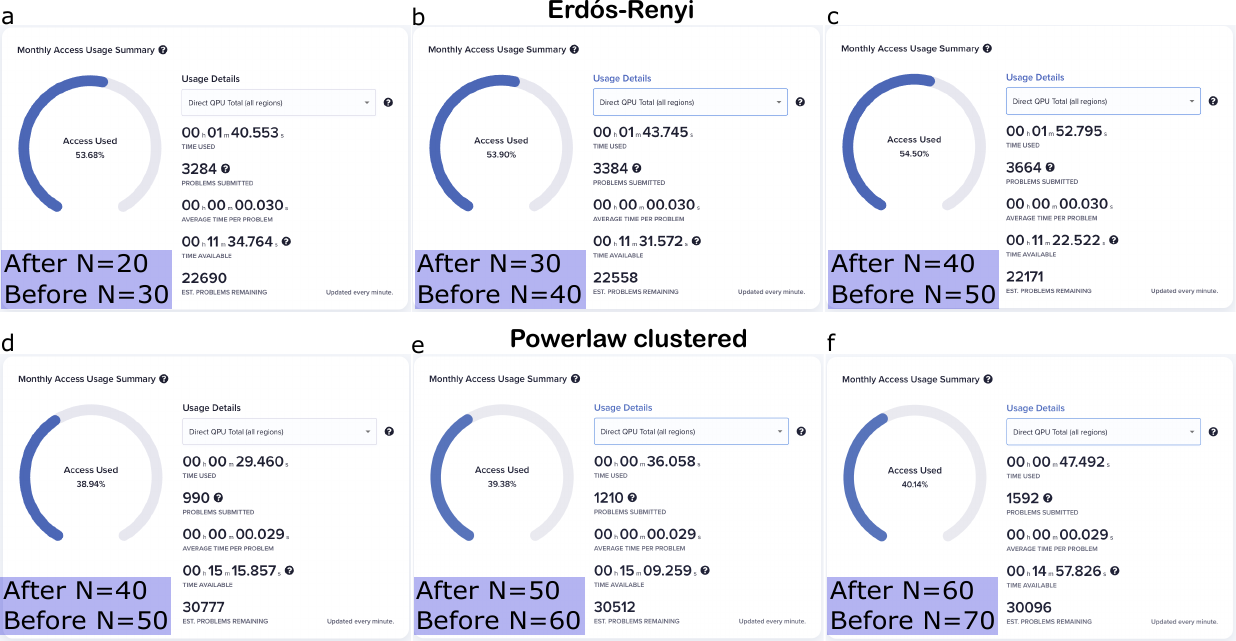}
    \caption{\textbf{Measurement of experimental QPU wall-clock times using the D-Wave Leap Dashboard}. Screenshots from the D-Wave Leap Dashboard illustrating the procedure used to measure wall-clock computing times for the experiments reported in Fig.~2. For each run (e.g., modularity maximization on a given network), we manually recorded the elapsed time displayed by the dashboard immediately before and after execution. Measurements were collected for both the DQM and hierarchical annealing workflows on the Advantage system; for simplicity, we report only screenshots from the Advantage system. Panels (\textbf{a–c}) and (\textbf{d–f}) show examples from Erd\H os–R'enyi and power-law clustered networks, respectively. The total computing time and the number of problems solved were obtained by subtracting the dashboard metrics recorded immediately after each experiment from those recorded immediately before. For instance, the time required to maximize the modularity of an Erd\H os–Renyi network with $N=40$ nodes corresponds to the difference between the values shown in \textbf{c} and \textbf{b}.}
    \label{fig:dwave_times}
\end{figure}

\begin{figure}[h]
    \centering
    \includegraphics[width=1\linewidth]{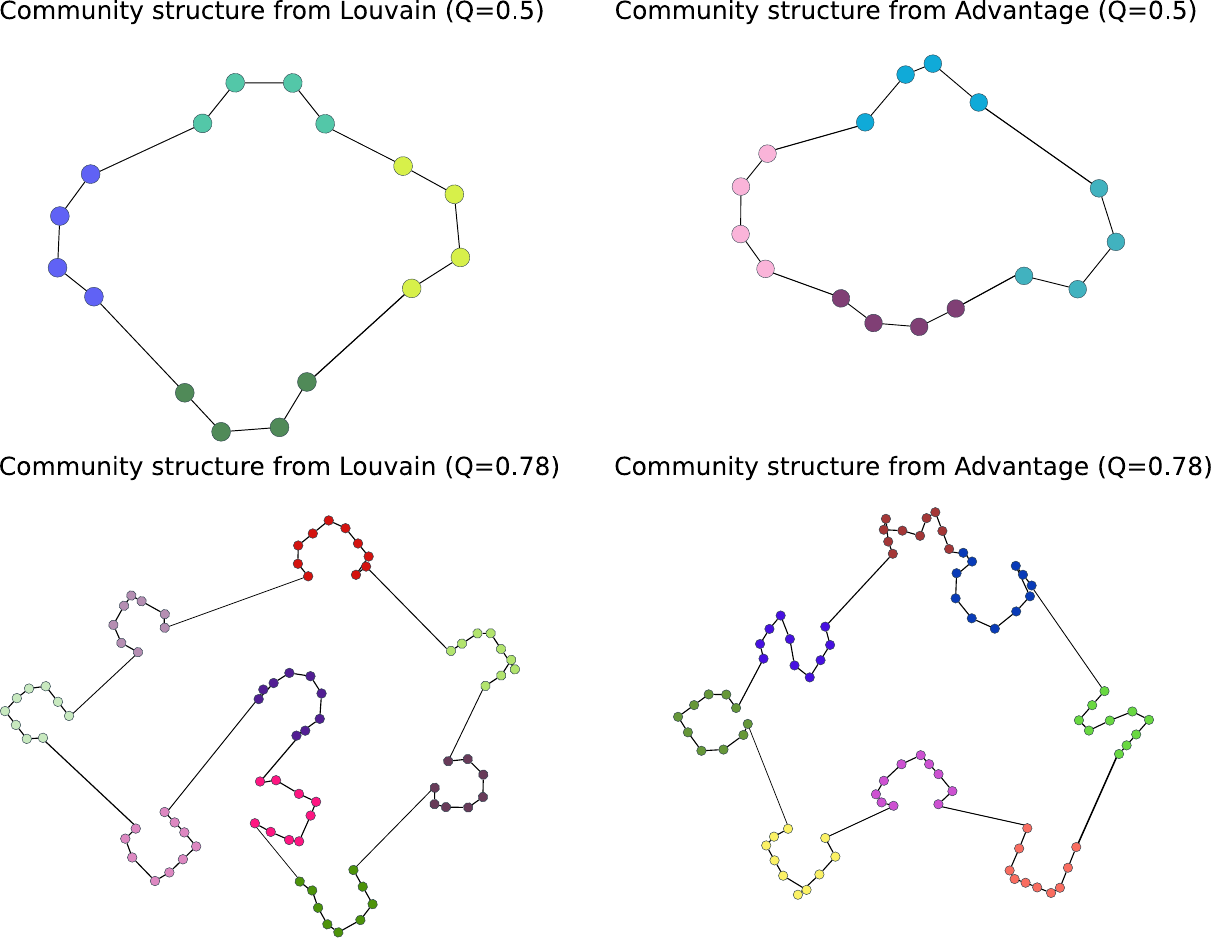}
    \caption{\textbf{Community structure of a chain of nodes.} Community structure and the resulting modularity discovered by the Louvain (left) and Hierarchical annealing (right) algorithms in a chain of 16 (top) and 81 (bottom) nodes. Even if both solutions from both algorithms are identical, the community structure appears somewhat arbitrary and represents an example of the resolution limit, where smaller communities cannot be resolved with the current resolution parameter (i.e., $\gamma=1$).}
    \label{fig:chain-nodes}
\end{figure}

\begin{figure*}
    \centering
    \includegraphics[width=1\linewidth]{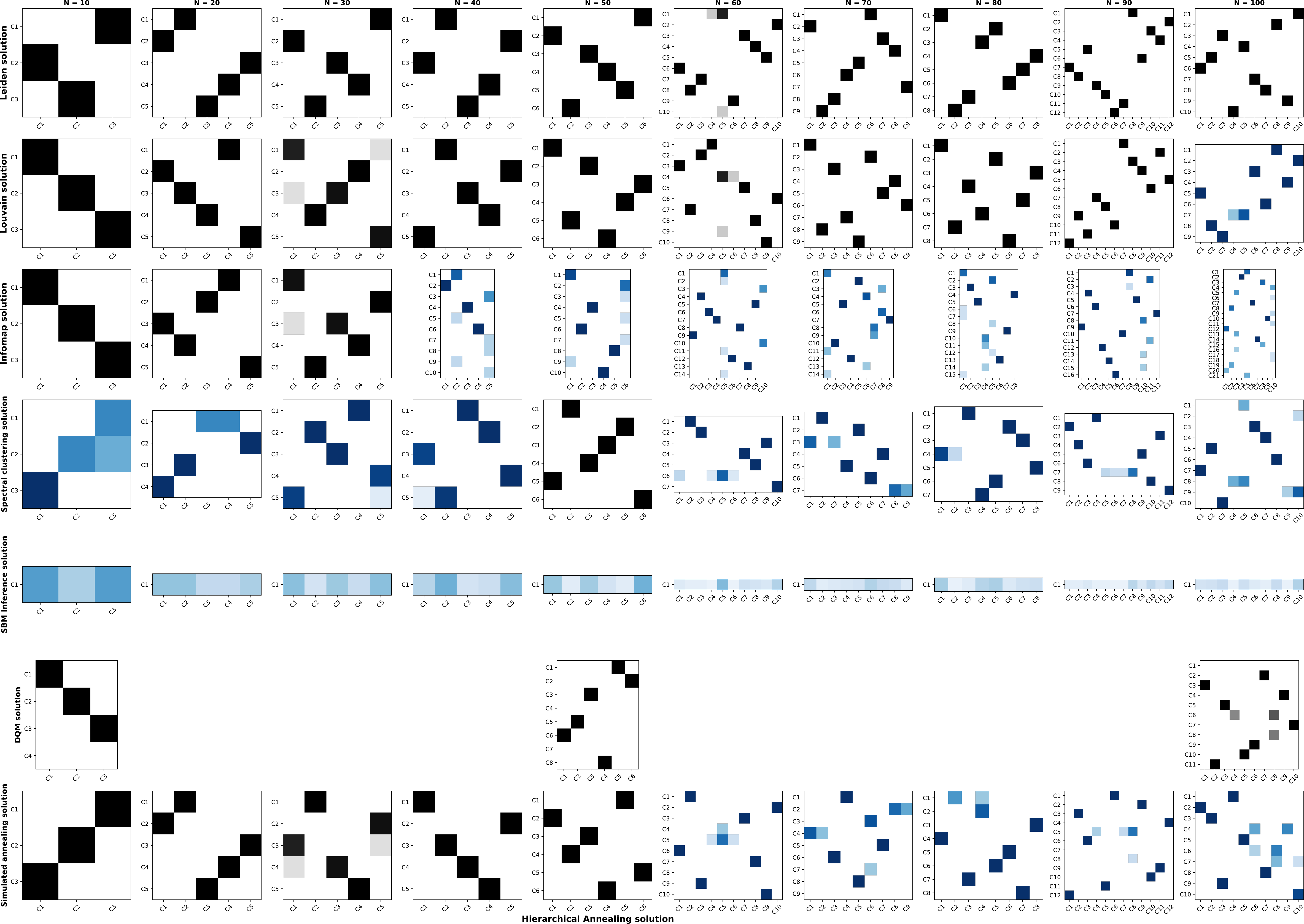}
    \caption{\textbf{Consensus matrices between the solutions found by the Hierarchical annealing and the alternative methods in power-law clustered networks of increasing sizes.} Communities identified by the hierarchical annealing procedure compared to those obtained using alternative algorithms. Opacity is proportional to the Dice similarity score between pairs of communities. Black, blue, and red outlines indicate cases in which hierarchical annealing achieved equal, higher, or lower modularity, respectively, relative to the corresponding alternative solution. White entries denote zero Dice similarity.}
    \label{fig:communities_PW}
\end{figure*}

\begin{figure*}
    \centering
    \includegraphics[width=1\linewidth]{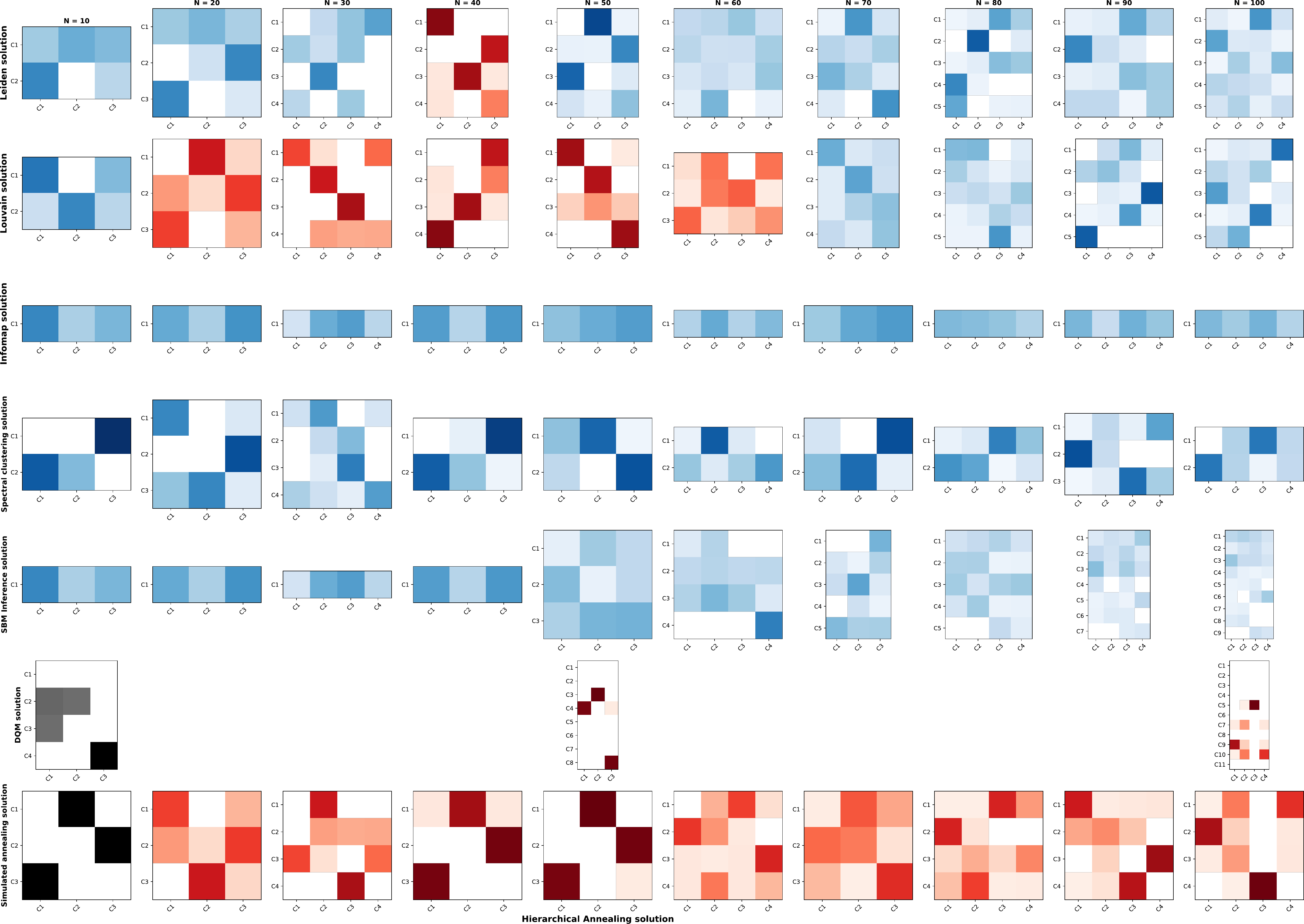}
    \caption{\textbf{Consensus matrices between the solutions found by the Hierarchical annealing and the alternative methods in Barabasi-Albert networks of increasing sizes.} Communities identified by the hierarchical annealing procedure compared to those obtained using alternative algorithms. Opacity is proportional to the Dice similarity score between pairs of communities. Black, blue, and red outlines indicate cases in which hierarchical annealing achieved equal, higher, or lower modularity, respectively, relative to the corresponding alternative solution. White entries denote zero Dice similarity.}
    \label{fig:communities_BA}
\end{figure*}

\begin{figure*}
    \centering
    \includegraphics[width=1\linewidth]{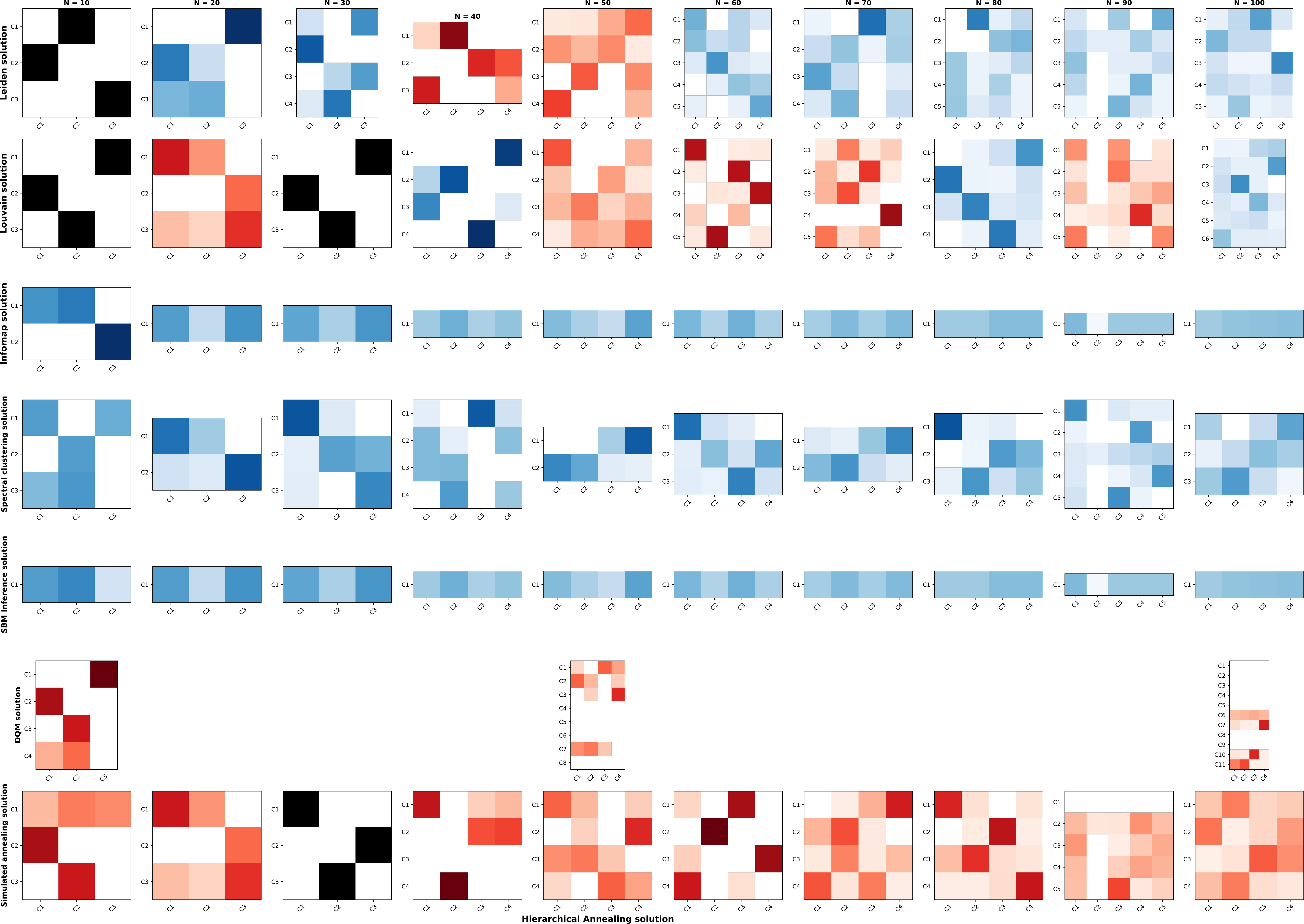}
    \caption{\textbf{Consensus matrices between the solutions found by the Hierarchical annealing and the alternative methods in Erd\H os-Renyi networks of increasing sizes.} Communities identified by the hierarchical annealing procedure compared to those obtained using alternative algorithms. Opacity is proportional to the Dice similarity score between pairs of communities. Black, blue, and red outlines indicate cases in which hierarchical annealing achieved equal, higher, or lower modularity, respectively, relative to the corresponding alternative solution. White entries denote zero Dice similarity.}
    \label{fig:communities_ER}
\end{figure*}

\begin{figure*}
    \centering
    \includegraphics[width=1\linewidth]{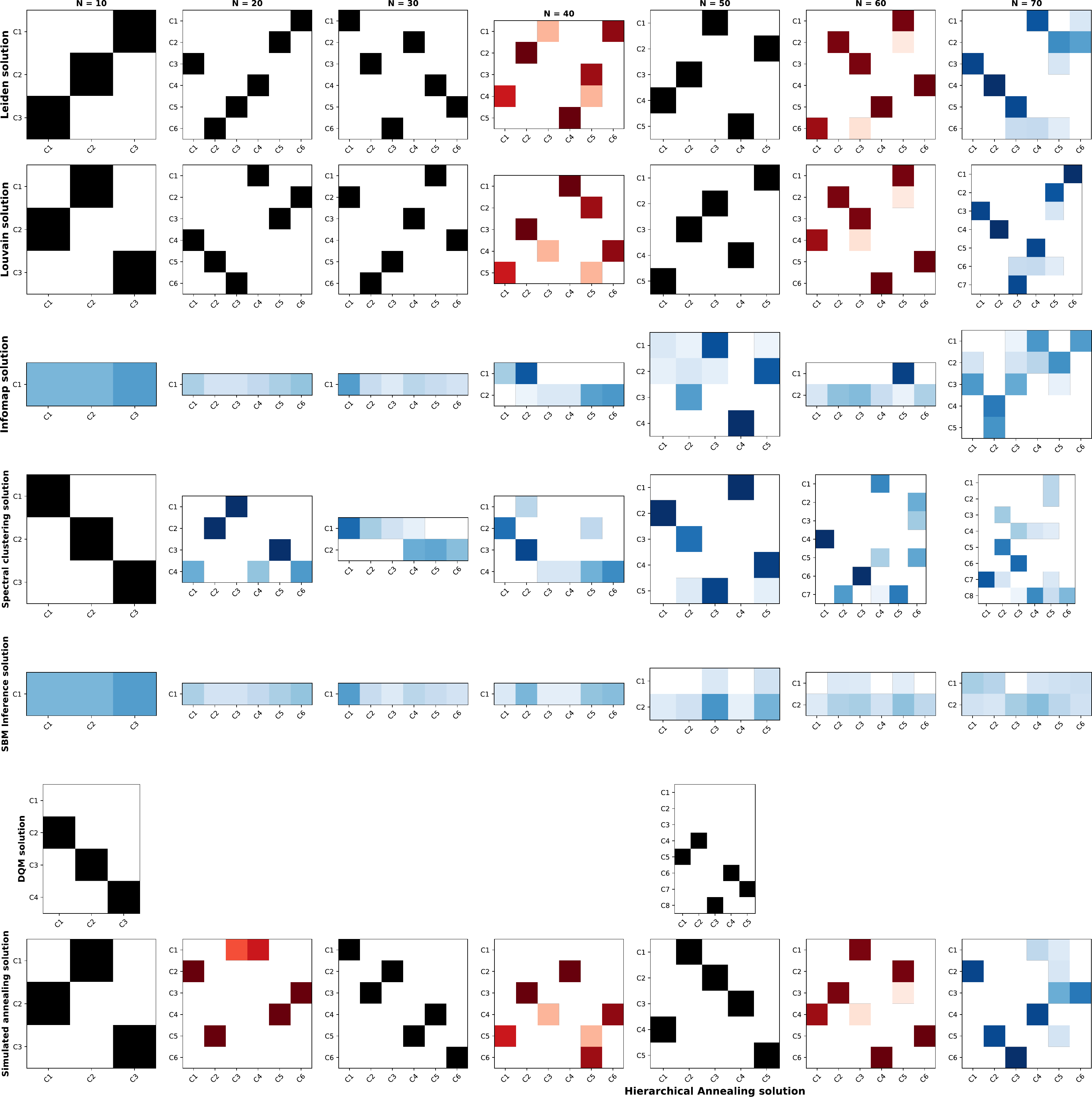}
    \caption{\textbf{Consensus matrices between the solutions found by the Hierarchical annealing and the alternative methods in directed scale-free networks of increasing sizes.} Communities identified by the hierarchical annealing procedure compared to those obtained using alternative algorithms. Opacity is proportional to the Dice similarity score between pairs of communities. Black, blue, and red outlines indicate cases in which hierarchical annealing achieved equal, higher, or lower modularity, respectively, relative to the corresponding alternative solution. White entries denote zero Dice similarity.}
    \label{fig:communities_DSF}
\end{figure*}

\begin{figure*}[h!]
    \centering
    \includegraphics[width=\linewidth]{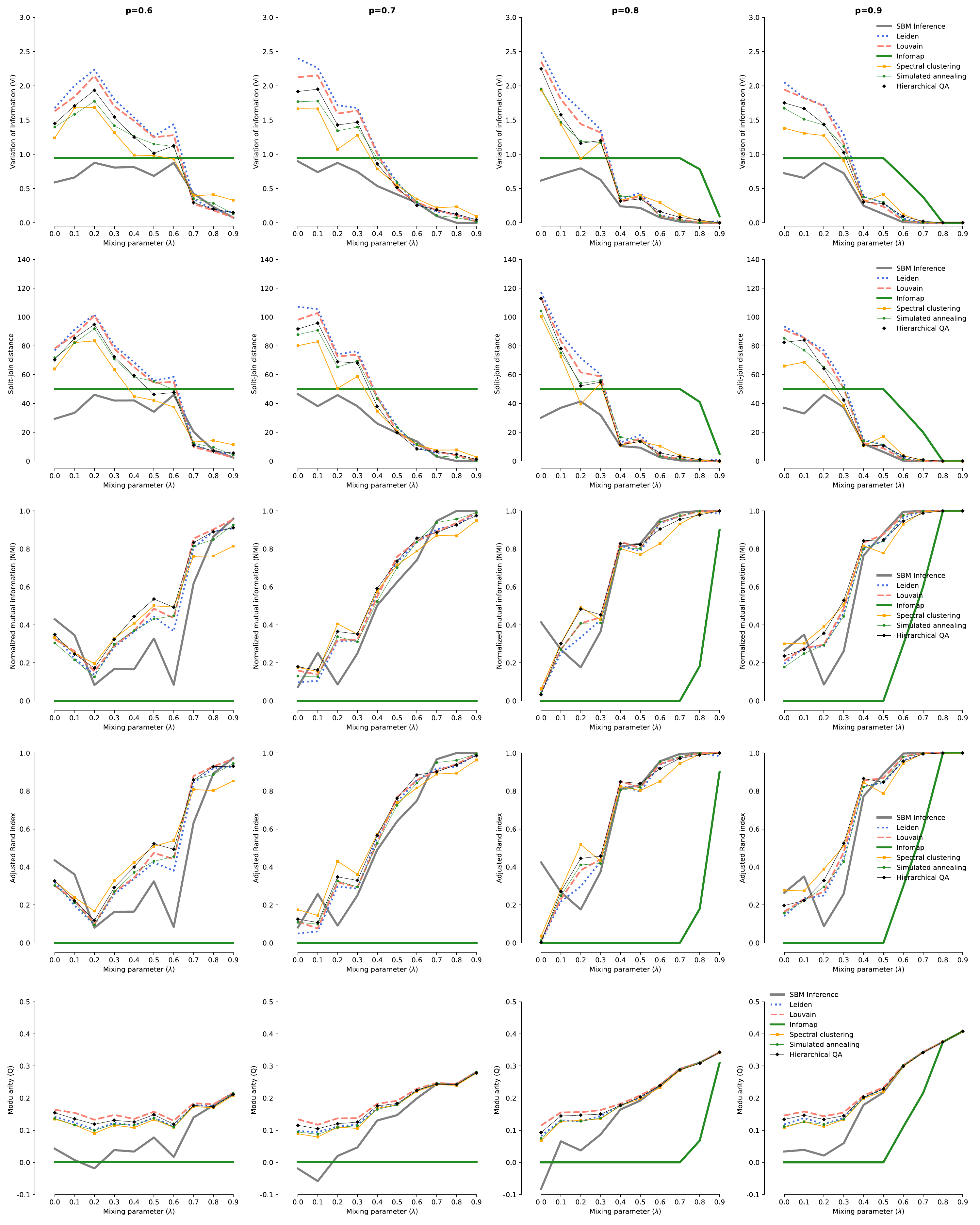}
    \caption{\textbf{Hierarchical annealing in mixed stochastic block models (SBM).} Performance of benchmark methods relative to the hierarchical annealing procedure on an SBM with three communities: $p_{in}=0.6,0.7,0.8, \text{ and } 0.9$, $p_{out}=\frac{1}{3}(1-p_{in})$. Recovery accuracy is assessed using four metrics against the ground truth, with the corresponding maximum modularity ($\max Q$) shown in the rightmost panels as a function of the mixing parameter $\lambda$. Metrics were averaged between the 10 networks generated for every pair of $p_{in}$ and $\lambda$. DQM was not studied due to computational constraints.}
    \label{fig:sbm_suppl}
\end{figure*}

\begin{figure}[h]
    \centering
    \includegraphics[width=1\linewidth]{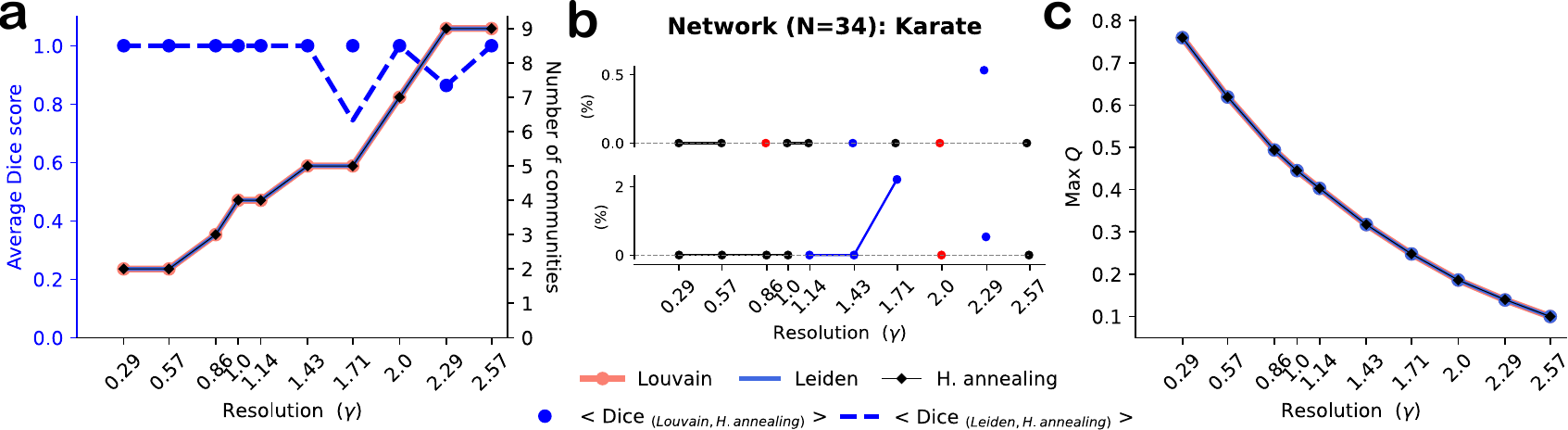}
    \caption{\textbf{Hierarchical annealing in the Karate Club network for different resolution parameters.} \textbf{a} Measure of the average overlap between the communities found by the Louvain and Hierarchical annealing algorithms (left axis, blue) and the number of communities (right axis, salmon and black) as a function of the resolution parameter $\gamma$. \textbf{b} Relative increase of the Hierarchical annealing measured w.r.t. the Louvain (top) and Leiden (bottom) solutions in \textbf{a}. Black, blue, and red symbols depict equal, better, and worse performance of the quantum method, respectively. \textbf{c} Maximum modularity per resolution value (same legend as in \textbf{a}).}
    \label{fig:resolution_karate}
\end{figure}

\begin{figure}[h]
    \centering
    \includegraphics[width=1\linewidth]{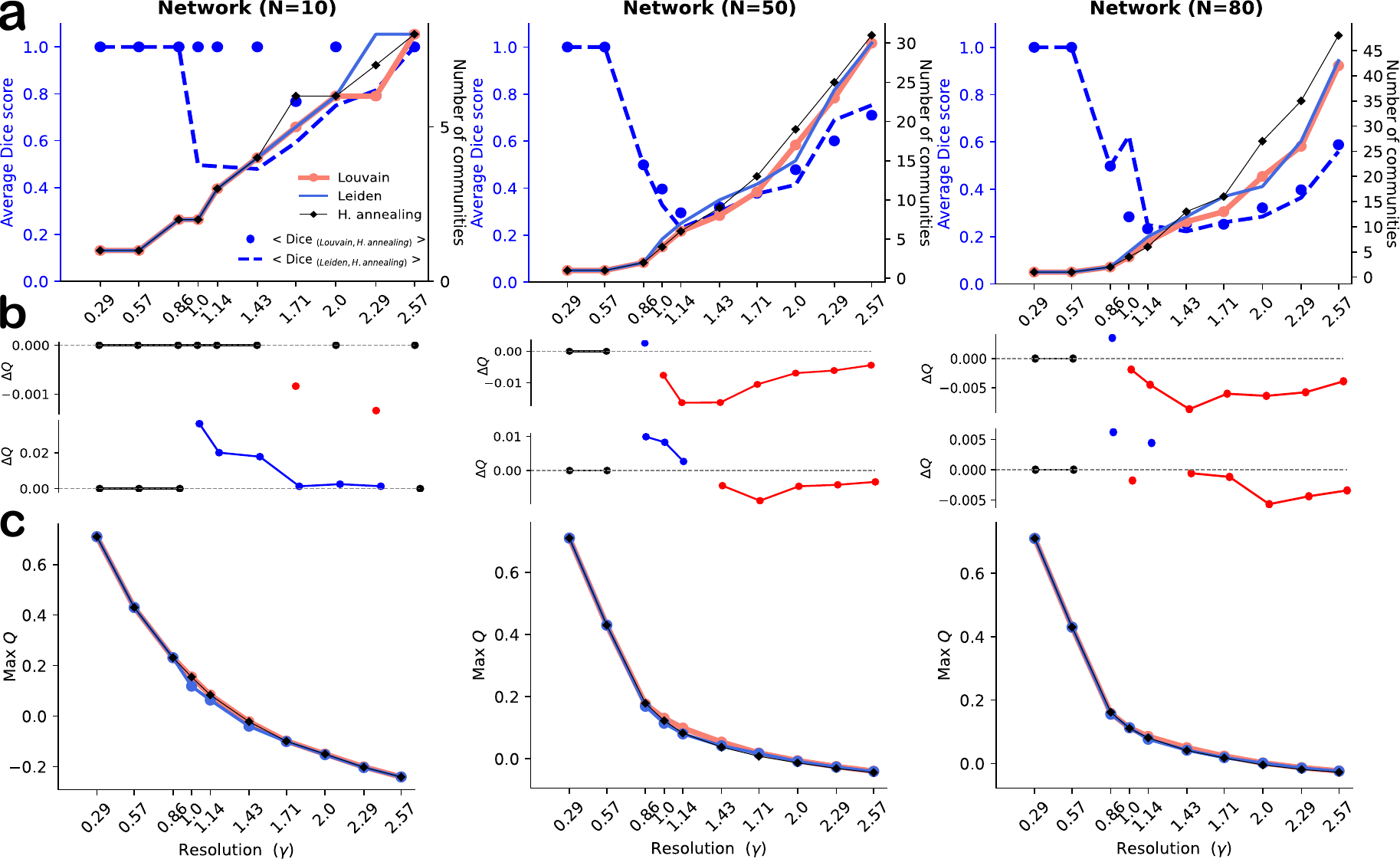}
    \caption{\textbf{Hierarchical annealing in different Erd\H{o}s-Renyi networks for different resolution parameters.} Each column depicts the results described below for a network of $N=10, 50, 80$ nodes, respectively. \textbf{a} Measure of the average overlap between the communities found by the Louvain and Hierarchical annealing algorithms (left axis, blue) and the number of communities (right axis, salmon and black) as a function of the resolution parameter $\gamma$. \textbf{b} Relative increase of the Hierarchical annealing measured w.r.t. the Louvain (top) and Leiden (bottom) solutions in \textbf{a}. Black markers correspond to identical solutions, blue markers correspond to better solutions from the annealer, and red markers indicate worse performances than the Louvain alternative. \textbf{c} Maximum modularity per resolution value (same legend as in \textbf{a}).}
    \label{fig:resolution_er}
\end{figure}

\begin{figure}[h]
    \centering
    \includegraphics[width=1\linewidth]{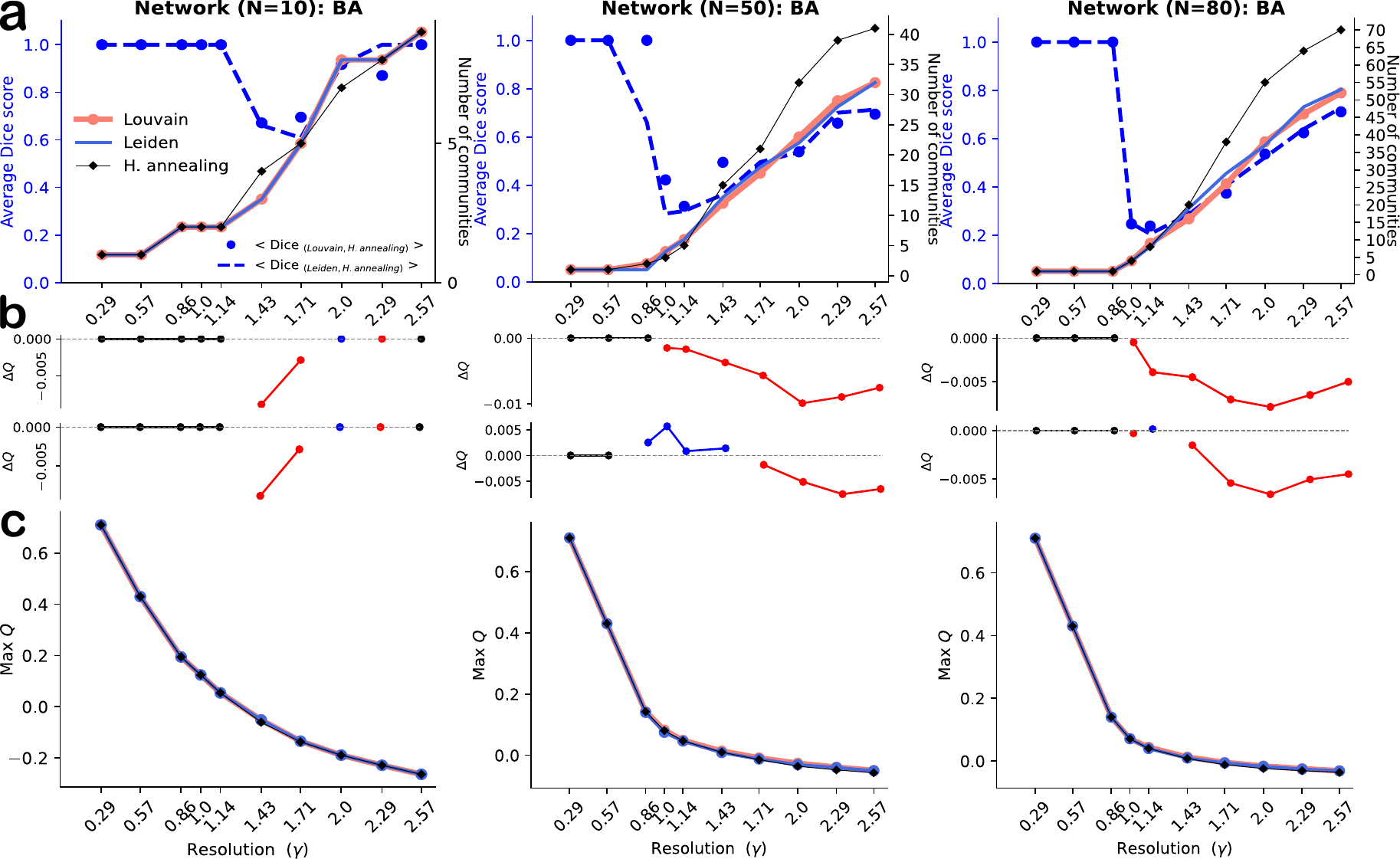}
    \caption{\textbf{Hierarchical annealing in different Barabasi-Albert networks for different resolution parameters.} Each column depicts the results described below for a network of $N=10, 50, 80$ nodes, respectively. \textbf{a} Measure of the average overlap between the communities found by the Louvain and Hierarchical annealing algorithms (left axis, blue) and the number of communities (right axis, salmon and black) as a function of the resolution parameter $\gamma$. \textbf{b} Relative increase of the Hierarchical annealing measured w.r.t. the Louvain (top) and Leiden (bottom) solutions in \textbf{a}. Black markers correspond to identical solutions, blue markers correspond to better solutions from the annealer, and red markers indicate worse performances than the Louvain alternative. \textbf{c} Maximum modularity per resolution value (same legend as in \textbf{a}).}
    \label{fig:resolution_ba}
\end{figure}

\begin{figure}[h]
    \centering
    \includegraphics[width=1\linewidth]{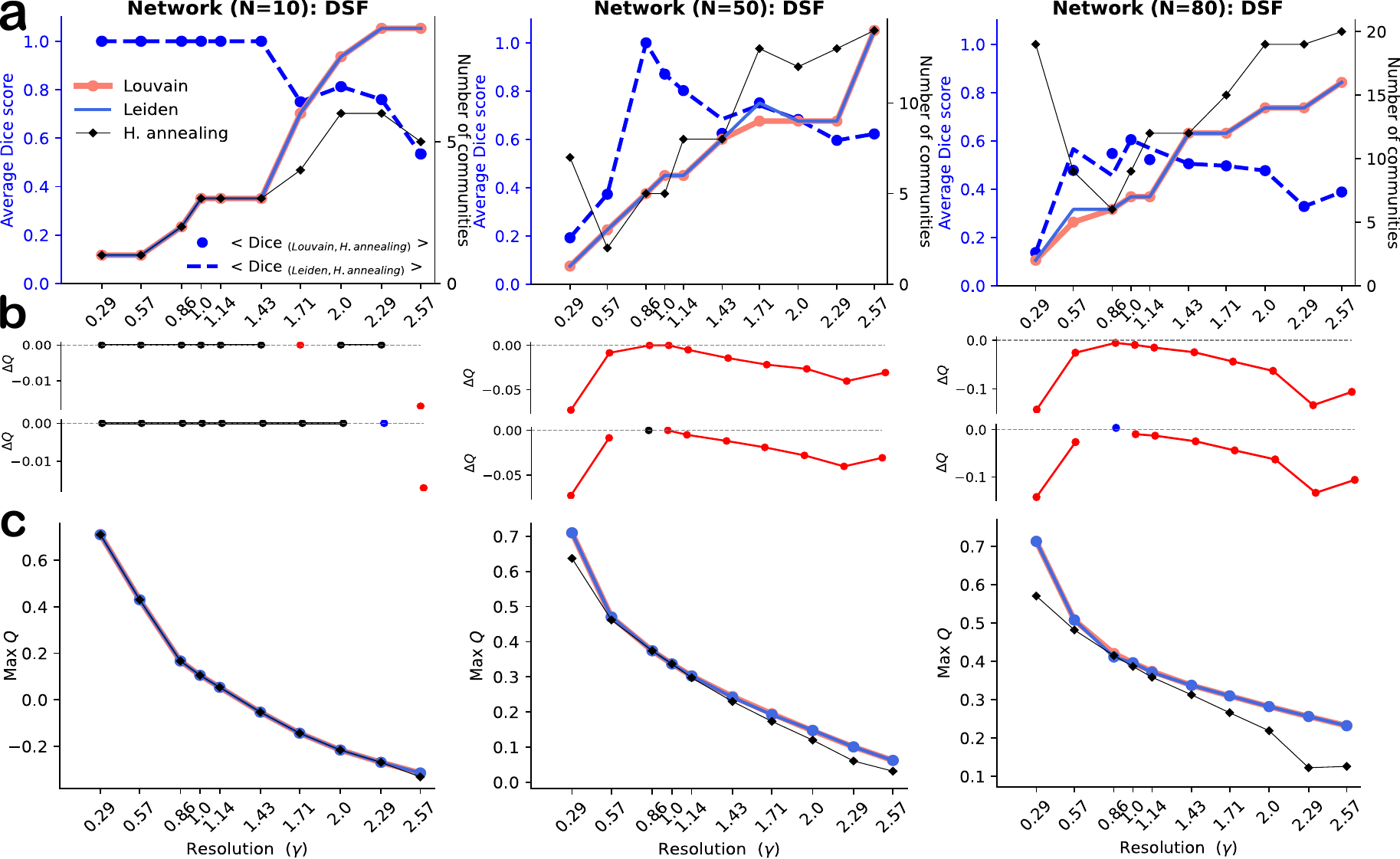}
    \caption{\textbf{Hierarchical annealing in different directed scale-free networks for different resolution parameters.} Each column depicts the results described below for a network of $N=10, 50, 80$ nodes, respectively. \textbf{a} Measure of the average overlap between the communities found by the Louvain and Hierarchical annealing algorithms (left axis, blue) and the number of communities (right axis, salmon and black) as a function of the resolution parameter $\gamma$. \textbf{b} Relative increase of the Hierarchical annealing measured w.r.t. the Louvain (top) and Leiden (bottom) solutions in \textbf{a}. Black markers correspond to identical solutions, blue markers correspond to better solutions from the annealer, and red markers indicate worse performances than the Louvain alternative. \textbf{c} Maximum modularity per resolution value (same legend as in \textbf{a}).}
    \label{fig:resolution_dsf}
\end{figure}

\begin{figure}[h]
    \centering
    \includegraphics[width=1\linewidth]{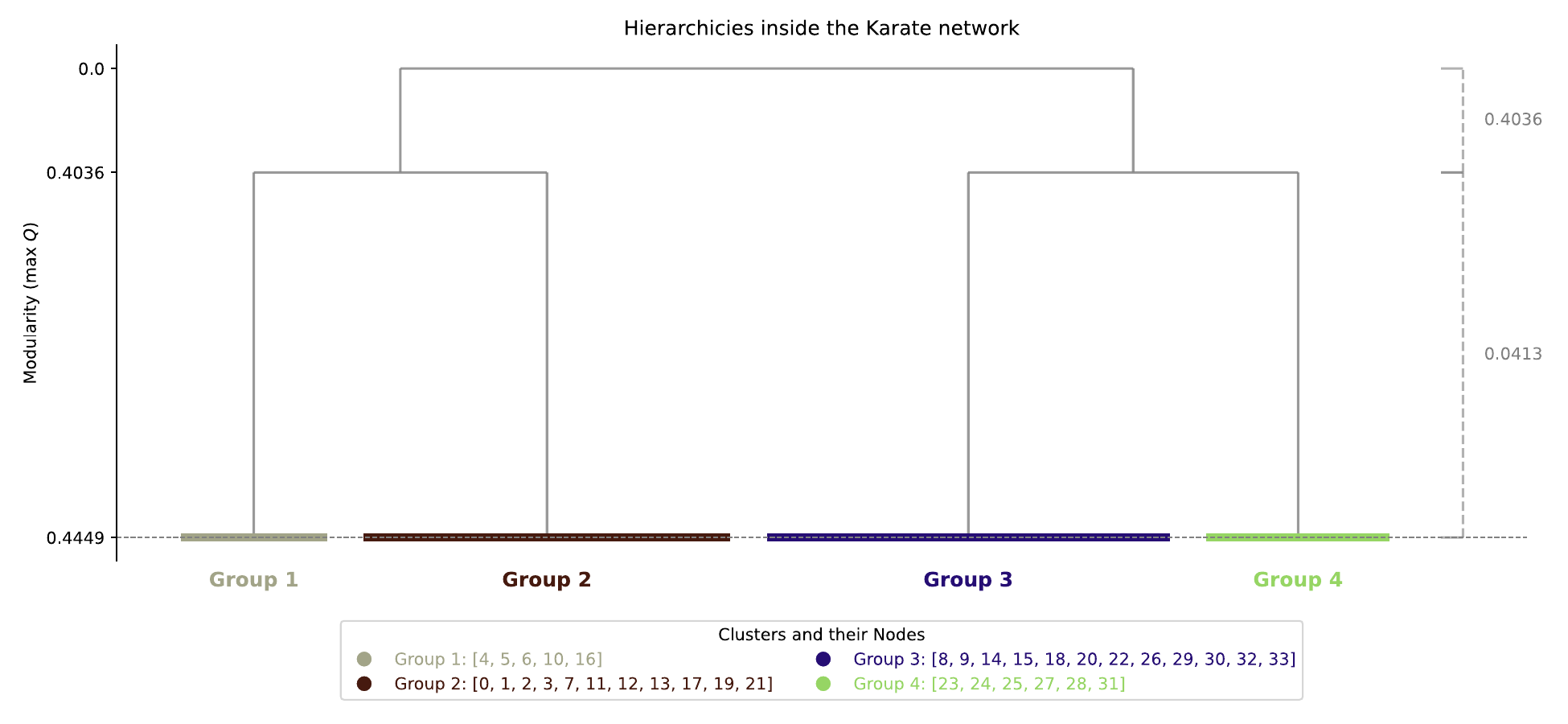}
    \caption{\textbf{Dendrogram of the Karate network \cite{zachary1977information}.} Hierarchical structure found by the H. annealing algorithm. The hierarchy corresponds to the one unraveled after maximizing the modularity of the network as described in the main text; that is, {\tt num\_runs = 50} and {\tt resolution = 1}. The left axis shows the modularity at each step of the process, while the right axis displays the corresponding increments. The bottom-most row represents the final output of the H. annealing algorithm.}
    \label{fig:dendro_karate}
\end{figure}

\begin{figure}[h]
    \centering
    \includegraphics[width=1\linewidth]{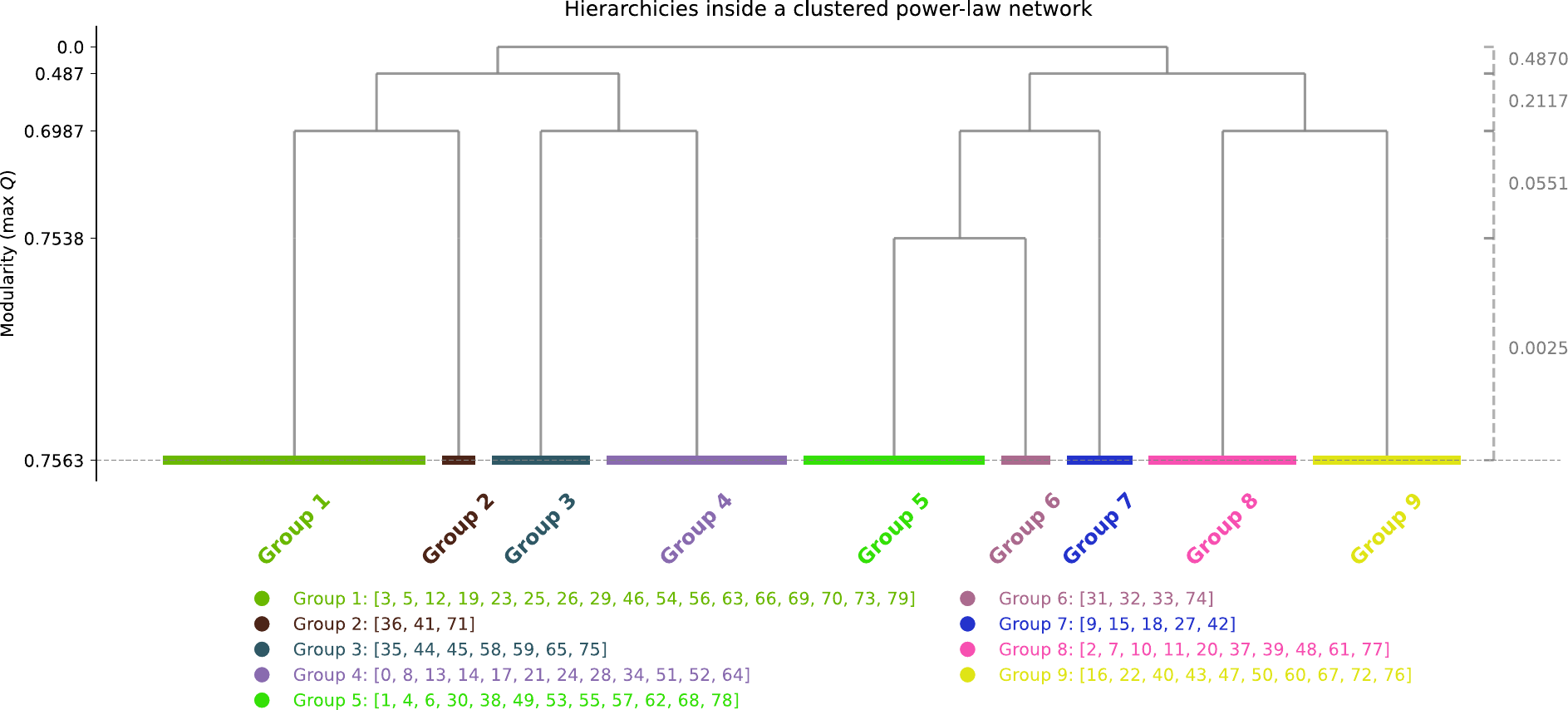}
    \caption{\textbf{Dendrogram of a clustered power-law network \cite{Holme2002} of N=80 nodes.} Hierarchical structure found by the H. annealing algorithm. The hierarchy corresponds to the one unraveled after maximizing the modularity of the network as described in the main text; that is, {\tt num\_runs = 10} and {\tt resolution = 1}. Both the Louvain and Leiden algorithms returned a maximum modularity of $Q=0.7563$. The left axis shows the modularity at each step of the process, while the right axis displays the corresponding increments. The bottom-most row represents the final output of the H. annealing algorithm.}
    \label{fig:dendro_pw_supp}
\end{figure}

\begin{figure}[h]
    \centering
    \includegraphics[width=1\linewidth]{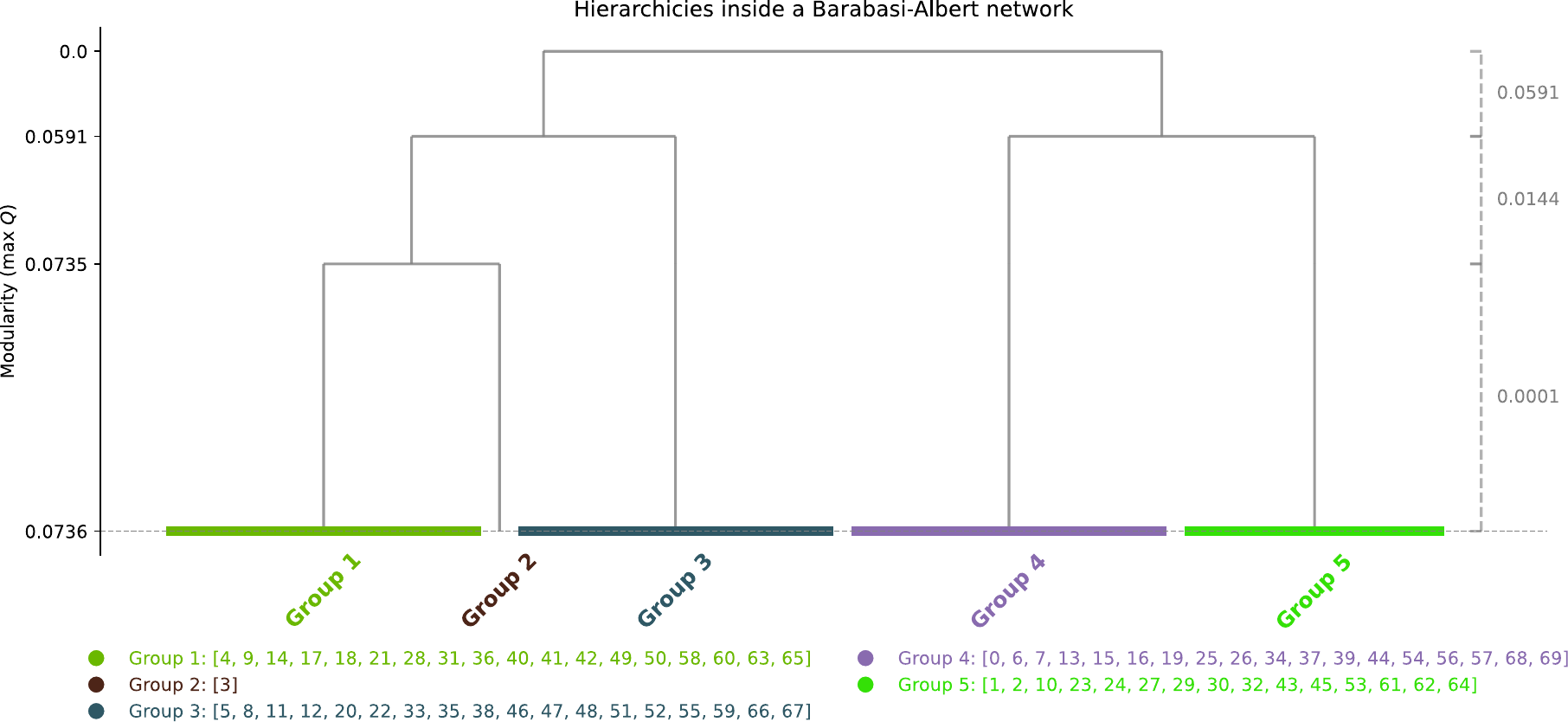}
    \caption{\textbf{Dendrogram of a Barabasi-Albert network \cite{Barabasi1999} of N=70 nodes.} Hierarchical structure found by the H. annealing algorithm. The hierarchy corresponds to the one unraveled after maximizing the modularity of the network as described in the main text; that is, {\tt num\_runs = 20} and {\tt resolution = 1}. The Louvain and Leiden algorithms returned maximum modularities of $Q=0.0760$ and $Q=0.0698$, respectively. The left axis shows the modularity at each step of the process, while the right axis displays the corresponding increments. The bottom-most row represents the final output of the H. annealing algorithm.}
    \label{fig:dendro_ba}
\end{figure}

\begin{figure}[h]
    \centering
    \includegraphics[width=1\linewidth]{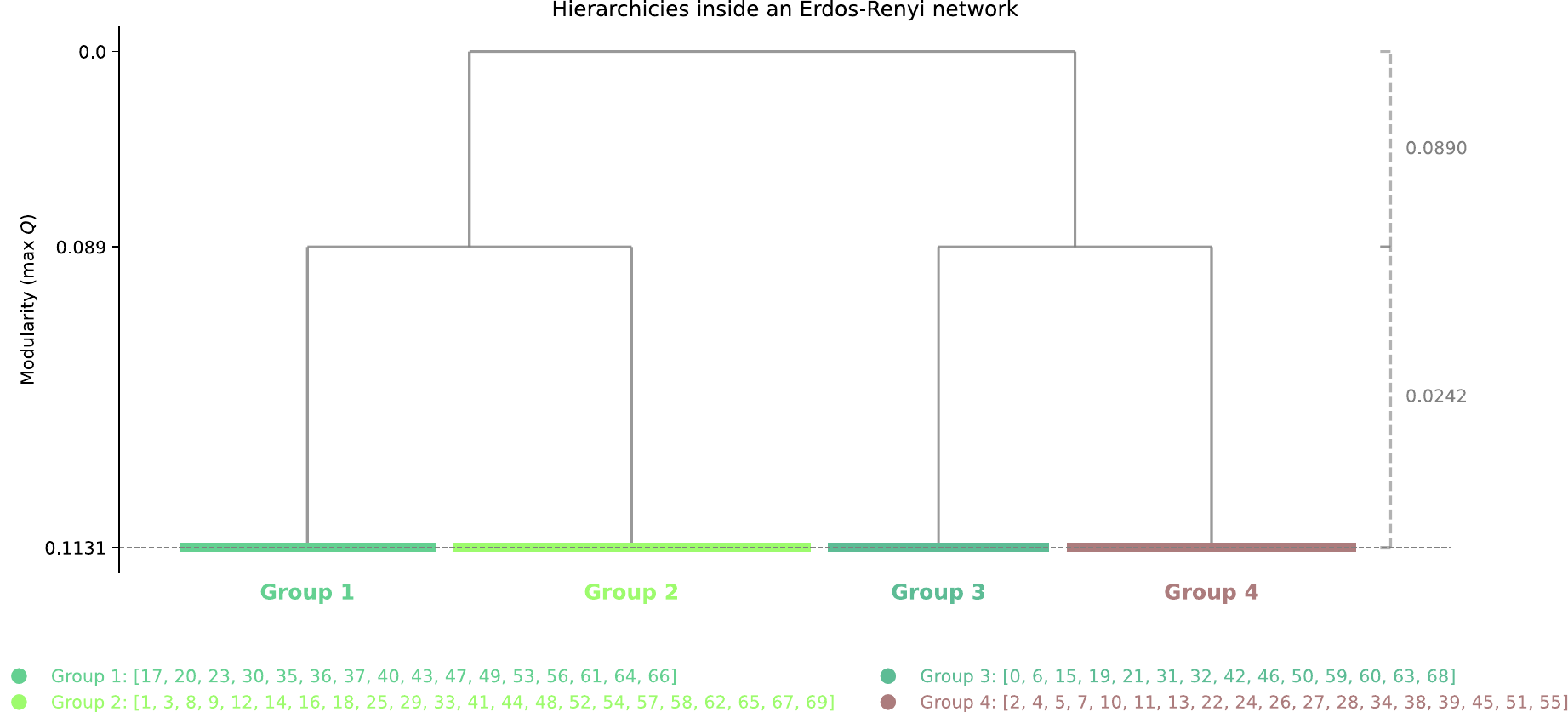}
    \caption{\textbf{Dendrogram of an Erdos-Renyi network \cite{erdds1959random} of N=70 nodes.} Hierarchical structure found by the H. annealing algorithm. The hierarchy corresponds to the one unraveled after maximizing the modularity of the network as described in the main text; that is, {\tt num\_runs = 50} and {\tt resolution = 1}. The Louvain and Leiden algorithms returned maximum modularities of $Q=0.1189$ and $Q=0.1021$ respectively. The left axis shows the modularity at each step of the process, while the right axis displays the corresponding increments. The bottom-most row represents the final output of the H. annealing algorithm.}
    \label{fig:dendro_er}
\end{figure}

\begin{figure}[h]
    \centering
    \includegraphics[width=1\linewidth]{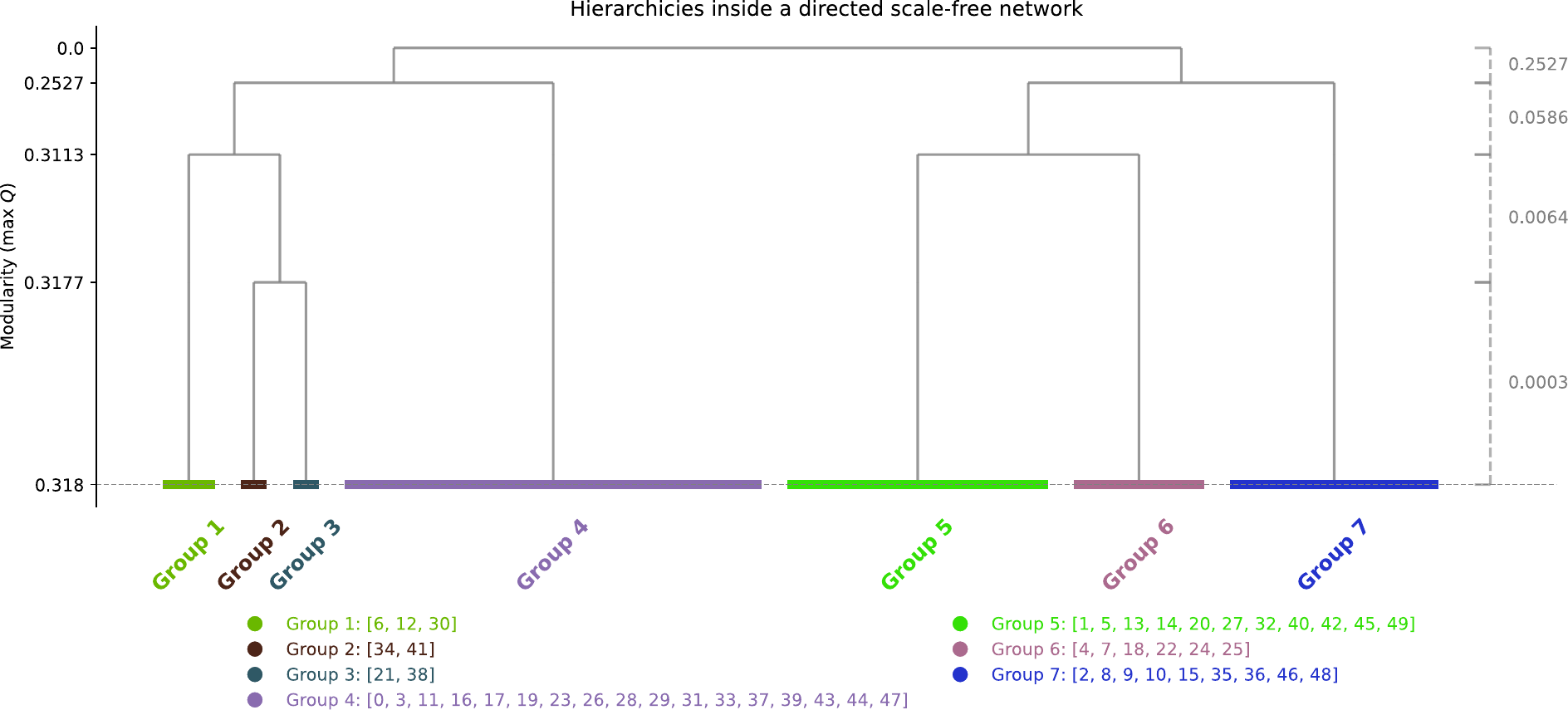}
    \caption{\textbf{Dendrogram of a directed sclae-free network \cite{bollobas2003} of N=50 nodes} Hierarchical structure found by the H. annealing algorithm. The hierarchy corresponds to the one unraveled after maximizing the modularity of the network as described in the main text; that is, {\tt num\_runs = 50} and {\tt resolution = 1}. Both the Louvain and Leiden algorithms returned a maximum modularity of $Q=0.31975$.}
    \label{fig:dendro_sf}
\end{figure}

\begin{figure}[h]
    \centering
    \includegraphics[width=1\linewidth]{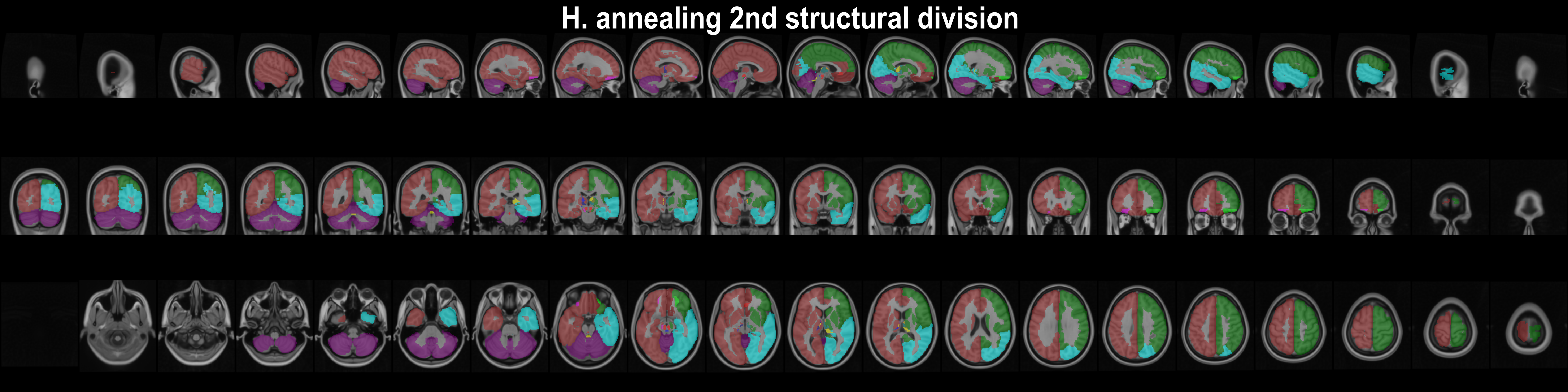}
    \caption{\textbf{Structural communities found at the early steps of the H. annealing process.} We plot the community structure discovered during the 2nd division of the hierarchical method described in the main text. It is visible how the hierarchy carries concise anatomical information, thus being a valid way to inspect hidden structures within complex networks through quantum annealing optimization.}
    \label{fig:hannealing_early_division}
\end{figure}

\begin{figure}[h]
    \centering
    \includegraphics[width=1\linewidth]{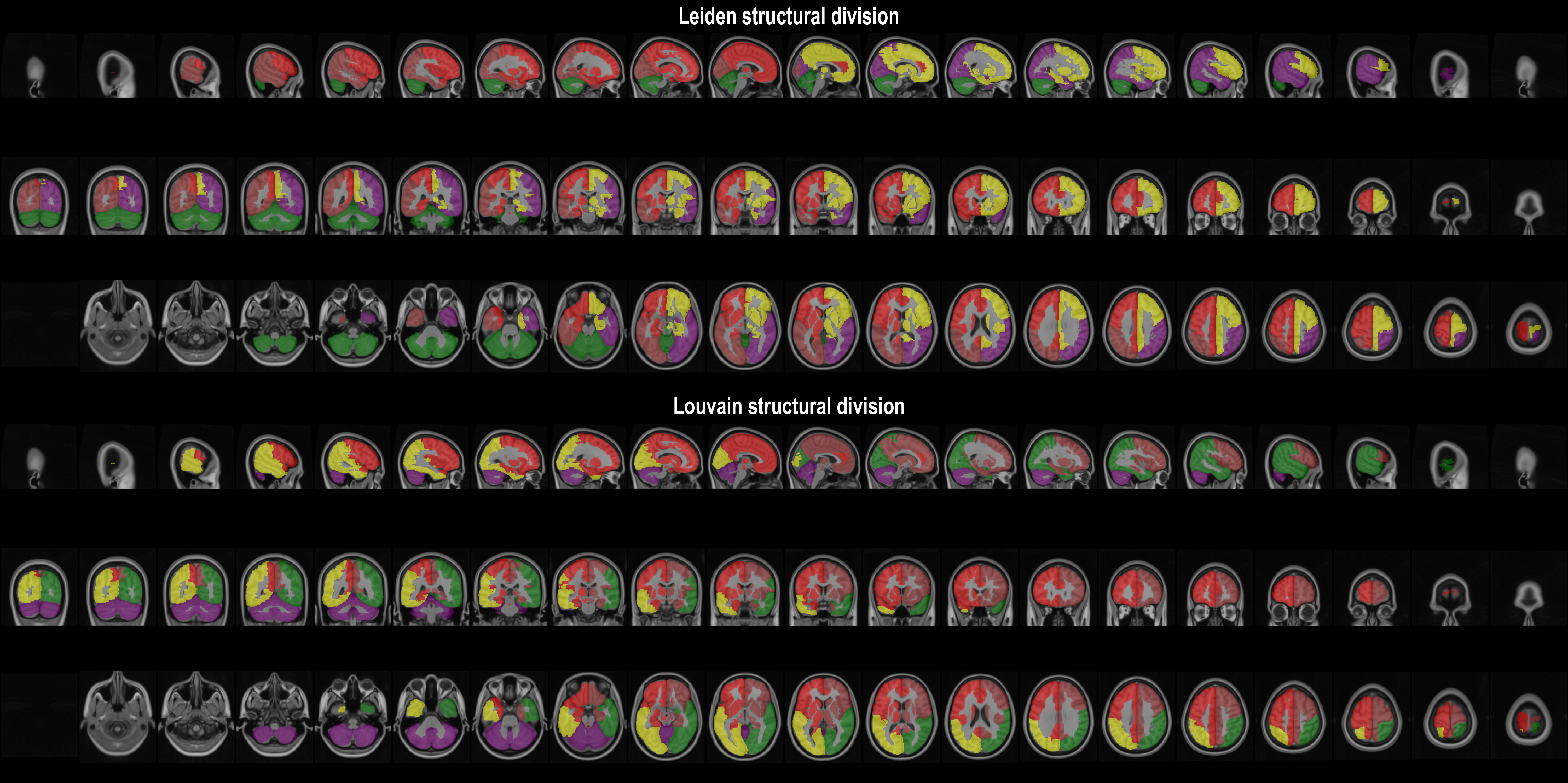}
    \caption{\textbf{Structural communities found by the Leiden and Louvain algorithms.} The resemblance between these communities and the ones found by the annealing process is strikingly high (see main text). The modularity found by the Leiden algorithm was $Q=0.610$, and the modularity returned by the Louvain algorithm was $Q=0.611$.}
    \label{fig:louva_leiden_structural}
\end{figure}

\begin{figure}[h]
    \centering
    \includegraphics[width=1\linewidth]{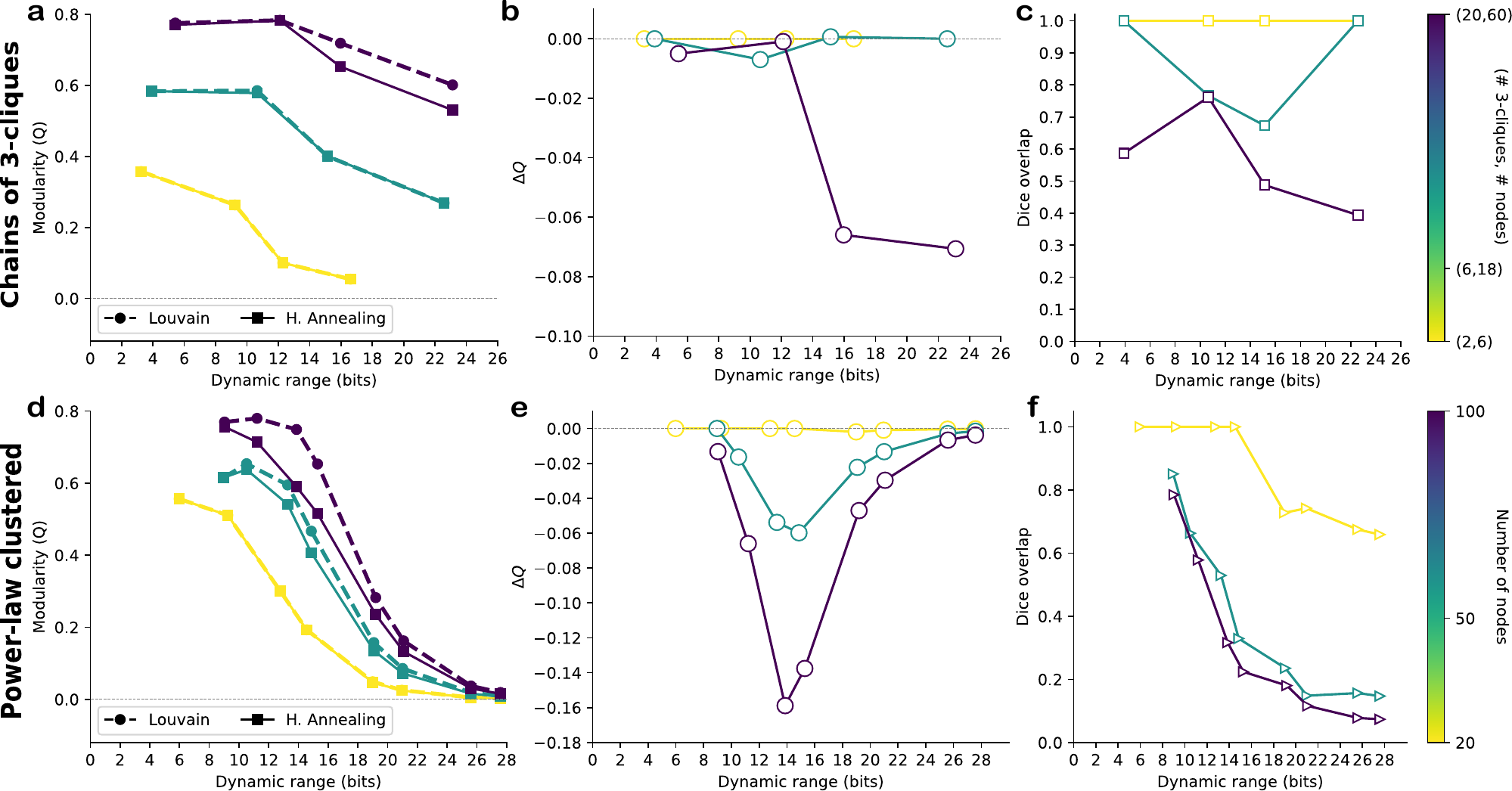}
    \caption{\textbf{Effect of the dynamical range ($\mathcal{DR}$) on the performance of the QA solver.} \textbf{a} Modularity ($Q$) found by the Louvain (circle-dashed lines) and Hierarchical annealing (square solid lines) algorithms for three different chains of 3-cliques. \textbf{b} Difference in the modularities ($\Delta Q \doteq Q_{HA}-Q_{Louvain}$) shown in \textbf{a}. \textbf{c} Average Dice score between the Louvain and Hierarchical solutions in \textbf{a}. The colorbar indicates the number of 3-cliques and the corresponding number of nodes in \textbf{a-c}. \textbf{d-f} Similarly, but for random networks generated with the preferential attachment and triadic formation mechanisms~\cite{Holme2002}. The colorbar in \textbf{f} indicates the number of nodes in the power-law clustered networks in \textbf{d-f}. The Louvain and Hierarchical annealing algorithms were run 100 and 20 times, respectively, and the solution with the highest modularity $Q$ was selected. See section S5 for details on the calculation of the $\mathcal{DR}$.}
    \label{fig:dynamic_range}
\end{figure}

\begin{figure}
    \centering
    \includegraphics[width=1\linewidth]{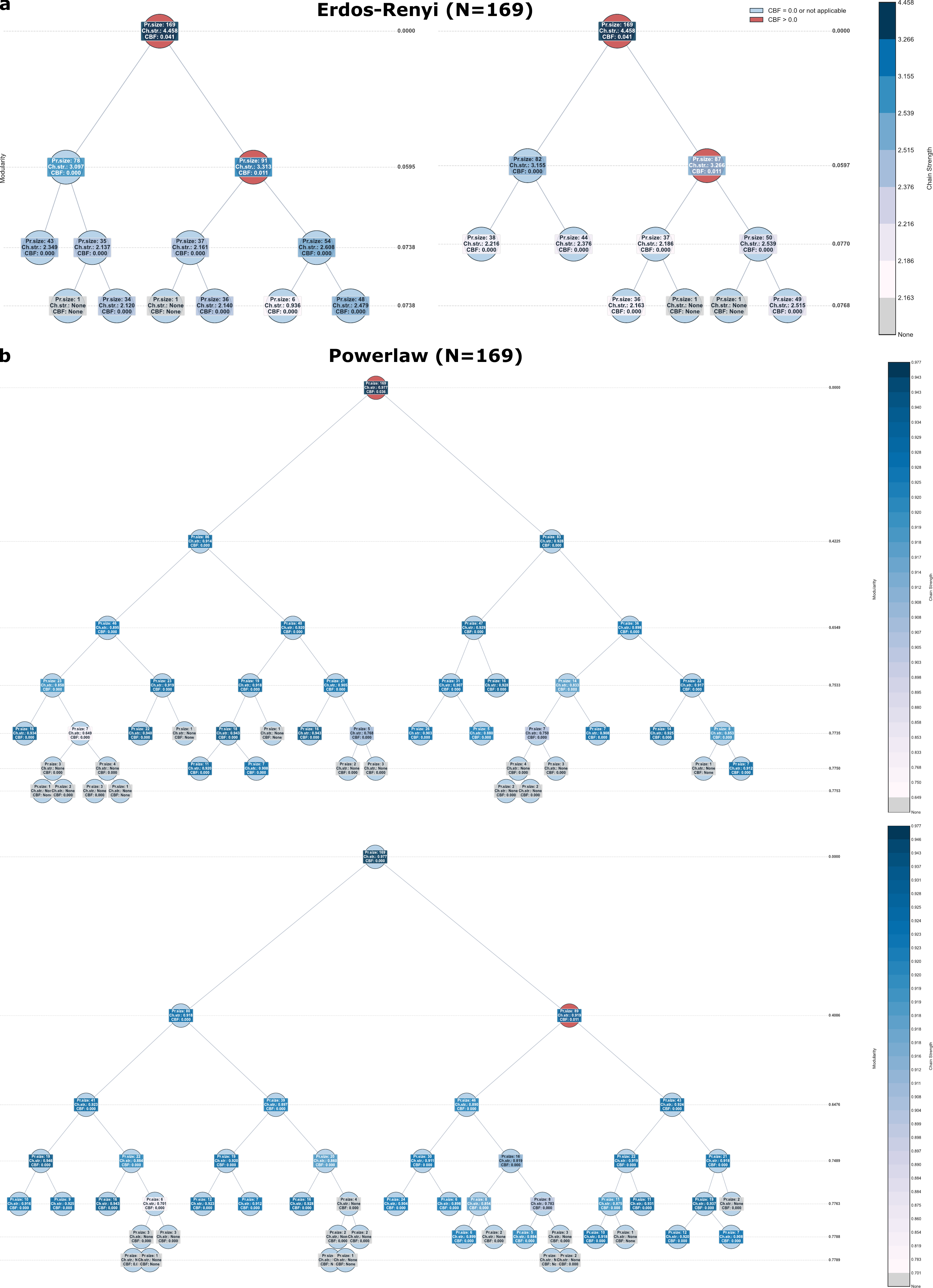}
    \caption{\textbf{Chain strength and break fractions along the hierarchy.} \textbf{a} Two randomly selected independent runs for an Erd\H os--Renyi network (left and right). Circles depict the number of nodes, the chain strength, and the observed chain break fraction at a given level. Circle color represents the chain break fraction, while text color indicates the chain strength. \textbf{b} Same as \textbf{a}, but for two randomly selected runs of a power-law network (top and bottom).}
    \label{fig:noise_hierarchy}
\end{figure}

\begin{figure}[h]
    \centering
    \includegraphics[width=.995\linewidth]{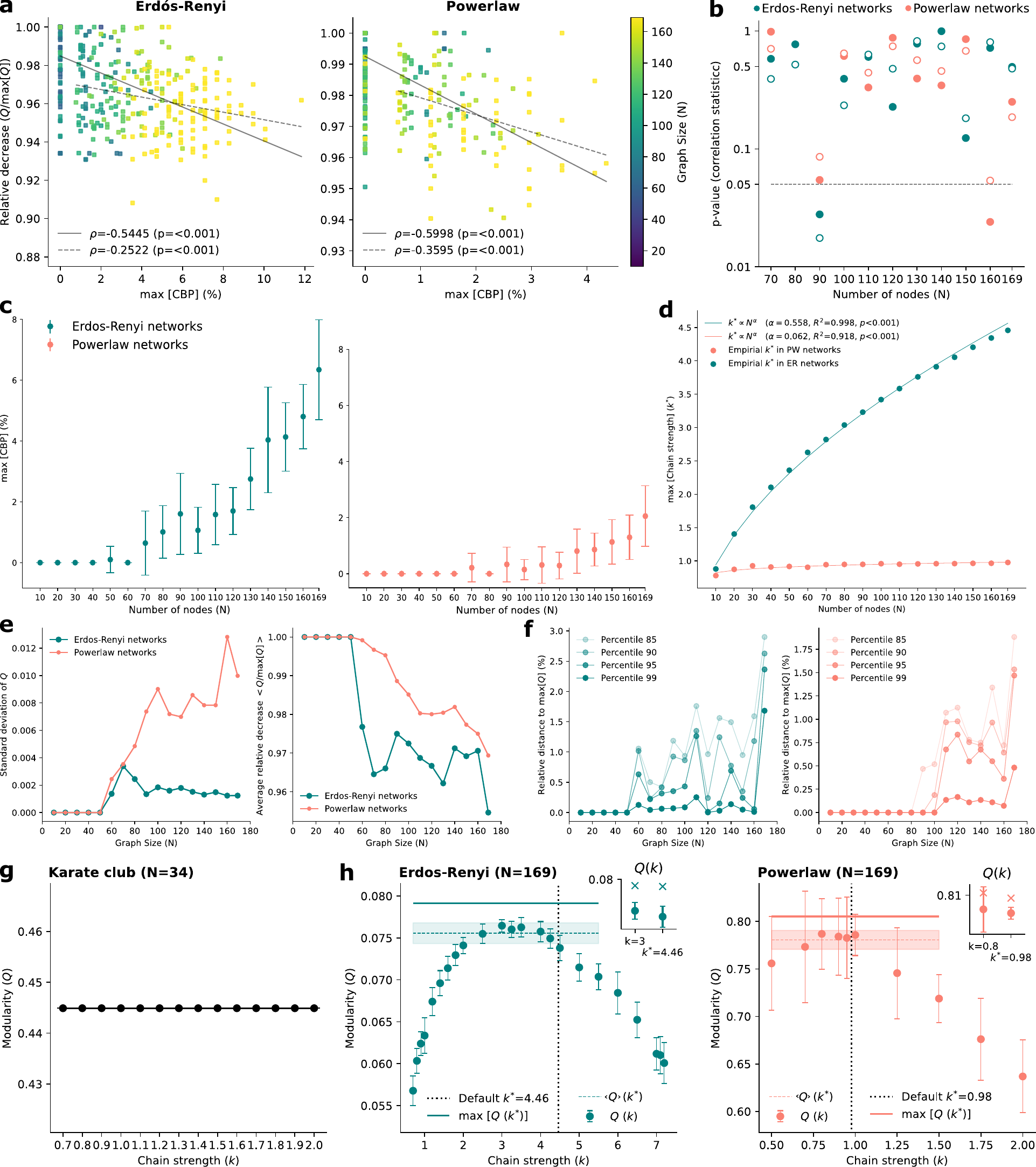}
    \caption{\textbf{Chain strength and noise in the quantum annealer.}
    \textbf{a} Relative decrease in modularity obtained at each independent run of Algorithm~1 with respect to the maximum modularity, shown as a function of the chain break probability ($CBP$) for \textbf{(left)} Erd\H{o}s--R\'enyi and \textbf{(right)} power-law networks. Linear models reveal a significant negative correlation between solution quality and $CBP$ (two-sided exact test; solid line includes all points; dashed line excludes points with $CBP=0$).
    \textbf{b} $p$-values of exact Pearson (filled circles) and Spearman (open circles) tests between the relative decrease defined in \textbf{a} and $CBP$ for networks of increasing size. No strong relationship is observed at the level of individual networks, suggesting that the global trend in \textbf{a} arises from aggregate effects.
    \textbf{c} Average $\max CBP$ along the hierarchical tree (first division) for \textbf{(left)} Erd\H{o}s--R\'enyi and \textbf{(right)} power-law networks.
    \textbf{d} Chain strength $k^{*}$ set by the default strategy. Solid lines show power-law fits to $k^{*}$.
    \textbf{e} \textbf{(left)} Standard deviation of modularity values obtained from the hierarchical annealing procedure over $N_{\text{runs}}=20$ as a function of network size. \textbf{(right)} Corresponding average relative decrease with respect to $\max Q$. Variability remains constant for Erd\H{o}s--R\'enyi networks but increases with size for power-law networks, leading to a progressive degradation in average performance.
    \textbf{f} Percentage distance between statistical percentiles of the modularity distribution and the best solution ($\max Q$) for \textbf{(left)} Erd\H{o}s--R\'enyi and \textbf{(right)} power-law networks.
    \textbf{g} $\max Q$ for varying chain strengths in the Karate Club network.
    \textbf{h} Modularity as a function of chain strength. Horizontal lines indicate the average (dashed) and best (solid) modularity obtained using the default chain strength from \textbf{d}. Insets show the average modularity and $\max Q$ (crosses) achieved by Algorithm~1. \\
    \textit{In all figures, error bars and shaded regions depict $\pm 1$ standard deviation.}}
    \label{fig:chain_strength}
\end{figure}

\clearpage
\section*{Supplementary tables}

\begin{table}[ht]
    \centering
    \scriptsize
    \resizebox{\columnwidth}{!}{%
        \begin{tabular}{ll|rr||rr|rrr}
        \hline
        \textbf{Solver} & \textbf{Type of network} & \# \textbf{Nodes} & \textbf{\texttt{num\_runs}} & $\mathbf{Q}$ & \(\mathbf{|C|}\) & \textbf{Wall Time in DWave portal (s)} & \# \textbf{Problems in DWave portal} & \textbf{Avg. time per problem (s)} \\
        \hline
        \hline
        DQM & Karate & 34 & 1 & 0.444904 & 4 & 5.000 & 1 & 5.000 \\
        \hline
        DQM & Erd\H os-Renyi & 10 & 5 & 0.153061 & 4 & 50.000 & 5 & 10.000 \\
        DQM & Erd\H os-Renyi & 50 & 5 & 0.145398 & 4 & 50.000 & 5 & 10.000 \\
        DQM & Erd\H os-Renyi & 100 & 5 & 0.105132 & 4 & 50.000 & 5 & 10.000 \\
        \hline
        DQM & Barabasi-Albert & 10 & 5 & 0.044271 & 3 & 50.000 & 5 & 10.000 \\
        DQM & Barabasi-Albert & 50 & 5 & 0.094261 & 3 & 50.000 & 5 & 10.000 \\
        DQM & Barabasi-Albert & 100 & 5 & 0.070131 & 4 & 50.000 & 5 & 10.000 \\
        \hline
        DQM & Directed scale-free & 10 & 5 & 0.375000 & 3 & 50.000 & 5 & 10.000 \\
        DQM & Directed scale-free & 50 & 5 & 0.265512 & 5 & 50.000 & 5 & 10.000 \\
        DQM & Directed scale-free & 100 & 5 & 0.405622 & 8 & 50.000 & 5 & 10.000 \\
        \hline
        DQM & Powerlaw & 10 & 5 & 0.425926 & 3 & 50.000 & 5 & 10.000 \\
        DQM & Powerlaw & 50 & 5 & 0.660558 & 6 & 50.000 & 5 & 10.000 \\
        DQM & Powerlaw & 100 & 5 & 0.783083 & 10 & 50.000 & 5 & 10.000 \\
        \hline
        \hline
        H. Annealing & Karate & 34 & 50 & 0.444904 & 4 & 10.924 & 350 & 0.03120 \\
        \hline
        H. Annealing & Erd\H os-Renyi & 10 & 20 & 0.140306 & 3 & 2.222 & 80 & 0.02776 \\
        H. Annealing & Erd\H os-Renyi & 20 & 20 & 0.185249 & 3 & 2.897 & 100 & 0.02897 \\
        H. Annealing & Erd\H os-Renyi & 30 & 20 & 0.173824 & 3 & 3.192 & 100 & 0.03192 \\
        H. Annealing & Erd\H os-Renyi & 40 & 20 & 0.150733 & 4 & 4.495 & 140 & 0.03210 \\
        H. Annealing & Erd\H os-Renyi & 50 & 20 & 0.135665 & 4 & 4.555 & 140 & 0.03253 \\
        H. Annealing & Erd\H os-Renyi & 60 & 20 & 0.121331 & 4 & 4.790 & 140 & 0.03421 \\
        H. Annealing & Erd\H os-Renyi & 70 & 20 & 0.122608 & 4 & 4.962 & 140 & 0.03544 \\
        H. Annealing & Erd\H os-Renyi & 80 & 20 & 0.115098 & 4 & 5.044 & 140 & 0.03603 \\
        H. Annealing & Erd\H os-Renyi & 90 & 20 & 0.103103 & 4 & 4.979 & 141 & 0.03531 \\
        H. Annealing & Erd\H os-Renyi & 100 & 20 & 0.102702 & 4 & 4.934 & 142 & 0.03475 \\
        \hline
        H. Annealing & Barabasi-Albert & 10 & 20 & 0.044271 & 3 & 2.292 & 100 & 0.02292 \\
        H. Annealing & Barabasi-Albert & 20 & 20 & 0.088162 & 3 & 2.333 & 100 & 0.02333 \\
        H. Annealing & Barabasi-Albert & 30 & 20 & 0.083998 & 4 & 3.293 & 140 & 0.02352 \\
        H. Annealing & Barabasi-Albert & 40 & 20 & 0.097280 & 3 & 2.842 & 160 & 0.01776 \\
        H. Annealing & Barabasi-Albert & 50 & 20 & 0.092511 & 3 & 2.459 & 100 & 0.02459 \\
        H. Annealing & Barabasi-Albert & 60 & 20 & 0.074836 & 3 & 3.335 & 136 & 0.02452 \\
        H. Annealing & Barabasi-Albert & 70 & 20 & 0.076799 & 3 & 3.442 & 140 & 0.02459 \\
        H. Annealing & Barabasi-Albert & 80 & 20 & 0.076184 & 4 & 3.506 & 141 & 0.02487 \\
        H. Annealing & Barabasi-Albert & 90 & 20 & 0.068596 & 4 & 3.679 & 146 & 0.02520 \\
        H. Annealing & Barabasi-Albert & 100 & 20 & 0.066951 & 4 & 4.005 & 154 & 0.02601 \\
        \hline
        H. Annealing & Directed scale-free & 10 & 20 & 0.37500 & 3 & 4.392 & 96 & 0.04575 \\
        H. Annealing & Directed scale-free & 20 & 20 & 0.327977 & 6 & 8.210 & 220 & 0.03732 \\
        H. Annealing & Directed scale-free & 30 & 20 & 0.249005 & 6 & 12.243 & 220 & 0.05565 \\
        H. Annealing & Directed scale-free & 40 & 20 & 0.297133 & 6 & 10.356 & 220 & 0.04707 \\
        H. Annealing & Directed scale-free & 50 & 40 & 0.265518 & 5 & 17.040 & 560 & 0.03043 \\
        H. Annealing & Directed scale-free & 60 & 40 & 0.368195 & 6 & 20.962 & 440 & 0.04764 \\
        H. Annealing & Directed scale-free & 70 & 40 & 0.391541 & 6 & 21.086 & 440 & 0.04792 \\
        \hline
        H. Annealing & Powerlaw & 10 & 20 & 0.425926 & 3 & 2.635 & 100 & 0.02640 \\
        H. Annealing & Powerlaw & 20 & 20 & 0.581717 & 5 & 5.031 & 180 & 0.02800 \\
        H. Annealing & Powerlaw & 30 & 20 & 0.629608 & 5 & 5.286 & 180 & 0.02940 \\
        H. Annealing & Powerlaw & 40 & 20 & 0.650230 & 5 & 5.584 & 180 & 0.03100 \\
        H. Annealing & Powerlaw & 50 & 20 & 0.660558 & 6 & 6.598 & 220 & 0.03000 \\
        H. Annealing & Powerlaw & 60 & 20 & 0.660011 & 10 & 11.434 & 382 & 0.02990 \\
        H. Annealing & Powerlaw & 70 & 20 & 0.750787 & 9 & 10.605 & 340 & 0.03120 \\
        H. Annealing & Powerlaw & 80 & 20 & 0.748838 & 8 & 9.986 & 302 & 0.03310 \\
        H. Annealing & Powerlaw & 90 & 20 & 0.756407 & 12 & 14.624 & 471 & 0.03100 \\
        H. Annealing & Powerlaw & 100 & 20 & 0.783083 & 10 & 12.727 & 399 & 0.03190 \\
        \hline
        \end{tabular}
    }
    \caption{\textbf{Wall times reported in the DWave Leap Portal for the different solvers, networks, and number of nodes.} Computing resources spent using DQM and Advantage for the experiments in Fig. 2 in the main text (see also Fig.~\ref{fig:dwave_times}). The time consumed by DQM to solve the 71 problems was 11.74 minutes, whereas the 7,880 problems were solved using Advantage in 4.38 minutes. This corresponded to an average of almost 300 more physical problems being solved per unit time using our approach. Such differences can substantially impact cost estimates when planning computational grants.}
    \label{tab:wall_times}
\end{table}

\begin{table}[h]
    \centering
    \resizebox{\columnwidth}{!}{%
        \begin{tabular}{l||c|c|c|c|c|c|c|c}
        Number of nodes                                    & 10                & 20                & 30                & 50                & 80                 & 100                & 120                & 140                \\
        Modularity                                         & 0.425926          & 0.526316          & 0.642687          & 0.697834          & 0.755969           & 0.766401           & 0.795389           & 0.821101           \\
        Total QPU sampling time (s)                        & 0.052296          & 0.105598          & 0.147786          & 0.208824          & 0.292442           & 0.329412           & 0.330046           & 0.418858           \\
        Total QPU anneal time per sample (s)               & 0.000100          & 0.000180          & 0.000220          & 0.000300          & 0.000380           & 0.000420           & 0.000380           & 0.000500           \\
        Total QPU readout time per sample (s)              & 0.000320          & 0.000691          & 0.001031          & 0.001480          & 0.002153           & 0.002442           & 0.002529           & 0.003174           \\
        \textbf{Total QPU access time (s)}                 & \textbf{0.131099} & \textbf{0.247444} & \textbf{0.321154} & \textbf{0.445232} & \textbf{0.591897}  & \textbf{0.660461}  & \textbf{0.629556}  & \textbf{0.812948}  \\
        Total QPU access overhead time (s)                 & 0.007715          & 0.014182          & 0.012381          & 0.019401          & 0.022146           & 0.025523           & 0.023243           & 0.027625           \\
        Total QPU programming time (s)                     & 0.078803          & 0.141846          & 0.173368          & 0.236408          & 0.299455           & 0.331049           & 0.299510           & 0.394090           \\
        Total QPU delay time per sample (s)                & 0.000103          & 0.000185          & 0.000226          & 0.000309          & 0.000391           & 0.000432           & 0.000391           & 0.000515           \\
        Total post-processing time (s)                     & 0.000101          & 0.000093          & 0.000047          & 0.000168          & 0.000215           & 0.000261           & 0.000290           & 0.000511           \\
        \textbf{Total cache-and-read embedding   time (s)} & \textbf{1.140593} & \textbf{0.380248} & \textbf{0.388070} & \textbf{0.685955} & \textbf{1.320070}  & \textbf{2.639409}  & \textbf{0.515561}  & \textbf{1.552527}  \\
        \textbf{Total solver communication   time (s)}     & \textbf{6.948981} & \textbf{8.597406} & \textbf{9.256390} & \textbf{9.308154} & \textbf{11.834502} & \textbf{11.629346} & \textbf{11.598674} & \textbf{14.987153}
        \end{tabular}
    }
    \caption{\textbf{Experimental computing times for the Powerlaw clustered networks and $\gamma=1$.} Running times across resolution parameters (see Methods). Bold rows indicate the subset summarizing overall recursive performance. "Total" is the sum of all substep durations within the recursion.}
\end{table}

\begin{table}[h]
    \begin{tabular}{l||c|c|c|c||r}
    Resolution                                & 0.5     & 0.75     & 1.25     & 1.5      & \multicolumn{1}{l}{Nodes} \\ \hline
    Total   QPU access time (s)               & 0.34396 & 0.388946 & 0.445299 & 0.496148 & \multirow{3}{*}{50}       \\
    Total cache-and-read embedding   time (s) & 1.30217 & 0.228922 & 0.373077 & 2.007349 &                           \\
    Total solver communication   time (s)     & 9.36499 & 8.738877 & 8.841599 & 12.29029 &                           \\ \hline
    Total   QPU access time (s)               & 0.43172 & 0.48114  & 0.658621 & 0.643321 & \multirow{3}{*}{80}       \\
    Total cache-and-read embedding   time (s) & 1.91894 & 0.655148 & 0.654045 & 1.334824 &                           \\
    Total solver communication   time (s)     & 11.1531 & 9.266652 & 11.69023 & 11.12751 &                           \\ \hline
    Total   QPU access time (s)               & 0.48956 & 0.489249 & 0.709041 & 0.76481  & \multirow{3}{*}{100}      \\
    Total cache-and-read embedding   time (s) & 1.001   & 1.461846 & 0.233278 & 0.309836 &                           \\
    Total solver communication   time (s)     & 8.88566 & 9.292936 & 12.88506 & 11.81102 &                                                   
    \end{tabular}
\caption{\textbf{Experimental computing times Powerlaw clustered networks across different resolution parameters.} Running times across resolution parameters (see Methods)."Total" is the sum of all substep durations within the recursion.}
\end{table}

\begin{table}[h]
    \centering
    \resizebox{\columnwidth}{!}{%
        \begin{tabular}{l||c|c|c|c|c|c|c|c}
        Number of nodes                                    & 10                & 20                & 30                & 50                & 80                & 100               & 120                & 140               \\
        Modularity                                         & 0.079861          & 0.110189          & 0.092432          & 0.081621          & 0.072629          & 0.064587          & 0.057521           & 0.048497          \\
        Total QPU sampling time (s)                        & 0.037396          & 0.072334          & 0.105932          & 0.110426          & 0.146730          & 0.134112          & 0.178538           & 0.198386          \\
        Total QPU anneal time per sample (s)               & 0.000060          & 0.000100          & 0.000140          & 0.000140          & 0.000140          & 0.000140          & 0.000180           & 0.000200          \\
        Total QPU readout time per sample (s)              & 0.000252          & 0.000520          & 0.000775          & 0.000820          & 0.001183          & 0.001057          & 0.001420           & 0.001578          \\
        \textbf{Total QPU access time (s)}                 & \textbf{0.084684} & \textbf{0.151147} & \textbf{0.216271} & \textbf{0.220758} & \textbf{0.257087} & \textbf{0.244463} & \textbf{0.320421}  & \textbf{0.356009} \\
        Total QPU access overhead time (s)                 & 0.003389          & 0.007431          & 0.008363          & 0.008620          & 0.009782          & 0.012430          & 0.014950           & 0.014678          \\
        Total QPU programming time (s)                     & 0.047288          & 0.078813          & 0.110339          & 0.110332          & 0.110357          & 0.110351          & 0.141883           & 0.157623          \\
        Total QPU delay time per sample (s)                & 0.000062          & 0.000103          & 0.000144          & 0.000144          & 0.000144          & 0.000144          & 0.000185           & 0.000206          \\
        Total post-processing time (s)                     & 0.000080          & 0.000005          & 0.000406          & 0.000114          & 0.000128          & 0.000113          & 0.000112           & 0.000301          \\
        \textbf{Total cache-and-read embedding   time (s)} & \textbf{0.815435} & \textbf{1.313140} & \textbf{1.473940} & \textbf{1.775991} & \textbf{1.229548} & \textbf{1.334796} & \textbf{2.530272}  & \textbf{0.455262} \\
        \textbf{Total solver communication   time (s)}     & \textbf{4.332876} & \textbf{6.487529} & \textbf{6.803421} & \textbf{7.032953} & \textbf{7.261384} & \textbf{6.872431} & \textbf{12.097399} & \textbf{9.158428}
        \end{tabular}
    }
    \caption{\textbf{Experimental computing times for the Barabasi-Albert networks and $\gamma=1$.} Running times across resolution parameters (see Methods). Bold rows indicate the subset summarizing overall recursive performance. "Total" is the sum of all substep durations within the recursion.}
\end{table}

\begin{table}[h]
    \begin{tabular}{l||c|c|c|c||r}
    Resolution                                & 0.5              & 0.75     & 1.25     & 1.5      & \multicolumn{1}{l}{Nodes} \\ \hline
    Total   QPU access time (s)               & 0.02939          & 0.029393 & 0.438861 & 0.775563 & \multirow{3}{*}{50}       \\
    Total cache-and-read embedding   time (s) & 0.13197          & 0.125277 & 0.685991 & 1.875352 &                           \\
    Total solver communication   time (s)     & 2.20161          & 2.152386 & 11.08863 & 20.46377 &                           \\ \hline
    Total   QPU access time (s)               & 0.03888          & 0.038878 & 0.623397 & 1.145491 & \multirow{3}{*}{80}       \\
    Total cache-and-read embedding   time (s) & 0.4043           & 0.409232 & 2.225447 & 1.221313 &                           \\
    Total solver communication   time (s)     & 2.44552 & 2.532007 & 11.58749 & 15.78885 &                           \\ \hline
    Total   QPU access time (s)               & 0.03983          & 0.03983  & 0.714871 & 1.352234 & \multirow{3}{*}{100}      \\
    Total cache-and-read embedding   time (s) & 0.31378          & 0.311219 & 2.587675 & 1.03531  &                           \\
    Total solver communication   time (s)     & 2.85523          & 3.034467 & 16.68066 & 16.17336 &                          
    \end{tabular}
\caption{\textbf{Experimental computing times Barabasi-Albert networks across different resolution parameters.} Running times across resolution parameters (see Methods)."Total" is the sum of all substep durations within the recursion.}
\end{table}

\begin{table}[h]
    \centering
    \resizebox{\columnwidth}{!}{%
        \begin{tabular}{l||c|c|c|c|c|c|c|c}
        Number of nodes                                    & 10                & 20                & 30                & 50                & 80                & 100               & 120               & 140               \\
Modularity                                         & 0.204082          & 0.164857          & 0.156025          & 0.130288          & 0.106642          & 0.109726          & 0.094313          & 0.087357          \\
Total QPU sampling time (s)                        & 0.035700          & 0.086848          & 0.105844          & 0.119666          & 0.142162          & 0.133570          & 0.145962          & 0.153868          \\
Total QPU anneal time per sample (s)               & 0.000060          & 0.000140          & 0.000140          & 0.000140          & 0.000140          & 0.000140          & 0.000140          & 0.000140          \\
Total QPU readout time per sample (s)              & 0.000235          & 0.000584          & 0.000774          & 0.000913          & 0.001138          & 0.001052          & 0.001176          & 0.001255          \\
\textbf{Total QPU access time (s)}                 & \textbf{0.082985} & \textbf{0.197195} & \textbf{0.216203} & \textbf{0.230025} & \textbf{0.252517} & \textbf{0.243930} & \textbf{0.256287} & \textbf{0.264223} \\
Total QPU access overhead time (s)                 & 0.004731          & 0.009224          & 0.008153          & 0.009745          & 0.012232          & 0.010596          & 0.011371          & \textbf{0.011914} \\
Total QPU programming time (s)                     & 0.047285          & 0.110347          & 0.110359          & 0.110359          & 0.110355          & 0.110360          & 0.110325          & 0.110355          \\
Total QPU delay time per sample (s)                & 0.000062          & 0.000144          & 0.000144          & 0.000144          & 0.000144          & 0.000144          & 0.000144          & 0.000144          \\
Total post-processing time (s)                     & 0.000030          & 0.000104          & 0.000007          & 0.000116          & 0.000042          & 0.000317          & 0.000093          & 0.000135          \\
\textbf{Total cache-and-read embedding   time (s)} & \textbf{0.108010} & \textbf{0.222964} & \textbf{0.336940} & \textbf{0.470641} & \textbf{0.634279} & \textbf{0.818109} & \textbf{0.184752} & \textbf{0.983994} \\
\textbf{Total solver communication   time (s)}     & \textbf{4.652850} & \textbf{7.061050} & \textbf{6.535298} & \textbf{6.668854} & \textbf{7.010201} & \textbf{6.827009} & \textbf{7.734851} & \textbf{7.010873}
        \end{tabular}
    }
    \caption{\textbf{Experimental computing times for the Erd\H{o}s-Renyi networks and $\gamma=1$.} Running times across resolution parameters (see Methods). Bold rows indicate the subset summarizing overall recursive performance. "Total" is the sum of all substep durations within the recursion.}
\end{table}

\begin{table}[h]
    \begin{tabular}{l||c|c|c|c||r}
    Resolution                                & 0.5              & 0.75     & 1.25     & 1.5      & \multicolumn{1}{l}{Nodes} \\ \hline
    Total   QPU access time (s)               & 0.02939          & 0.029388 & 0.330299 & 0.475241 & \multirow{3}{*}{50}       \\
    Total cache-and-read embedding   time (s) & 0.09356          & 0.086068 & 0.399439 & 0.836026 &                           \\
    Total solver communication   time (s)     & 3.77651          & 2.521583 & 10.08389 & 11.96028 &                           \\ \hline
    Total   QPU access time (s)               & 0.03887          & 0.038872 & 0.538843 & 0.73034  & \multirow{3}{*}{80}       \\
    Total cache-and-read embedding   time (s) & 0.27529          & 0.284979 & 1.674121 & 2.242139 &                           \\
    Total solver communication   time (s)     & 2.51419 & 2.605492 & 14.61886 & 20.78053 &                           \\ \hline
    Total   QPU access time (s)               & 0.03983          & 0.039832 & 0.494737 & 0.918594 & \multirow{3}{*}{100}      \\
    Total cache-and-read embedding   time (s) & 0.04732          & 0.052486 & 4.394353 & 0.913664 &                           \\
    Total solver communication   time (s)     & 2.66458          & 2.796214 & 9.762317 & 14.52301 &                          
    \end{tabular}
\caption{\textbf{Experimental computing times Erd\H{o}s-Renyi networks across different resolution parameters.} Running times across resolution parameters (see Methods)."Total" is the sum of all substep durations within the recursion.}
\end{table}

\begin{table}[h]
    \centering
    \resizebox{\columnwidth}{!}{%
        \begin{tabular}{l||c|c|c|c|c|c|c}
        Resolution                                         & 0.5               & 0.75               & 1                  & 1.25               & 1.5                & 1.75               & 2                  \\
        Modularity                                         & 0.654195          & 0.540177           & 0.444904           & 0.369982           & 0.298209           & 0.238243           & 0.186016            \\
        Total QPU sampling time (s)                        & 0.058968          & 0.084012           & 0.109100           & 0.109100           & 0.128392           & 0.128392           & 0.172253             \\
        Total QPU anneal time per sample (s)               & 0.000060          & 0.000100           & 0.000140           & 0.000140           & 0.000180           & 0.000180           & 0.000260            \\
        Total QPU readout time per sample (s)              & 0.000468          & 0.000637           & 0.000807           & 0.000807           & 0.000919           & 0.000919           & 0.001195            \\
        \textbf{Total QPU access time (s)}                 & \textbf{0.106251} & \textbf{0.162813}  & \textbf{0.219421}  & \textbf{0.219433}  & \textbf{0.270250}  & \textbf{0.270232}  & \textbf{0.377168}  \\
        Total QPU access overhead time (s)                 & 0.002773          & 0.005505           & 0.007915           & 0.007280           & 0.009618           & 0.010887           & 0.016028            \\
        Total QPU programming time (s)                     & 0.047283          & 0.078801           & 0.110321           & 0.110333           & 0.141858           & 0.141840           & 0.204915            \\
        Total QPU delay time per sample (s)                & 0.000062          & 0.000103           & 0.000144           & 0.000144           & 0.000185           & 0.000185           & 0.000268             \\
        Total post-processing time (s)                     & 0.000060          & 0.000062           & 0.000089           & 0.000090           & 0.000115           & 0.000126           & 0.000197            \\
        \textbf{Total cache-and-read embedding   time (s)} & \textbf{0.207172} & \textbf{0.614797}  & \textbf{0.746770}  & \textbf{0.749034}  & \textbf{1.989081}  & \textbf{1.273854}  & \textbf{3.788942}  \\
        \textbf{Total solver communication   time (s)}     & \textbf{7.136631} & \textbf{11.398048} & \textbf{16.256874} & \textbf{16.292811} & \textbf{20.882102} & \textbf{21.455918} & \textbf{30.844166}
        \end{tabular}
    }
    \caption{\textbf{Experimental computing times for the Karate network.} Running times across resolution parameters (see Methods). Bold rows indicate the subset summarizing overall recursive performance. "Total" is the sum of all substep durations within the recursion.}
\end{table}

\begin{table}[]
    \scriptsize
    \resizebox{\columnwidth}{!}{%
        \begin{tabular}{l||c|c|c|c|c|c|c|c|c|c}
        \textit{\textbf{Powerlaw}}           & \textbf{N=10} & \textbf{N=20} & \textbf{N=30} & \textbf{N=40} & \textbf{N=50} & \textbf{N=60} & \textbf{N=70} & \textbf{N=80} & \textbf{N=90} & \textbf{N=100} \\
        \hline
        Min Q                                & 0.425926      & 0.581717      & 0.629608      & 0.650230      & 0.660558      & 0.647084      & 0.747007      & 0.738103      & 0.737218      & 0.769003       \\
        Max Q                                & 0.425926      & 0.581717      & 0.629608      & 0.650230      & 0.660558      & 0.660011      & 0.750788      & 0.748838      & 0.756407      & 0.783083       \\
        Frequency Max Q                      & 20            & 20            & 20            & 20            & 20            & 19            & 8             & 19            & 6             & 7              \\
        \hline
        \textit{\textbf{N resamples =   20}} &               &               &               &               &               &               &               &               &               &                \\
        Arg Max LH                           & 0.425926      & 0.581717      & 0.629608      & 0.650230      & 0.660558      & 0.660011      & 0.747423      & 0.748838      & 0.754748      & 0.782148       \\
        LH Max Q (\%)                        & 100.000       & 100.000       & 100.000       & 100.000       & 100.000       & 94.940        & 39.990        & 94.650        & 29.580        & 34.540         \\
        Max LH (\%)                          & 100.000       & 100.000       & 100.000       & 100.000       & 100.000       & 94.940        & 61.910        & 94.650        & 30.680        & 34.980         \\
        \hline
        \textit{\textbf{N resamples = 100}}  &               &               &               &               &               &               &               &               &               &                \\
        Arg Max LH                           & 0.425926      & 0.581717      & 0.629608      & 0.650230      & 0.660558      & 0.660011      & 0.747060      & 0.748838      & 0.756398      & 0.783083       \\
        LH Max Q (\%)                        & 100.000       & 100.000       & 100.000       & 100.000       & 100.000       & 94.960        & 39.782        & 95.002        & 30.118        & 34.846         \\
        Max LH (\%)                          & 100.000       & 100.000       & 100.000       & 100.000       & 100.000       & 94.960        & 60.314        & 95.002        & 30.118        & 34.846         \\
        \hline
        \textit{\textbf{N resamples = 500}}  & \textbf{}     & \textbf{}     & \textbf{}     & \textbf{}     & \textbf{}     & \textbf{}     & \textbf{}     & \textbf{}     & \textbf{}     & \textbf{}      \\
        Arg Max LH                           & 0.425926      & 0.581717      & 0.629608      & 0.650230      & 0.660558      & 0.660011      & 0.747007      & 0.748838      & 0.756407      & 0.783083       \\
        LH Max Q (\%)                        & 100.000       & 100.000       & 100.000       & 100.000       & 100.000       & 95.049        & 39.908        & 94.958        & 30.063        & 35.066         \\
        Max LH (\%)                          & 100.000       & 100.000       & 100.000       & 100.000       & 100.000       & 95.049        & 60.092        & 94.958        & 30.063        & 35.066         \\
        \hline
        \textit{\textbf{N resamples = 1000}} & \textbf{}     & \textbf{}     & \textbf{}     & \textbf{}     & \textbf{}     & \textbf{}     & \textbf{}     & \textbf{}     & \textbf{}     & \textbf{}      \\
        Arg Max LH                           & 0.425926      & 0.581717      & 0.629608      & 0.650230      & 0.660558      & 0.660011      & 0.747007      & 0.748838      & 0.756407      & 0.783083       \\
        LH Max Q (\%)                        & 100.000       & 100.000       & 100.000       & 100.000       & 100.000       & 94.991        & 39.997        & 95.067        & 30.013        & 34.991         \\
        Max LH (\%)                          & 100.000       & 100.000       & 100.000       & 100.000       & 100.000       & 94.991        & 60.003        & 95.067        & 30.013        & 34.991        
        \end{tabular}
    }
    \caption{\textbf{Sampling characteristics of the hierarchical annealing algorithm for power-law clustered networks.} After running the proposed procedure 20 times independently, we recorded $\min Q$, $\max Q$, and the frequency with which the maximum modularity was observed. We then applied bootstrapping to assess whether the estimated likelihood of observing $\max Q$ remained stable as a function of the bootstrap sample size. This approach provides a proxy for the distribution that would be obtained if the hierarchical annealing algorithm were run an arbitrary number of times. Notably, the most likely modularity did not differ substantially from the maximum value, indicating robust behavior of the annealing procedure.}
    \label{tab:powerlaw}
\end{table}

\begin{table}[]
    \scriptsize
    \resizebox{\columnwidth}{!}{%
        \begin{tabular}{l||c|c|c|c|c|c|c|c|c|c}
            \textit{\textbf{Erd\H os-Renyi}}        & \textbf{N=10}        & \textbf{N=20}        & \textbf{N=30}        & \textbf{N=40}        & \textbf{N=50}        & \textbf{N=60}        & \textbf{N=70}        & \textbf{N=80}        & \textbf{N=90}        & \textbf{N=100}       \\
            \hline
Min Q                                & 0.140306             & 0.185249             & 0.173824             & 0.150733             & 0.135665             & 0.114545             & 0.113360             & 0.106708             & 0.097805             & 0.096991             \\
Max Q                                & 0.140306             & 0.185249             & 0.173824             & 0.150733             & 0.135665             & 0.121331             & 0.122608             & 0.115098             & 0.103103             & 0.102702             \\
Frequency Max Q                      & 20                   & 20                   & 20                   & 20                   & 20                   & 2                    & 1                    & 1                    & 1                    & 1                    \\
\hline
\textit{\textbf{N resamples =   20}} & \textbf{}            & \textbf{}            & \textbf{}            & \textbf{}            & \textbf{}            & \textbf{}            & \textbf{}            & \textbf{}            & \textbf{}            & \textbf{}            \\
Arg Max LH                           & 0.140306             & 0.185249             & 0.173824             & 0.150733             & 0.135665             & 0.117207             & 0.120152             & 0.109206             & 0.100496             & 0.098548             \\
LH Max Q (\%)                        & 100.000              & 100.000              & 100.000              & 100.000              & 100.000              & 9.940                & 5.100                & 5.320                & 4.810                & 4.750                \\
Max LH (\%)                          & 100.000              & 100.000              & 100.000              & 100.000              & 100.000              & 24.770               & 45.480               & 17.960               & 17.270               & 16.430               \\
\hline
\textit{\textbf{N resamples = 100}}  &                      &                      &                      &                      &                      &                      &                      &                      &                      &                      \\
Arg Max LH                           & 0.140306             & 0.185249             & 0.173824             & 0.150733             & 0.135665             & 0.117238             & 0.120111             & 0.109302             & 0.101220             & 0.098850             \\
LH Max Q (\%)                        & 100.000              & 100.000              & 100.000              & 100.000              & 100.000              & 9.940                & 5.100                & 5.320                & 4.810                & 4.750                \\
Max LH (\%)                          & 100.000              & 100.000              & 100.000              & 100.000              & 100.000              & 24.770               & 45.480               & 17.960               & 17.270               & 16.430               \\
\hline
\textit{\textbf{N resamples = 500}}  & \textbf{}            & \textbf{}            & \textbf{}            & \textbf{}            & \textbf{}            & \textbf{}            & \textbf{}            & \textbf{}            & \textbf{}            & \textbf{}            \\
Arg Max LH                           & 0.140306             & 0.185249             & 0.173824             & 0.150733             & 0.135665             & 0.116864             & 0.119993             & 0.109294             & 0.101732             & 0.098998             \\
LH Max Q (\%)                        & 100.000              & 100.000              & 100.000              & 100.000              & 100.000              & 9.882                & 4.959                & 4.963                & 5.010                & 5.032                \\
Max LH (\%)                          & 100.000              & 100.000              & 100.000              & 100.000              & 100.000              & 19.986               & 40.221               & 10.728               & 10.033               & 6.971                \\
\hline
\textit{\textbf{N resamples = 1000}} & \textbf{}            & \textbf{}            & \textbf{}            & \textbf{}            & \textbf{}            & \textbf{}            & \textbf{}            & \textbf{}            & \textbf{}            & \textbf{}            \\
Arg Max LH                           & 0.140306             & 0.185249             & 0.173824             & 0.150733             & 0.135665             & 0.116836             & 0.119941             & 0.109308             & 0.101759             & 0.098933             \\
LH Max Q (\%)                        & 100.000              & 100.000              & 100.000              & 100.000              & 100.000              & 9.912                & 5.222                & 4.958                & 5.124                & 5.014                \\
Max LH (\%)                          & 100.000              & 100.000              & 100.000              & 100.000              & 100.000              & 21.026               & 41.304               & 12.094               & 11.056               & 9.542              
        \end{tabular}
    }
    \caption{\textbf{Sampling characteristics of the hierarchical annealing algorithm for Erd\H os-Renyi networks.} After running the proposed procedure 20 times independently, we recorded $\min Q$, $\max Q$, and the frequency with which the maximum modularity was observed. We then applied bootstrapping to assess whether the estimated likelihood of observing $\max Q$ remained stable as a function of the bootstrap sample size. This approach provides a proxy for the distribution that would be obtained if the hierarchical annealing algorithm were run an arbitrary number of times. Notably, the most likely modularity did not differ substantially from the maximum value, indicating robust behavior of the annealing procedure.}
    \label{tab:erdosrenyi}
\end{table}

\begin{table}[]
    \scriptsize
    \resizebox{\columnwidth}{!}{%
        \begin{tabular}{l||c|c|c|c|c|c|c|c|c|c}
            \textit{\textbf{Barabasi-Albert}}    & \textbf{N=10}        & \textbf{N=20}        & \textbf{N=30}        & \textbf{N=40}        & \textbf{N=50}        & \textbf{N=60}        & \textbf{N=70}        & \textbf{N=80}        & \textbf{N=90}        & \textbf{N=100}       \\
\hline
Min Q                                & 0.044271             & 0.088162             & 0.083998             & 0.097280             & 0.089488             & 0.071198             & 0.072995             & 0.065422             & 0.063278             & 0.058137             \\
Max Q                                & 0.044271             & 0.088162             & 0.083998             & 0.097280             & 0.092511             & 0.074836             & 0.076799             & 0.071529             & 0.068596             & 0.066951             \\
Frequency Max Q                      & 20                   & 20                   & 20                   & 20                   & 1                    & 1                    & 1                    & 1                    & 1                    & 1                    \\
\hline
\textit{\textbf{N resamples =   20}} &\textbf{}            & \textbf{}            & \textbf{}            & \textbf{}            & \textbf{}            & \textbf{}            & \textbf{}            & \textbf{}            & \textbf{}            & \textbf{} \\
Arg Max LH                           & 0.044271             & 0.088162             & 0.083998             & 0.097280             & 0.090154             & 0.074397             & 0.074165             & 0.067249             & 0.065310             & 0.061155             \\
LH Max Q (\%)                        & 100.000              & 100.000              & 100.000              & 100.000              & 5.190                & 2.500                & 5.030                & 4.590                & 5.180                & 5.310                \\
Max LH (\%)                          & 100.000              & 100.000              & 100.000              & 100.000              & 80.200               & 29.250               & 21.530               & 16.130               & 16.290               & 15.940               \\
\hline
\textit{\textbf{N resamples = 100}}  &                      &                      &                      &                      &                      &                      &                      &                      &                      &                      \\
Arg Max LH                           & 0.044271             & 0.088162             & 0.083998             & 0.097280             & 0.090154             & 0.074461             & 0.074346             & 0.067513             & 0.065519             & 0.061623             \\
LH Max Q (\%)                        & 100.000              & 100.000              & 100.000              & 100.000              & 4.998                & 2.454                & 4.928                & 5.000                & 5.180                & 5.042                \\
Max LH (\%)                          & 100.000              & 100.000              & 100.000              & 100.000              & 80.064               & 25.878               & 17.076               & 9.552                & 9.610                & 9.630                \\
\hline
\textit{\textbf{N resamples = 500}}  & \textbf{}            & \textbf{}            & \textbf{}            & \textbf{}            & \textbf{}            & \textbf{}            & \textbf{}            & \textbf{}            & \textbf{}            & \textbf{}            \\
Arg Max LH                           & 0.044271             & 0.088162             & 0.083998             & 0.097280             & 0.090154             & 0.074465             & 0.074361             & 0.067594             & 0.065651             & 0.061721             \\
LH Max Q (\%)                        & 100.000              & 100.000              & 100.000              & 100.000              & 5.010                & 2.448                & 4.998                & 4.998                & 4.942                & 4.955                \\
Max LH (\%)                          & 100.000              & 100.000              & 100.000              & 100.000              & 80.126               & 24.929               & 15.932               & 6.942                & 6.952                & 6.982                \\
\hline
\textit{\textbf{N resamples = 1000}} & \textbf{}            & \textbf{}            & \textbf{}            & \textbf{}            & \textbf{}            & \textbf{}            & \textbf{}            & \textbf{}            & \textbf{}            & \textbf{}            \\
Arg Max LH                           & 0.044271             & 0.088162             & 0.083998             & 0.097280             & 0.090154             & 0.074466             & 0.074385             & 0.067766             & 0.065734             & 0.062008             \\
LH Max Q (\%)                        & 100.000              & 100.000              & 100.000              & 100.000              & 5.031                & 2.501                & 5.016                & 5.050                & 5.002                & 4.970                \\
Max LH (\%)                          & 100.000              & 100.000              & 100.000              & 100.000              & 80.041               & 24.930               & 15.721               & 6.353                & 6.379                & 6.386               
        \end{tabular}
    }
    \caption{\textbf{Sampling characteristics of the hierarchical annealing algorithm for Barabasi-Albert.} After running the proposed procedure 20 times independently, we recorded $\min Q$, $\max Q$, and the frequency with which the maximum modularity was observed. We then applied bootstrapping to assess whether the estimated likelihood of observing $\max Q$ remained stable as a function of the bootstrap sample size. This approach provides a proxy for the distribution that would be obtained if the hierarchical annealing algorithm were run an arbitrary number of times. Notably, the most likely modularity did not differ substantially from the maximum value, indicating robust behavior of the annealing procedure.}
    \label{tab:barabasialbert}
\end{table}

\begin{table}[]
    \scriptsize
    \resizebox{\columnwidth}{!}{%
        \begin{tabular}{l||c|c|c|c|c|c|c}
            \textit{\textbf{Directed   scale-free}} & \textbf{N=10}        & \textbf{N=20}        & \textbf{N=30}        & \textbf{N=40}        & \textbf{N=50}        & \textbf{N=60}        & \textbf{N=70}        \\
            \hline
Min Q                                   & 0.000000             & 0.327977             & 0.294005             & 0.297133             & 0.265518             & 0.365142             & 0.382186             \\
Max Q                                   & 0.375000             & 0.327977             & 0.294005             & 0.297133             & 0.265518             & 0.368195             & 0.391541             \\
Frequency Max Q                         & 19                   & 20                   & 20                   & 40                   & 40                   & 11                   & 1                    \\
\hline
\textit{\textbf{N resamples =   20}}    & \textbf{}            & \textbf{}            & \textbf{}            & \textbf{}            & \textbf{}            & \textbf{}            & \textbf{} \\
Arg Max LH                              & 0.375000             & 0.327977             & 0.294005             & 0.297133             & 0.265518             & 0.365160             & 0.389130             \\
LH Max Q (\%)                           & 94.900               & 100.000              & 100.000              & 100.000              & 100.000              & 27.030               & 2.630                \\
Max LH (\%)                             & 94.900               & 100.000              & 100.000              & 100.000              & 100.000              & 70.610               & 94.870               \\
\hline
\textit{\textbf{N resamples = 100}}     &                      &                      &                      &                      &                      &                      &                      \\
Arg Max LH                              & 0.375000             & 0.327977             & 0.294005             & 0.297133             & 0.265518             & 0.365142             & 0.389130             \\
LH Max Q (\%)                           & 94.886               & 100.000              & 100.000              & 100.000              & 100.000              & 27.946               & 2.580                \\
Max LH (\%)                             & 94.886               & 100.000              & 100.000              & 100.000              & 100.000              & 69.534               & 94.944               \\
\hline
\textit{\textbf{N resamples = 500}}     & \textbf{}            & \textbf{}            & \textbf{}            & \textbf{}            & \textbf{}            & \textbf{}            & \textbf{}            \\
Arg Max LH                              & 0.375000             & 0.327977             & 0.294005             & 0.297133             & 0.265518             & 0.365142             & 0.389130             \\
LH Max Q (\%)                           & 95.036               & 100.000              & 100.000              & 100.000              & 100.000              & 27.391               & 2.506                \\
Max LH (\%)                             & 95.036               & 100.000              & 100.000              & 100.000              & 100.000              & 70.089               & 94.989               \\
\hline
\textit{\textbf{N resamples = 1000}}    & \textbf{}            & \textbf{}            & \textbf{}            & \textbf{}            & \textbf{}            & \textbf{}            & \textbf{}            \\
Arg Max LH                              & 0.375000             & 0.327977             & 0.294005             & 0.297133             & 0.265518             & 0.365142             & 0.389130             \\
LH Max Q (\%)                           & 95.024               & 100.000              & 100.000              & 100.000              & 100.000              & 27.483               & 2.520                \\
Max LH (\%)                             & 95.024               & 100.000              & 100.000              & 100.000              & 100.000              & 70.034               & 95.002              
\end{tabular}
    }
    \caption{\textbf{Sampling characteristics of the hierarchical annealing algorithm for directed scale-free networks.} After running the proposed procedure 20 times independently, we recorded $\min Q$, $\max Q$, and the frequency with which the maximum modularity was observed. We then applied bootstrapping to assess whether the estimated likelihood of observing $\max Q$ remained stable as a function of the bootstrap sample size. This approach provides a proxy for the distribution that would be obtained if the hierarchical annealing algorithm were run an arbitrary number of times. Notably, the most likely modularity did not differ substantially from the maximum value, indicating robust behavior of the annealing procedure. For networks of size 40, 50, 60, and 70, we set \texttt{num\_runs}$ = 40$.}
    \label{tab:directedscalefree}
\end{table}

\end{document}